\documentclass[reprint,superscriptaddress,amsmath,amssymb,aps,pre,nofootinbib]{revtex4-2}
\usepackage{graphicx}
\usepackage{dcolumn}
\usepackage{bm}
\usepackage{xcolor}
\usepackage{units}
\usepackage{wasysym}
\usepackage{MnSymbol}
\usepackage{comment}
\usepackage{times}
\usepackage{amsmath}
\usepackage{upgreek}
\usepackage{amssymb}
\usepackage{float}
\usepackage[unicode=true,pdfusetitle,bookmarks=false,colorlinks=true,citecolor=blue,urlcolor=blue,linkcolor=blue]{hyperref}
\begin{document}

\title{Dynamic Permeability in Metastable Droplet Interfacial Bilayers}

\author{Nivedina A. Sarma}
\thanks{These authors contributed equally to this work}
\affiliation{Department of Materials Science and Engineering, University of California, Berkeley, California 94720, USA}
\affiliation{Materials Sciences Division, Lawrence Berkeley National Laboratory, Berkeley, California, 94720, USA}
\thanks{These authors contributed equally to this work}

\author{David A. King}
\thanks{These authors contributed equally to this work}
\affiliation{Department of Materials Science and Engineering, University of California, Berkeley, California 94720, USA}
\affiliation{Materials Sciences Division, Lawrence Berkeley National Laboratory, Berkeley, California, 94720, USA}

\author{Xuefei Wu}
\affiliation{Materials Sciences Division, Lawrence Berkeley National Laboratory, Berkeley, California, 94720, USA}

\author{Brett A. Helms}
\affiliation{The Molecular Foundry, Lawrence Berkeley National Laboratory, Berkeley, California, 94720, USA}
\affiliation{Materials Sciences Division, Lawrence Berkeley National Laboratory, Berkeley, California, 94720, USA}

\author{Paul D. Ashby}
\affiliation{The Molecular Foundry, Lawrence Berkeley National Laboratory, Berkeley, California, 94720, USA}
\affiliation{Materials Sciences Division, Lawrence Berkeley National Laboratory, Berkeley, California, 94720, USA}

\author{Thomas P. Russell}
\affiliation{Polymer Science and Engineering Department, University of Massachusetts, Amherst 01003, USA}
\affiliation{Materials Sciences Division, Lawrence Berkeley National Laboratory, Berkeley, California, 94720, USA}

\author{Ahmad K. Omar}
\email{aomar@berkeley.edu}
\affiliation{Department of Materials Science and Engineering, University of California, Berkeley, California 94720, USA}
\affiliation{Materials Sciences Division, Lawrence Berkeley National Laboratory, Berkeley, California, 94720, USA}

\begin{abstract}
Membrane pores are implicated in several critical functions, including cell fusion and the transport of signaling molecules for intercellular communication. 
However, these structural features are often difficult to probe directly.
Droplet interfacial bilayers offer a synthetic platform to study such membrane properties.
We develop a theory that links  size-selective transport across a metastable membrane with its transient structural properties.
The central quantity of our theory is a dynamic permeability that depends on the mechanism of pore growth, which controls the transient distribution of pore sizes in the membrane.
We present a mechanical perspective to derive pore growth dynamics and the resulting size distribution for growth \textit{via} Ostwald ripening and discuss how these dynamics compare to other growth mechanisms such as coalescence and growth through surfactant desorption.
We find scaling relations between the transported particle size, the pore growth rate, and the time for a given fraction of particles to cross the membrane, from which one may deduce the dominant mechanism of pore growth, as well as material properties and structural features of the membrane.
Finally, we suggest experiments using droplet interfacial bilayers to validate our theoretical predictions. 
\end{abstract}

\maketitle

\section{Introduction}
When two droplets that are each coated with a tightly packed surfactant monolayer are brought together, a \textit{bilayer} is formed at the interface.
These droplet interfacial bilayers (DIBs) offer synthetic platforms for studying structural and dynamic properties of bilayer membranes~\cite{Tsofina1966ProductionSolution, Hwang2008AsymmetricBilayers, Holden2007FunctionalDroplets, Mruetusatorn2014DynamicBilayers, Guiselin2018DynamicBilayers, Bird2009CriticalDroplets} due to their resemblance to the lipid bilayers that regulate communication and transport between distinct chemical environments in biological systems~\cite{Almeida2005ThermodynamicsDomains, Yang2016LineFusion, Phillips2012PhysicalCell}. 
In contrast to other model systems commonly used in membrane research, such as giant unilamellar vesicles~\cite{Moscho1996RapidVesicles, Wesoowska2009GiantSystems, Kahya2010Protein-proteinVesicles, Jrgensen2017MembraneTechniques}, black lipid membranes~\cite{Bach1980GlycerylSolutions, Hladky1982ThicknessMembranesb, Horn2005PhotocurrentsMembranesb, Nomoto2018EffectsMembranesb}, and supported lipid bilayers~\cite{Steinem1996ImpedanceTechniquesb, Richter2006FormationViewb, Seu2007EffectBilayersb, ElKirat2010NanoscaleMicroscopyb, Cheney2017SingleCurvatureb, Glazier2017SupportedMechanobiologyb}, DIBs often offer more control over bilayer composition and geometry, and can be more easily probed to access a suite of membrane properties.
As such, DIBs represent a promising platform for obtaining valuable insight into lipid bilayer properties.

Over the last 50 years, DIBs have evolved from a simple experimental setup into a widely used model system to study protein functions and interfacial stability across length scales.
Tsofina~et al. first introduced DIBs as artificial protein--phospholipid bilayers~\cite{Tsofina1966ProductionSolution} and measured capacitance and resistance across membranes for different lipid compositions.
Since then, to better model biological systems, much work has been done to create DIBs that approach cellular length scales~\cite{Mruetusatorn2014DynamicBilayers} and identify why small DIBs are often short-lived~\cite{Bird2009CriticalDroplets, Guiselin2018DynamicBilayers}.
As simplified biological membranes, DIBs have been used to study membrane proteins in controlled environments.
For instance, proteins such as ion channels, which selectively allow charged species to cross lipid bilayers, can be reconstituted in DIBs~\cite{Funakoshi2006LipidAnalysis,Hwang2008AsymmetricBilayers}.
This allows them to be electrically characterized with single-channel recordings without the interference from other proteins that would occur in native membrane environments~\cite{Heron2007DirectBilayers, Holden2007FunctionalDroplets}.
Networks of larger droplets have even been used to model biological circuits such as the sinoatrial node (the cluster of cells that initiates electrical signals for heartbeats) as each of the DIBs can propagate electrical signals between droplets~\cite{Bayley2008DropletBilayers, Maglia2009DropletPropertiesb, Stephenson2022ChallengesBilayers}.
Understanding species transport and how membrane structure affects signal propagation could further enable the design of more sophisticated DIB systems.

Each of the examples noted above consisted of DIBs comprised of molecular surfactants. 
The resulting DIBs are somewhat delicate. 
They can rupture rapidly, within minutes or seconds~\cite{Villar2011FormationEnvironmentsb}, and even if they are stable for several days~\cite{Bayley2008DropletBilayers}, the high fluidity and mobility of the constituent lipids reduces their stability in the face of in-plane stresses or applied electric fields~\cite{Jalali2024ExplorationSimulation,Huang2024InterfacialMicroscopy}.
In contrast, DIBs formed with nanoparticle (or colloidal) surfactants exhibit longer lifetimes and greater stability to external perturbation, as their larger size leads to higher binding energies and slower desorption kinetics and in-plane diffusion compared to smaller molecular surfactants~\cite{Forth2018ReconfigurableLiquidsb, Huang2017Self-RegulatedLiquidsb, Toor2017EffectDropletsb, Wu2023BallisticAssemblies, Yang2022ReconfigurableLiquidsb}.
Nanoparticle surfactant DIBs may therefore offer the same tunability and ease of measurement as molecular DIBs, while additionally providing a new platform for studying structural dynamics and transport over longer timescales.
Indeed, recent work has explored how larger surfactants consisting of co-assembled microgels and polymer can be used to make reconfigurable DIB networks that support size-selective adsorption and permeability~\cite{Guan2025DynamicInterfaces}.
However, while nanoparticle surfactants may represent a  new frontier for longer-lived DIBs, it remains challenging to directly and non-destructively probe the structural features of a bilayer between two droplets~\cite{Fink2024MixedInterfaceb,Wu2024OversaturatingNanoparticleSurfactantsb, Chai2020DirectInterfacec}.

\begin{figure}
\includegraphics[width=.5\textwidth]{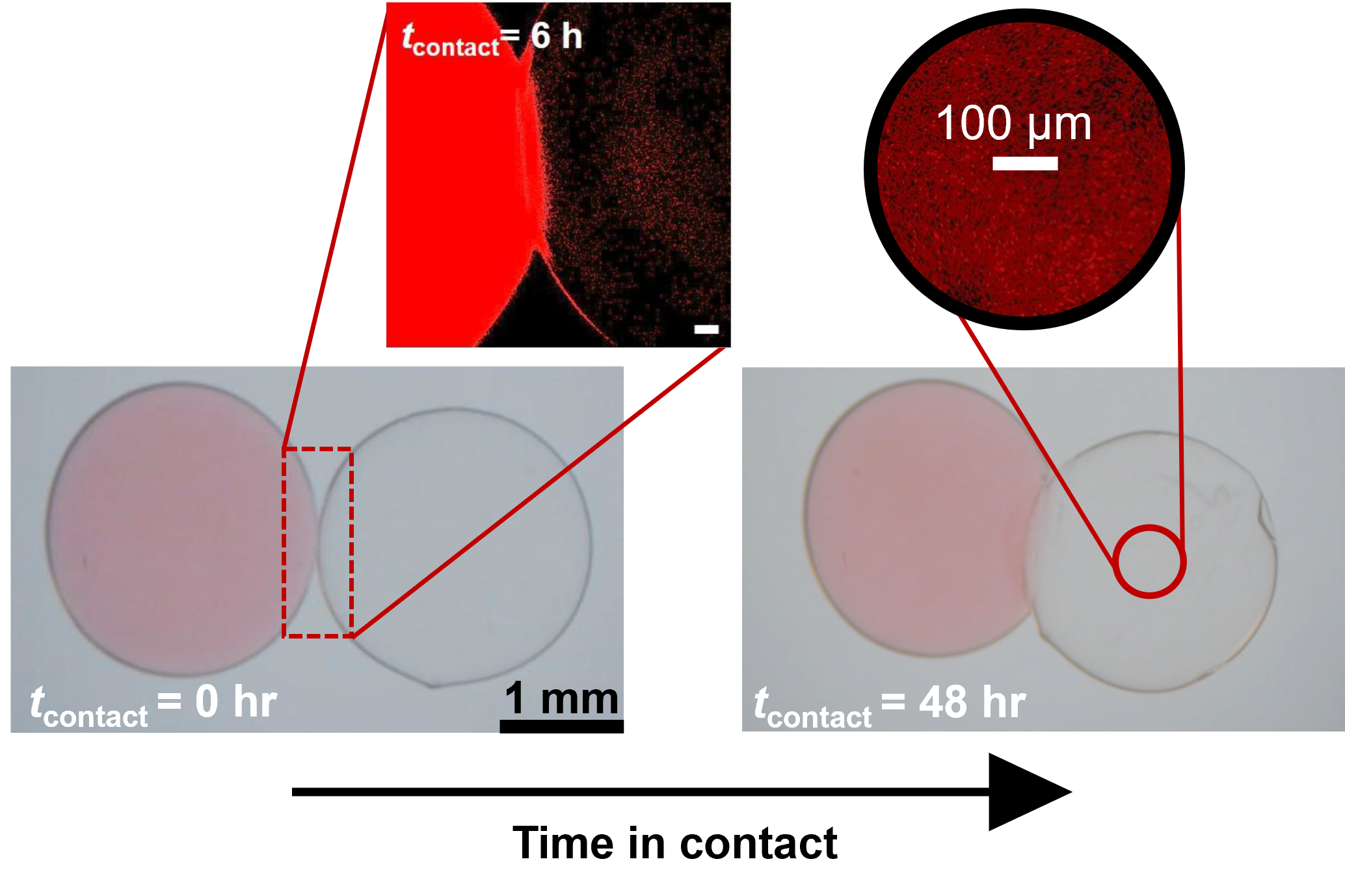}
\caption{Dye transport across a nanoparticle--surfactant droplet interface bilayer (NPS--DIB)~\cite{Wu2025Submitted}.
Optical images show two aqueous droplets—one containing Rhodamine~6G dye (pink) and one initially dye-free—at the moment of contact ($t_{\rm{contact}} = 0~\rm{hr}$) and after 48~hours. 
Dye transport across the interfacial barrier is slow; little is visible in the initially empty droplet by eye after two days. 
Confocal fluorescence microscopy confirms that dye transfer does occur, as shown by the red emission signal detected in the receiving droplet (upper right inset). 
Imaging the droplet junction after six hours of contact (upper left inset, scale bar $100~\upmu \rm m$) shows a bridge connecting the droplets, consistent with the formation of a dense NPS DIB that impedes but does not completely block transport.
}
\label{fig:ExpSchematic}
\end{figure}

In some of our recent experiments, we have begun to address this challenge by directly monitoring species transport through nanoparticle--surfactant DIBs~\cite{Wu2025Submitted}. 
In these studies, aqueous droplets containing silica nanoparticles were dispersed in a silicone oil phase containing amine-functionalized surfactants. 
Nanoparticle--surfactant (NPS) complexes adsorb at the oil--water interface, producing mechanically robust, jammed interfacial layers whose surface coverage increased with droplet aging. 
The two such droplets, aged for one hour, were brought into contact, but were not observed to fuse into a single droplet, even after days.
Introducing a fluorescent dye (Rhodamine~6G) into one droplet and tracking its fluorescence across the pair revealed slow but measurable transport. 
These findings suggest that the interfacial layers of the two droplets merged on contact to form a stable NPS bilayer that acted as a mechanical barrier yet remained partially permeable. 
Figure~\ref{fig:ExpSchematic} shows optical and laser scanning confocal microscopy images of the droplets when they are first brought into contact and after two days. 
Even after this long wait, little dye is visible in the initially empty droplet to the naked eye, yet confocal fluorescence microscopy confirms that transfer occurs. 
Imaging the junction between droplets after six hours shows a distinct bridge connecting them which, given how strongly it impedes transport, is consistent with a dense NPS bilayer. 

To further probe this transport, one droplet containing poly(ethylene glycol)-functionalized gold nanoparticles (Au--PEG) of defined sizes was placed atop an initially particle-free aqueous phase to form a DIB~\cite{Wu2025Submitted}. 
Using \textit{in situ} UV-vis absorbance spectroscopy to monitor the gold plasmonic signal in the acceptor phase revealed strongly size- and time-dependent permeability: smaller nanoparticles ($\approx$1.8~nm) crossed the bilayer within hours, whereas larger ones (5--10~nm) required days.

\begin{figure*} 
	\centering
	\includegraphics[width=1\textwidth]{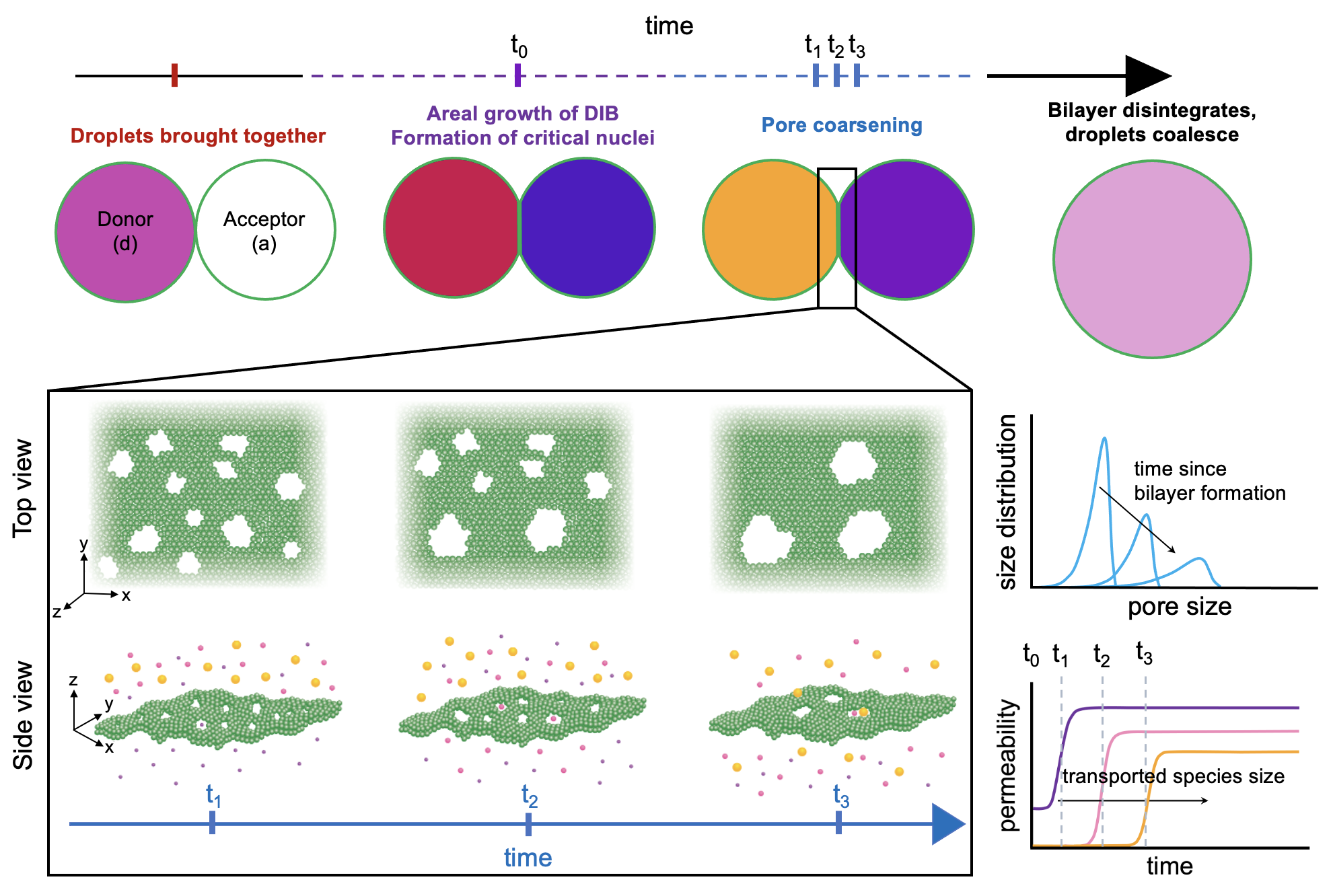}
	\caption{Schematic of a DIB formed between two droplets each coated with a surfactant monolayer (green) and with one droplet containing dye particles of different sizes.
    Droplets are brought together and a DIB forms upon contact.
    Over time, the DIB grows and critical nuclei form.
    The smallest dye particles cross the bilayer through these small pores.
    As the pores grow, the membrane becomes permeable to larger particles.
    Pore growth causes the bilayer to degrade until, eventually, the two droplets coalesce.
    The zoomed-in panel depicts pore growth through a mechanism that conserves the total pore area faction (top and side view are of the same time points).
    The top panel highlights that the pore distribution shifts to larger pores.
    The bottom panel emphasizes that the membrane becomes permeable to larger particles over time.
    }
    \label{fig:modelsystem}
\end{figure*}

In this Article, we present a theoretical framework that links readily measurable DIB transport properties to the underlying bilayer structure.
We do so by describing the nucleation and growth of a population of pores within the membrane and explicitly couple this to solute exchange across the membrane by introducing a dynamic permeability. 
Our perspective is based on the view that DIBs—whether formed from molecular or nanoparticle-based surfactants—are inherently metastable membranes that will ultimately rupture as droplets coalesce either of their own accord or when driven by external fields~\cite{Glaser1988ReversiblePores,Zhelev1993Tension-stabilizedTension} (see Fig.~\ref{fig:modelsystem}). 
In this picture, the same thermal pore nucleation events that initiate coalescence also act as the primary conduits for species transport, leading to enhanced permeability compared to previously studied insertion-based mechanisms~\cite{Marqusee1986SoluteMicellesb, Mitragotri1999AnBilayers, Xiang1997PermeabilityTheoryb}. 
By measuring size-selective transport, one can therefore infer key structural features of the membrane, including the pore number and size distribution, and how these evolve with time. 
In this way, while previous work has probed membrane structure to understand transport, our framework reverses this logic by using transport to reveal membrane structural evolution.

We begin in Section~\ref{sec:ModelSystem} by presenting a simple model system to probe the transport of ``dye'' particles from one droplet to another, across a DIB.
There, we define the \textit{dynamic permeability} of the membrane in terms of the distribution of its pores. 
To understand how these pores grow, we introduce a mechanical picture of the in-plane bilayer dynamics in Section~\ref{sec:InPlane}.
We then apply this framework to a canonical mechanism of domain growth, Ostwald ripening, and re-derive the classical results of Lifshitz, Slyozov and Wagner (LSW)~\cite{Lifshitz1961TheSolutions, Wagner1961TheorieOstwaldReifung}, modifying them appropriately for our finite, two-dimensional system.
We consider that pores may also grow through mechanisms such as coalescence and surfactant desorption and briefly discuss the timescales of these processes.
We conclude that the typical behavior for coalescence is similar to Ostwald ripening, while for desorption it is more relevant to inherently unstable bilayers that disappear rapidly.

In Section~\ref{sec:Perm} we solve our model to find the dynamic permeability and dye density for the Ostwald ripening mechanism.
These results establish clear, experimentally testable predictions.
In particular, we find that for dye particles of different sizes passing through the same membrane, the time for a fraction of the dye to be transported scales with the \textit{square} of the dye size, indicating strong size selectivity of the membrane.
Section~\ref{sec:Exp} discusses possible experimental tests of these predictions before we present our conclusions in Section~\ref{sec:Conc}.
In comparing the pore growth and dye transport timescales, our theory is general to DIBs made with either molecular or nanoparticle surfactants.
We find that tracking the membrane permeability over time not only offers fundamental insight into membrane structure, but may also inform design principles for creating responsive membranes.

\section{Model System}
\label{sec:ModelSystem}
Consider the situation shown in Fig.~\ref{fig:modelsystem} where two droplets, each coated with surfactants (shown as green spherical particles), make contact. 
This initial contact is frequently made in experiments by slowly pushing the droplets towards one another using pipette tips~\cite{Bayley2008DropletBilayers}. 
The monolayer surfaces flatten upon contact to form an approximately planar bilayer.
It is natural to expect that the two droplets will ultimately coalesce to minimize the ratio of surface area to volume (and thus the overall surface tension) in the system. 
For this fusion to occur, the bilayer must rupture and dissolve.
Such a process suggests that the bilayer is not thermodynamically stable.
Rather, it is likely a \textit{metastable} membrane.
Inspired by studies of bilayer fusion in biological systems~\cite{Kozlov1989EuropeanContents, Kozlovsky2002LipidDiaphragm, Kozlovsky2002StalkCrisis, Markin1984OnMechanism, Markin2002MembraneRevisited, Ting2011MinimumRupture}, we expect that droplet coalescence occurs predominantly through the formation of pores in the bilayer interior.
These pores nucleate with a critical radius $R^{\dagger}$, set by the material properties of the DIB, and subsequently grow as the membrane collapses~\cite{Frolov2006EntropicNanodomains, Ting2011MinimumRupture}.
The presence of pores provides passage across the membrane to large particles that are otherwise unable to overcome the insertion barrier. 
Our goal is to understand how the growth dynamics of these pores affect the transport of species from one droplet to the other, by means of a simplified model system.

Our model system involves two spherical droplets of equal and constant volume, $V$.
One, the ``donor'', is initially loaded with a mixture of dye particles of several different species $\alpha$ each with a different radius $r^{\alpha}$.
We take the initial number densities of different species in the donor to be ${\rho_d^{\alpha}(t=0) = 2\rho^{\alpha}}$, with the factor of two introduced for later convenience.
This initial difference in dye concentration will of course lead to differences in the osmotic pressure between the donor and acceptor droplets.
Assuming that the DIB is more permeable to the solvent than to the dye, solvent will transfer from the acceptor to the donor, resulting in swelling (contraction) of the donor (acceptor) before the dye has had time to cross the bilayer.
We neglect these volume changes, which will eventually reverse as the dye concentration increases in the acceptor, under the assumption that the initial dye concentration in the acceptor is sufficiently small.

To probe the effects of the growing pores, we suppose that the dye particles are larger than the initial pores, ${r^{\alpha} > R^{\dagger}}$, so that the DIB is initially impermeable and no species can pass to the other ``acceptor'' droplet which is initially devoid of dye: ${\rho_a^{\alpha}(t=0)= 0, \ \forall \alpha}$.
For simplicity, we shall suppose that the dyes can be treated as ideal solutes that spread rapidly throughout the acceptor droplet after crossing the DIB.
These assumptions ensure that the dye density in each droplet is spatially uniform and a function of time only.
Furthermore, we assume that the DIB is perfectly planar and neglect thermal fluctuations about this planar state.
We note that such fluctuations can affect thermodynamics and dynamics of intra-membrane phase transitions, including arresting the coarsening of two-dimensional pores at a finite length-scale~\cite{Yu2025Pattern}.
However, since we aim to describe metastable DIBs that collapse en route to droplet coalescence, we expect pores to eventually grow to sizes spanning the entire DIB, making them largely insensitive to the arrest of geometric fluctuations.

The number density of a particular species in the acceptor droplet as a function of time naturally depends on the permeability of the membrane to that species.
This permeability, in turn, is related to the distribution of pore sizes in the membrane.
In this way, measuring the density of dye transferred to the acceptor serves as an indirect measure of the structure of the DIB as it dissolves.
To see this, let us construct the evolution equation for $\rho^{\alpha}_a(t)$.

First, we note that the dye density in a given droplet may only change due to the flux of molecules across the DIB.
This flux must be proportional to the density difference between the two droplets, ${j^{\alpha} = \kappa^{\alpha}(\rho_d^{\alpha} - \rho_a^{\alpha})}$, and the constant of proportionality \textit{defines} the permeability of the DIB to species $\alpha$~\cite{Phillips2012PhysicalCell}.
Integrating this flux over the area of the DIB, which under our assumptions amounts to multiplying $j^{\alpha}$ by the bilayer area $S$, gives the rate of change of the total amount of dye in the acceptor:
\begin{equation}
\label{eq:drhodt}
    \frac{\partial }{\partial t}\left(V \rho_a^{\alpha}(t)\right)=V\frac{\partial \rho_a^{\alpha}}{\partial t} = S(t) \kappa^{\alpha}(t) \left[\rho_d^{\alpha}(t)-\rho_a^{\alpha}(t)\right].
\end{equation}
Here, for generality, we have maintained the time dependence of the DIB area, $S(t)$, which could grow as the droplets coalesce~\cite{Mruetusatorn2014DynamicBilayers, Bird2009CriticalDroplets}.
However, from here onward, we shall assume that the timescale of this area change is much longer that of particle diffusion so that $S$ does not measurably change in the regime of interest.
The time dependence of the membrane permeability, $\kappa^{\alpha}$, on the other hand, cannot be discounted because it changes as the pores in the membrane grow.
This is the central relationship we explore in this work.
As we have written it, Eq.~\eqref{eq:drhodt} is general to any mode of transport across a membrane.
Solving the transport behavior for a particular system requires specifying the relationship between membrane structure and permeability.

We now establish this relationship for our problem.
We take the perspective that a particle only moves across the bilayer if the bilayer features a large enough pore through which the particle can travel.
When such a pore exists, the particle is taken to diffuse freely across the membrane, as if in bulk solution, with diffusion coefficient $D_0^{\alpha}$.
This approximation neglects confinement, chemical, and hydrodynamic effects that arise when a particle moves through a pore comparable in size to itself~\cite{Zwanzig1992DiffusionBarrier, Kalinay2006CorrectionsEquation, Kalinay2008ApproximationsEquation, Martens2011EntropicEquation, Dorfman2014AssessingEquation, ValeroValdes2014Fick-JacobsCurves, Ledesma-Duran2016GeneralizedConditions}.
While such effects can be significant for individual particle–pore pairs, we assume (and later find) that the DIB hosts a broad population of pores, making close size matches rare enough that their influence on overall transport is negligible.
Crucially, though, this approximation captures the need for sufficiently many large pores to be present in the DIB to allow transfer of a given dye size.

If we define the number of pores per unit area of the DIB with radii between $R$ and ${R+dR}$ at a given time $t$ as $n(R,t/\tau_P)$, then we can write the permeability as
\begin{equation}\label{eq:dimensionalperm}
    \kappa^\alpha (t/\tau_P) = \frac{D_0^{\alpha}}{h} \displaystyle \int_{r^{\alpha}}^\ell dR \ \pi R^2 \ n(R, t/\tau_P).
\end{equation}
Here $\ell \sim \sqrt{S}$ is the bilayer radius, taken to be much larger than any particle size, and $h$ is the bilayer thickness, which may be approximated as twice the length of the surfactants comprising the monolayers.
The integral in Eq.~\eqref{eq:dimensionalperm} naturally measures the fraction of the bilayer area covered by pores large enough to permit passage of dye particles at time $t$.
In writing the permeability in this form, we assume the existence of a natural timescale for pore growth, $\tau_P$, set by the specific growth mechanism, against which all times are measured.
Having established the fundamental relationship between membrane permeability and structure, which is encoded by $n(R,t)$, we now look to understand the different mechanisms governing the evolution of $n$.
Before doing so, we will first discuss each of the dimensionless numbers controlling the problem.

We non-dimensionalize the permeability such that its initial value is unity if all pores present at $t=0$ are larger than species $\alpha$. 
We define the initial area fraction of pores as
\begin{equation}
\label{eq:areafrac}
    \phi_0 = \displaystyle \int_{0}^{\ell} dR \ \pi R^2 \ n(R,0),
\end{equation}
resulting in the dimensionless permeability ${\overline{\kappa}^{\alpha} = \kappa^{\alpha} h/(D_0^{\alpha}}\phi_0)$.
The factor of $\phi_0$ in the denominator accounts for less barrier to transport across the DIB when there are more pores.
It also allows $\overline{\kappa}^{\alpha}$ to measure departures from a membrane with a fixed area fraction of perfectly permeable pores as a function of time.
Introducing our dimensionless permeability into Eq.~\eqref{eq:drhodt} allows us to identify the natural timescale for dye diffusion ${\tau_D = h V/(2 S \phi_0 D_0^{\alpha})}$.
With these definitions, and using per-species mass conservation, we arrive at the dimensionless evolution equation for the number density of dye in the acceptor droplet:
\begin{equation}
\label{eq:dimlessdens}
    \frac{\partial \overline{\rho}_{a}^{\alpha}}{\partial \overline{t}} =  \overline{\kappa}^\alpha\left(\overline{t}  \ \mathrm{Pe}\right)( 1 - \overline{\rho}_{a}^{\alpha}),
\end{equation}
where we have introduced the dimensionless time ${\overline{t}=t/\tau_D}$ and density, ${\overline{\rho}^{\alpha}_a = \rho^{\alpha}_a/\rho^{\alpha}}$.
The fact that the bilayer permeability evolves with the pore growth timescale, $\tau_P$, while the dye transport is governed by a separate time scale, $\tau_D$, gives rise to the dimensionless P\'eclet number ${\mathrm{Pe}=\tau_D/\tau_P}$ in Eq.~\eqref{eq:dimlessdens}.
When the timescale for dye transport is much longer than that for pore growth, the P\'eclet number is large and we consider such transport ``dye diffusion limited''. 
The converse situation, in which pore growth occurs on time scales much longer than the dye transport corresponds to small  P\'eclet number and is ``pore growth limited''.
We emphasize that the permeability depends on the product of $\overline{t}$ and $\mathrm{Pe}$ as ${t/\tau_p \equiv \overline{t} \ \mathrm{Pe}}$.

Figure~\ref{fig:modelsystem} illustrates the connection between pore growth and the membrane's time-dependent size selectivity captured by Eqs.~\eqref{eq:dimensionalperm}  and~\eqref{eq:dimlessdens}.
We hypothesize that shortly after the two droplets are brought together, small pores of size $R^{\dagger}$ form and begin to coarsen.
Only the particles smaller than these very small pores can cross from the donor to the acceptor. 
Over time, the pores grow so that larger species travel across the membrane at later times.
The smallest species therefore have non-zero initial permeability while larger species have zero permeability until large enough pores form.
To use Eqs.~\eqref{eq:dimensionalperm} and~\eqref{eq:dimlessdens} to predict the time-dependent permeability of the bilayer, we must first determine the pore distribution.
The distribution will depend on the mechanism of pore growth.
In our treatment of this problem, we consider three such mechanisms: Ostwald ripening, coalescence, and desorption.
During Ostwald ripening, larger pores grow at the expense of smaller pores, which shrink to reduce the energetic cost associated with the line tension acting on the pore perimeter.
If the total area fraction of pores remains constant at $\phi_0$, $\overline{\kappa}^{\alpha}$ approaches unity as the average size of pores grow.
During coalescence, pores diffuse in the plane of the membrane and merge upon contact to create larger pores that also conserve the total pore area fraction.
Finally, pores may also grow simply through the desorption of surfactants from the bilayer into the bulk.
To apply these equations and find transport trends for different pore growth dynamics, we will now consider the thermodynamics and mechanics of the DIB.

\section{In-plane Bilayer Thermodynamics and Mechanics}
\label{sec:InPlane}
We consider a bilayer surface that is in principle described by a two-dimensional curvilinear coordinate $\mathbf{x}$. 
We assume throughout this work that the membrane is quite rigid and we can, therefore, neglect membrane curvature and fluctuations for simplicity.
The local area fraction of surfactants in each of the two layers is denoted as $\varphi_1(\mathbf{x})$ and  $\varphi_2(\mathbf{x})$.
Here, we assume that the free energy penalty for ${\varphi_1 \neq \varphi_2}$ is suitably strong such that we can safely take ${\varphi_1 = \varphi_2 \equiv \varphi(\mathbf{x})}$.
Accordingly, the following analysis is equally applicable to monolayers or to systems with any number of layers for which a single composition field suffices to describe the surface.
For a one-component bilayer, ${1-\varphi(\mathbf{x})}$ represents the local unoccupied area fraction of the bilayer.
Therefore, a hole in the bilayer is represented by a region of space in which ${\varphi(\mathbf{x})=0}$.

The simplest free energy functional describing the in-plane thermodynamics of the surfactants comprising the bilayer takes the form:
\begin{equation}
\label{eq:freeenergyfunctional}
\mathcal{F}[\varphi(\mathbf{x})] = \displaystyle \int_S d\mathbf{x}\left(\upgamma(1-\varphi) +  u\varphi + f(\varphi) + \frac{k}{2}\left|\boldsymbol{\nabla} \varphi \right|^2\right),
\end{equation}
where $S$ is again the total bilayer surface area. 
The first and second terms in Eq.~\eqref{eq:freeenergyfunctional} represent, respectively, the areal free energy density of exposed and occupied regions.
The free energy of an exposed region is set by $\upgamma$.
For a monolayer dividing immiscible liquids, this corresponds to the liquid–liquid surface tension, whereas for a bilayer with the same liquid on both sides, $\upgamma$ is negligible, if not identically zero.
The energy per unit area, $u$, represents the free energy associated with surfactant-solvent interactions. 
The free energy density associated with the in-plane surfactant configurations and interactions is denoted as $f(\varphi)$, which depends solely on the local area fraction. 
The total bulk in-plane free energy density of the membrane is thus ${f^{\rm mem}(\varphi) = \upgamma(1-\varphi) +  u\varphi + f(\varphi)}$.
The final term in Eq.~\eqref{eq:freeenergyfunctional} penalizes spatial gradients (${\boldsymbol{\nabla} \equiv \partial/\partial \mathbf{x}}$) in the composition with $k (\varphi) > 0 $.
The square-gradient coefficient, $k$, may also depend on the local composition. 

From the free energy functional, we can determine the (exchange) chemical potential through functional differentiation ${\mu = \delta F/\delta \varphi}$.
We also determine the reversible in-plane stress (up to an additive divergenceless field), $\boldsymbol{\Sigma}$,  using the Gibbs--Duhem relation ${\boldsymbol{\nabla}\cdot \boldsymbol{\Sigma} = -\varphi \boldsymbol{\nabla} \mu}$ finding:
\begin{equation}
\label{eq:stress}
    \boldsymbol{\Sigma} = \upgamma_{\rm eff}\mathbf{I} + \left(\frac{1}{2} \frac{\partial ( k \varphi )}{\partial \varphi} \left|\boldsymbol{\nabla} \varphi \right|^2 + k \varphi \boldsymbol{\nabla}^2 \varphi \right) \mathbf{I} 
    - k \boldsymbol{\nabla}\varphi \boldsymbol{\nabla}\varphi \ ,
\end{equation}
where $\mathbf{I}$ is the two-dimensional identity tensor. 
Here, we have defined the \textit{effective bulk tension} as ${\upgamma_{\rm eff}(\varphi) = \upgamma - p(\varphi)}$ where ${p(\varphi) = -f(\varphi)+\varphi \partial f/ \partial \varphi}$ is the in-plane bulk thermodynamic pressure of the surfactants. 

The free energy of a uniform system with composition $\varphi$ is simply ${\mathcal{F} = Sf^{\rm mem}(\varphi)}$.
The bilayer is free to exchange surfactants with the bulk fluid, which we take to be a surfactant reservoir with constant chemical potential, $\mu^{\rm sur}$.
Therefore, the relevant thermodynamic potential is the grand potential, $\mathcal{W}$.
The grand potential areal density takes the form ${w \equiv \mathcal{W}/S =  f^{\rm mem}(\varphi) - \mu^{\rm sur}\varphi = \upgamma_{\rm eff}}(\varphi)$. 
Thus, the thermodynamic stability of a membrane composition is entirely encoded in the effective tension.
Locally stable compositions will be those that satisfy ${\partial \upgamma_{\rm eff} / \partial \varphi = 0}$ and ${\partial^2 \upgamma_{\rm eff} / \partial \varphi^2 > 0}$.
The globally stable composition will be the one that satisfies these stability criteria with the lowest free energy or, equivalently, the lowest effective tension.

\subsection{Nucleation of Critical Pores}\label{sec:PoreNuc}
The free energy landscape will be controlled by the natural variables of the effective tension, namely $\mu^{\rm sur}$ and the temperature $T$. 
We consider fixed temperature throughout the present study and turn our focus to $\mu^{\rm sur}$. 
We anticipate that decreasing $\mu^{\rm sur}$ will drive surfactant desorption from the membrane until an intact membrane is no longer the globally stable state (i.e., the globally stable state has $\varphi \rightarrow 0$). 
However, the intact membrane may remain a \textit{metastable} state that requires a nucleation event to trigger this membrane dissolution.

We consider such a membrane with composition $\varphi_0^{\rm mem}$.
The globally stable state has composition $\varphi^{\rm pore}$ (we will ultimately take $\varphi^{\rm pore} = 0$) with ${\upgamma_{\rm eff}(\varphi^{\rm pore}) < \upgamma_{\rm eff}(\varphi_0^{\rm mem})}$.
We consider the nucleation of a single circular pore of radius $R$ with a uniform composition of $\varphi^{\rm pore}$ and a composition of $\varphi_0^{\rm mem}$ on the remaining surface.
Note that in order for this pore to form, the overall surface coverage of surfactants must decrease through their desorption from the surface.
This is an important point to consider when examining later-stage growth dynamics.

The only spatial variation in membrane composition occurs rapidly at the pore boundary. 
For a membrane with a single circular pore of radius $R$, the total grand potential energy can be written as the sum of contributions from the pore interior, the surrounding membrane, and the interface between these regions. 
The resulting free energy change relative to the intact membrane is:
\begin{equation}
\label{eq:DeltaW}
\Delta\mathcal{W} = \pi R^2\left[\upgamma_{\rm eff}(\varphi^{\rm pore}) - \upgamma_{\rm eff}(\varphi_0^{\rm mem})\right] + 2 \pi R \sigma,
\end{equation}
where $\sigma$ is the (positive) line tension\footnote{We will strictly refer to $\sigma$ as a line tension and reserve ``tension'' for $\upgamma_{\rm eff}$.}, and $\varphi^{\rm pore}$ and $\varphi_0^{\rm mem}$ are again the compositions inside and outside the pore, respectively.

Equation~\eqref{eq:DeltaW} shows that pore formation can only reduce the free energy if the pore has lower tension than the surrounding membrane: ${\upgamma_{\rm eff}(\varphi^{\rm pore}) < \upgamma_{\rm eff}(\varphi_0^{\rm mem})}$.
This aligns with the physical expectation that pore nucleation in bilayers is driven by an associated reduction in tension and free energy.
In contrast, for a monolayer separating immiscible liquids, forming a hole typically \textit{increases} the local effective tension, making pore nucleation thermodynamically unfavorable.
Finding the radius at which $\Delta \mathcal{W}$ reaches a local maximum allows us to identify the critical radius as:
\begin{equation}
\label{eq:criticalradius}
R^{\dagger} = \frac{2\sigma}{\upgamma_{\rm eff}(\varphi_0^{\rm mem}) - \upgamma_{\rm eff}(\varphi^{\rm pore})}.
\end{equation}

The preceding derivation of $R^{\dagger}$ followed the standard approach of classical nucleation theory.
We can arrive at an identical expression from purely mechanical arguments~\cite{Langford2025TheMatterb}, beginning from the static momentum balance: ${\boldsymbol{\nabla}\cdot \boldsymbol{\Sigma} = \mathbf{0}}$. 
If we again consider a single circular pore centered at $r = 0$ in polar coordinates, the radial-component of the static mechanical balance takes the form: 
\begin{equation}
\label{eq:radialstaticmech}
\frac{\partial \Sigma_{rr}}{\partial r} - \frac{2}{r}\ \left (\Sigma_{tt} - \Sigma_{rr} \right) = 0,
\end{equation}
where $\Sigma_{rr}$ and $\Sigma_{tt}$ are the radial and tangential components of the stress. 
Direct integration from a location within the pore, ${R_{-}^{\dagger}}$, to a location outside the pore, ${R_{+}^{\dagger}}$, results in:
\begin{subequations}
\label{eq:stressjump}
\begin{equation}
\Sigma_{rr}(r={R_{+}^{\dagger}}) - \Sigma_{rr}(r={R_{-}^{\dagger}}) - \frac{2}{R^{\dagger}}\sigma + \mathcal{O}({R^{\dagger}}^{-2}) =0, 
\end{equation}
where the line tension is defined through the stress anisotropy at the pore boundary,
\begin{equation}
\sigma = \displaystyle \int_{{R_{-}^{\dagger}}}^{{R_{+}^{\dagger}}}  dr \left(\Sigma_{tt} - \Sigma_{rr} \right),
\end{equation}
\end{subequations}
and we have assumed that the normal stress difference is only finite near the pore radius (see Supporting Information (SI) for details).

As the local compositions at ${R_{-}^{\dagger}}$ and ${R_{+}^{\dagger}}$ are $\varphi^{\rm pore}$ and $\varphi_0^{\rm mem}$, respectively, and spatially uniform at these locations, we can identify ${\Sigma_{rr}(r={R_{-}^{\dagger}})=\upgamma_{\rm eff}(\varphi^{\rm pore})}$ and ${\Sigma_{rr}(r={R_{+}^{\dagger}})=\upgamma_{\rm eff}(\varphi_0^{\rm mem})}$.
Substituting these values into Eq.~\eqref{eq:stressjump} gives the critical radius as seen in Eq.~\eqref{eq:criticalradius}.
We can now appreciate that the line tension allows the pore to sustain a lower tension than the surrounding membrane; the two-dimensional analogue of the Young--Laplace equation.

As discussed above, fixing the surface composition during pore nucleation ensures that any increase in the overall surface area occupied by pores must result in an increase in the membrane density outside of the pores, $\varphi^{\rm mem}$, with ${\varphi^{\rm mem} > \varphi_0^{\rm mem}}$. 
This increased density acts to reduce the membrane tension as the in-plane surfactant pressure is expected to increase with area fraction.
This pathway for pore growth and tension reduction, in the absence of surfactant desorption, is isomorphic to well-studied nucleation and growth pathways in the canonical ensemble~\cite{Desai2009DynamicsStructures}.
Frolov~et al. highlight that nucleation occurs on very short timescales and can thus be distinguished from the later stages of coarsening~\cite{Frolov2006EntropicNanodomains}.
We now proceed to consider the growth dynamics of pores beyond this initial nucleation stage, after which all the pores are at least as large as the critical radius, given by Eq.~\eqref{eq:criticalradius}, and no new pores nucleate.

\subsection{Pore Growth Mechanisms}
In this Section, we will discuss how the pore distribution evolves over time after the nucleation stage.
To do so, we will first describe the mechanics of a membrane with a single pore and use that mechanical picture to derive the radial pore growth dynamics, which are expressed through the radial velocity of the pore boundary $v(R)$ or, equivalently, the rate of growth of the size of a pore.
This velocity can then be used to determine the dynamics of the pore distribution, $n(R,t)$. 
Recall that $n$ describes the number of pores per-unit area of the membrane with sizes in the window between $R$ and ${R+dR}$. 
This number can change in two ways.
First, pores may grow or shrink due to the velocity $v(R)$, causing their sizes to move into or out of the interval ${[R,R+dR]}$.
Second, new pores may be created or destroyed in discontinuous events within that window.
These discontinuous events include: nucleation (creation of new pores), coalescence (merging of smaller pores into a larger one) and fission (division of large pores into smaller ones).

These two mechanisms of changing the number of pores of a certain size are naturally captured in a continuity equation for $n(R,t)$:
\begin{equation}
\label{eq:nucleusdist}
    \frac{\partial n}{\partial t} + \frac{\partial}{\partial R}\left(v(R)n(R,t) \right) = \mathcal{C}(R,t).
\end{equation}
The left-hand side expresses the deterministic redistribution of pores by the velocity $v(R)$, while the right-hand side takes into account discontinuous changes through nucleation or coalescence through the source (or sink) term ${\mathcal{C}(R,t)}$.
Here, we will determine $n$ in detail  for the specific mechanism of Ostwald ripening.
We will briefly comment on the timescale of pore growth for coalescence, and we refer readers to the SI for a derivation of the pore distribution for desorption.
This discussion aims to illustrate how these additional mechanisms affect the dynamics of $n(R, t)$ and to highlight when their inclusion becomes necessary.

In the classical treatment of Ostwald ripening in the LSW theory, it is assumed that pores are homogeneously distributed on the membrane and that nucleation and coalescence \textit{do not occur}; ${\mathcal{C}=0}$~\cite{Lifshitz1961TheSolutions, Wagner1961TheorieOstwaldReifung, Frolov2006EntropicNanodomains, Arnold2023ActiveDomains}.
In this case, it appears at first glance that the total number density of pores on the membrane, which is simply the integral of $n$ over all $R$, is conserved.
This is at odds with the expectation sketched in Fig.~\ref{fig:modelsystem}, which shows that the number density shrinks over time. 
However, the treatment we shall outline here is strictly valid only \textit{after} pores have nucleated at size $R^{\dagger}$ and pores smaller than this size have closed. 
Thus, if the radial velocities of pores at size $R^{\dagger}$ are \textit{negative} then we expect the number density of pores to \textit{decrease} as a function of time and, as shown schematically in Fig.~\ref{fig:modelsystem}, the bilayer to be occupied by smaller number of increasingly larger pores. 
This is the case for the LSW theory in three-dimensions and we shall show that it remains so when the theory is adapted to two-dimensional bilayers. 

\subsubsection{Ostwald ripening}\label{sec:Ost}
We begin our discussion of Ostwald ripening with a description of the in-plane surfactant dynamics. 
In the absence of surfactant adsorption or desorption, the local area fraction satisfies the following continuity equation:
\begin{equation}
\label{eq:continuity}
\frac{\partial \varphi}{\partial t} = -\boldsymbol{\nabla}\cdot \mathbf{j},
\end{equation}
where $\mathbf{j}$ is the in-plane surfactant flux.
Linear irreversible thermodynamics asserts that the flux takes the form ${\mathbf{j} = - L \boldsymbol{\nabla} \mu}$ where $L$ is the Onsager mobility (assuming spatial isotropy).
We again invoke the Gibbs--Duhem relation to express this flux in terms of the reversible stress with:
\begin{equation}
\label{eq:inplaneflux}
    \mathbf{j} = M \boldsymbol{\nabla} \cdot \boldsymbol{\Sigma},
\end{equation}
where ${M = L/\varphi}$ represents a single-particle mobility. 
For a pore the size of the critical radius, ${R = R^{\dagger}}$, the static mechanical balance is expected to be satisfied at all locations.
As a result, from Eq.~\eqref{eq:inplaneflux}, the flux is identically zero.
We expect that those pores larger than $R^{\dagger}$ grow, while those smaller shrink.
In this way, the critical radius represents an unstable dynamical fixed point.

To connect our constitutive relation for the in-plane surfactant dynamics to pore growth rates, we again consider a pore of radius $R$ centered at $r = 0$.
We take the composition to be uniformly ${\varphi(r) = \varphi^{\rm pore}}$ for ${r < R}$ and approximate ${\mathbf{j} \approx \mathbf{0}}$ near ${r=R}$, such that the mechanical balance locally holds near the pore interface.
This allows us to derive the tension difference across the interface through integration of Eq.~\eqref{eq:radialstaticmech} from a location just within the pore at $R-\varepsilon$ and one just outside at $R+\varepsilon$, with $\varepsilon$ much smaller than the length scales of interest.
We arrive at: 
\begin{equation}
\label{eq:stressjumpgeneral}
\Sigma_{rr}(R+\varepsilon) - \Sigma_{rr}(R-\varepsilon) - \frac{2}{R}\sigma + \mathcal{O}(R^{-2}) = 0,
\end{equation} 
where $\sigma$ takes the form found in Eq.~\eqref{eq:stressjump}.
The stress within the pore is again simply ${\Sigma_{rr}(R-\varepsilon)=\upgamma_{\rm eff}(\varphi^{\rm pore})}$.
For distances just beyond the interface, we assume the composition variation is weak and take the limit $\varepsilon \to 0$ such that ${\Sigma_{rr}(R+\varepsilon) \to \upgamma_{\rm eff}(r=R)}$. 
From this, we can see that Eq.~\eqref{eq:stressjumpgeneral} results in a Gibbs--Thomson relation:
\begin{equation}
\label{eq:gt}
\upgamma_{\rm eff}(r=R) = \upgamma_{\rm eff}(\varphi^{\rm pore}) + \frac{2}{R}\sigma.
\end{equation}

If we consider length scales much larger than the width of the interface at the pore boundary, the Gibbs--Thomson relation can be used as a boundary condition on the effective tension at ${r=R}$. 
By assuming a uniform pore with a sharp interface that satisfies the static momentum balance, the in-plane flux of surfactants for ${r<R}$ is identically zero. 
To determine the flux of surfactants in the remainder of the membrane (${r\ge R}$), we must determine the radial dependence of the stress throughout this region.
We can rearrange our constitutive equation for the flux to arrive at a \textit{dynamic} mechanical balance, ${  \boldsymbol{\nabla} \cdot \boldsymbol{\Sigma} - M^{-1}\mathbf{j}}=\mathbf{0}$.
Assuming weak composition variations for ${r>R}$ such that ${\boldsymbol{\Sigma} \approx \upgamma_{\rm eff}\mathbf{I}}$, the dynamic balance takes the form:
\begin{equation}
\label{eq:dynamicmech}
\boldsymbol{\nabla}\upgamma_{\rm eff} - M_0^{-1}\mathbf{j}=\mathbf{0} ,
\end{equation}
where $M_0$ is now a constant mobility (see SI for details). 
If we assume that the time-variation of the membrane composition is negligible during the timescales of interest, we have ${\partial \varphi / \partial t = -\boldsymbol{\nabla}\cdot \mathbf{j} \approx 0}$.
We apply this assumption by taking the divergence of the mechanical balance to arrive at:
\begin{equation}
\label{eq:radialdynamicmech}
\nabla^2 \upgamma_{\rm eff}=0.
\end{equation}
The Gibbs--Thomson relation, found in Eq.~\eqref{eq:gt}, provides one of the two boundary conditions needed to find $\upgamma_{\rm eff}(r)$.
The remaining boundary condition is taken to be the membrane tension at the edge of the bilayer with ${\upgamma_{\rm eff}(r = \ell) = \upgamma^{\infty}}$.
As previously alluded to, this tension is expected to be less than that of the original metastable membrane, ${\upgamma^{\infty} < \upgamma_{\rm eff}(\varphi_0^{\rm mem})}$, and to decrease monotonically with time. 
The tension then follows as:
\begin{equation} 
\label{eq:tensionprofile}
\upgamma_{\rm eff}(r) = \upgamma_{\rm eff}(\varphi^{\rm pore}) + \frac{2}{R}\sigma + \frac{\upgamma^{\infty} - \upgamma_{\rm eff}(\varphi^{\rm pore}) - \frac{2}{R}\sigma }{\ln(\ell/R)}\ln(r/R).
\end{equation}
\begin{figure}
	\includegraphics[width=.5\textwidth]{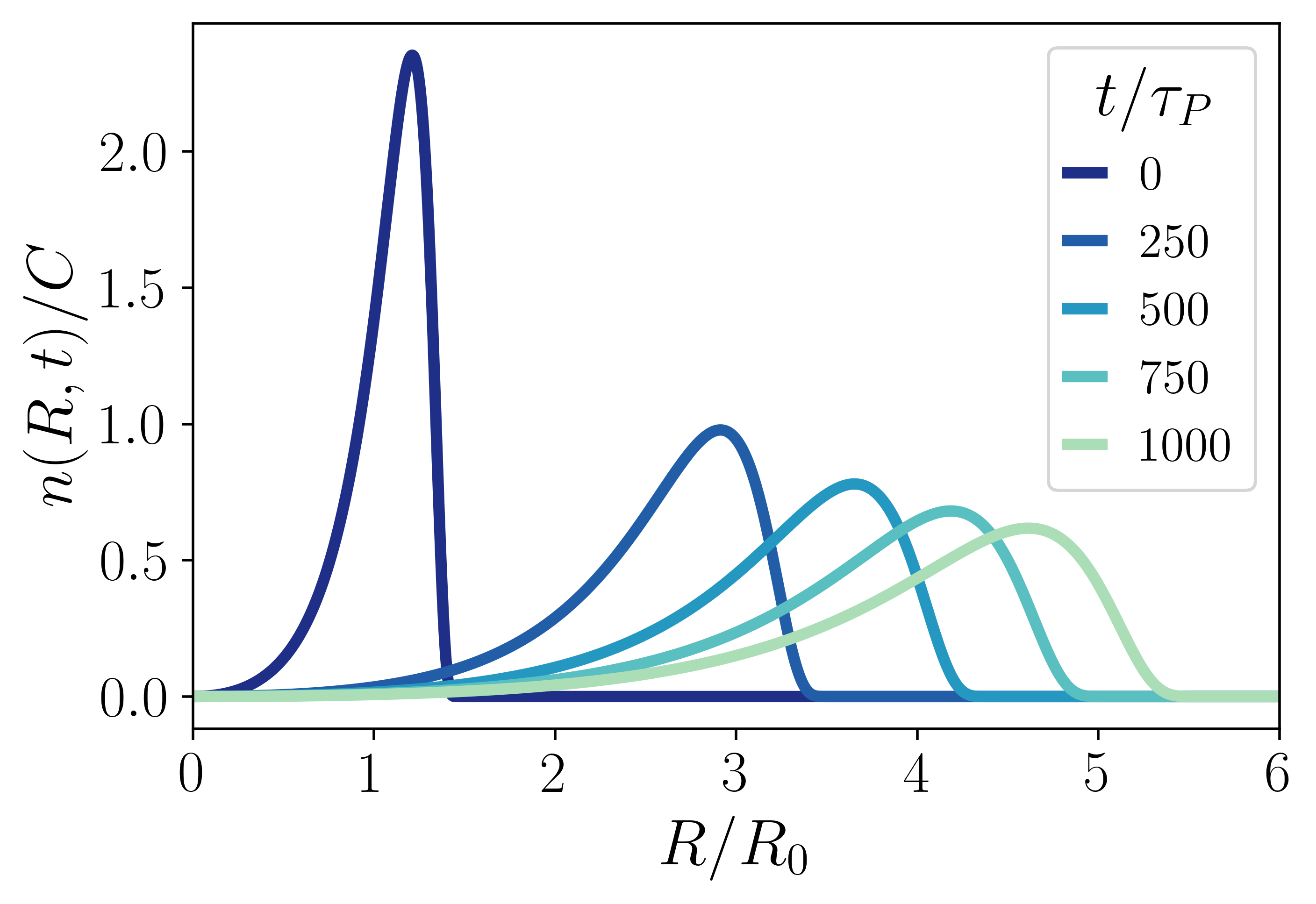}
	\caption{Plots of the pore distribution for the Ostwald ripening mechanism, normalized so that the area under the curves is unity, $n(R,t)/C$ with ${C = \int n(R,t) d R/R_0}$, for various times non-dimensionalized by the pore-growth time scale, $t/\tau_p$. At early times the bilayer features several small pores and as $t/\tau_P$ increases the distribution widens such that the bilayer contains fewer but larger pores.}
	\label{fig:dist}
\end{figure}

We define the critical radius associated with a membrane with tension $\upgamma^{\infty}$ as $R_c$ such that,
\begin{equation}
\frac{2 \sigma}{R_c} = \upgamma^{\infty} - \upgamma_{\rm eff}(\varphi^{\rm pore}).
\end{equation}
With this we can finally obtain the flux outside the pore using Eq.~\eqref{eq:dynamicmech} (which only has a finite radial component):
\begin{equation}
\label{eq:fluxprofile}
j_{r}(r) =  \frac{2M_0\sigma}{r\ln(\ell/R)}\left(\frac{1}{R_c} - \frac{1}{R}\right),
\end{equation}
which allows us to determine the pore growth dynamics.
The time variation of the pore area is equivalent to the flux from its circumference, ${\partial (\pi R^2)/\partial t = 2\pi R j_r(r=R)}$, from which we obtain the rate of change of its radius:
\begin{equation}
\label{eq:velocity}
\frac{\partial R}{\partial t} \equiv v(R) =  \frac{2M_0\sigma}{R^2\ln(\ell/R)}\left(\frac{R}{R_c} - 1\right).
\end{equation}
As anticipated, for pores larger (smaller) than $R_c$, the flux of surfactants is away from (towards) the pore, resulting in the pore expanding (shrinking).
The growth rate is linearly proportional to both the line tension---for a fixed areal coverage of pores the line tension would drive the growth dynamics to reduce overall line energy---and, intuitively, the in-plane surfactant mobility.

For the form of the pore distribution in the Ostwald ripening mechanism, we refer to LSW theory~\cite{Lifshitz1961TheSolutions, Wagner1961TheorieOstwaldReifung}, which introduced a fundamental approach for understanding domain growth kinetics in three-dimensional systems. 
The key aspects of LSW theory are: that there are no discontinuous changes in pore sizes [$\mathcal{C}=0$ in Eq.~\eqref{eq:nucleusdist}], and that we may make the ansatz of self-similar growth for the form of $n(R,t)$:

\begin{equation}\label{eq:ansatz}
    n(R,t) = \frac{\mathcal{N}(x)}{R_c^{d+1}(t)}.
\end{equation}
Here, $d$ is the dimensionality of the system ($d=2$ in the present context) and $\mathcal{N}(x)$ is a ``universal scaling function'' of the dimensionless variable ${x=R/R_c(t)}$, written in terms of the time-dependent critical radius $R_c(t)$.
From the continuity equation [Eq.~\eqref{eq:nucleusdist}], after a careful analysis to find a form of $\mathcal{N}(x)$ that satisfies conservation of matter, one finds the \textit{only} stable distribution for domain growth and the canonical scaling result ${R_c \sim t^{1/3}}$ (see SI for details). 

Applying LSW theory to two-dimensional systems presents challenges due to singularities in the general form of the membrane tension and in the form of the velocity in Eq.~\eqref{eq:velocity}~\cite{Marqusee1984DynamicsDimensionsb, Marqusee1984Theory, Desai2009DynamicsStructures}.
Our solution avoids this first challenge by having boundary conditions applied at the edge of the bilayer when solving for $\upgamma_\mathrm{eff}(r)$ in Eq.~\eqref{eq:tensionprofile}. 
We also take the limit that the pores remain significantly smaller than the DIB radius $(R \ll \ell)$, from which it is possible to show how pore growth scales with time for a two-dimensional system. 
Defining ${\overline{R}_c(t)=R_c(t)/R_0}$, where $R_0$ is the initial critical pore radius, we can write:
\begin{equation}
\label{eq:2Dcriticalradius}
    \begin{split}
        3 \overline{R}_c^3\left(t/\tau_p\right) \left[3 \ln \overline{R}_c(t/\tau_p) + k \right] = - 4 \ t/\tau_p + 3 k,
    \end{split}
\end{equation}
where ${\tau_P = R_0^3/M_0 \sigma}$ is the pore growth timescale for Ostwald ripening, and ${k = 3 \ln(R_0/\ell) -1}$ is a constant for a fixed $R_0$.
We aim to describe the dynamics of the population of pores present in the membrane after an initial nucleation phase. 
Therefore, it is natural to take the initial critical pore radius to be greater than that of a single isolated pore; ${R_0 \gtrsim R^{\dagger}}$.  

Under our assumption that pore size is small relative to the bilayer radius, we find the same universal scaling function, $\mathcal{N}(x)$, as LSW theory (see SI for details). 
The precise form is non-trivial, but its key features are shown in Fig.~\ref{fig:dist}, which illustrates that the distribution is left skewed and the mean pore size increases with time. 
For all times, the scaling function is only finite for ${0<x<3/2}$; i. e. at any given time, there are \textit{no pores larger than} $3 R_c(t)/2$.
Consequently, since the critical radius grows with time according to Eq.~\eqref{eq:2Dcriticalradius}, the scaling function broadens as its peak shifts towards larger values. 
Physically, this is represented by the membrane featuring fewer, larger pores as time passes, as shown in Fig.~\ref{fig:modelsystem}.

\subsubsection{Coalescence}
\label{sec:Coal}
The above growth dynamics are unique to the Ostwald ripening mechanism of pore growth, in which pores shrink or grow depending on their size relative to a critical radius.
Pores may also grow through coalescence as they diffuse through the membrane and merge upon contact~\cite{Bird2009CriticalDroplets}.
Merging will occur concurrently with fission, in which a larger pore separates into two smaller ones.
To find the pore growth timescale for coalescence we require the pore velocity, $v(R)$, and the source term $\mathcal{C}(R,t)$ in order to obtain the form of $n(R,t)$ from Eq.~\eqref{eq:nucleusdist}.
While we applied the ansatz of a self-similar probability distribution that evolves with the critical radius to describe Ostwald ripening, it is not immediately clear whether or not the coalescence size distribution is self-similar.
Self-similarity has been observed in collections of coalescing drops~\cite{Strutt1879I.Drops, Rayleigh1882FurtherJets, Blaschke2012BreathDroplets} and following Meakin~et al.~\cite{Meakin1992DropletCoalescence} one may define a natural length scale, such as the mean radius, that varies with time in the coalescing system.
While Ostwald ripening and coalescence involve different dynamics, both have been found to exhibit the classical result that the average domain size $\langle R\rangle \sim t^{1/3}$, with several theoretical, computational, and experimental studies showing that this growth law holds for two-dimensional systems~\cite{Marqusee1984Theory, Marqusee1984DynamicsDimensions, Frolov2006EntropicNanodomains, Stanich2013CoarseningMembranes, Arnold2023ActiveDomains}.
We refer to Frolov's~\cite{Frolov2006EntropicNanodomains} result that the mean radius changes like ${\langle R \rangle \sim t^{1/3}}$ during coalescence, albeit with a higher amplitude than in Ostwald ripening and expect that if domains grow through coalescence, transport across the membrane will exhibit similar scaling laws as in the Ostwald ripening mechanism.

\subsubsection{Surfactant Desorption}
\label{sec:Des}
Finally, we consider the simplest possible mechanism: pore growth by surfactants desorbing from the bilayer into the bulk, as detailed in the SI.
The thermodynamic driving force for desorption is set by the elevated chemical potential of surfactants within the membrane relative to that of surfactants in the bulk reservoir. 
We suppose that only those surfactants at the pore perimeter desorb due to our physical expectation that the barrier to desorption is far greater in the bulk of the membrane compared to the pore perimeter, where surfactants have fewer neighbors.
We also assume that this desorption occurs at a constant rate ${1/\tau_P^{\mathrm{des}}}$.
This naturally leads to the exponential growth of pores, as the rate at which the perimeter increases is proportional to the perimeter itself. 
This exponential growth is the consequence of a feedback mechanism wherein the surfactants that desorb from the perimeter into the bulk introduce additional perimeter surfactants. 
The newly exposed surfactants desorb to reduce their chemical potential, leading to a runaway effect as the process continues at increasingly faster rates until the entire bilayer dissolves.
The exponential growth of pores inevitably leads to a rapid increase of the DIB permeability. 
It follows that the dye transport in this case is very fast for all but the slowest desorption rates.
In fact, it is possible to show (see SI for details) that for dye transport to take place over the course of days, the rate of desorption would only need to be on the order of one particle per day.
This suggests desorption is a more fitting model for membranes that collapse quickly, while Ostwald ripening and coalescence more appropriately describe membranes that remain stable over several days.

In Section~\ref{sec:Perm}, we focus on Ostwald ripening as the dominant mechanism of pore evolution.
This choice is motivated by our interest in long-lived membranes that exhibit strong size selectivity.
While desorption leads to rapid membrane breakdown unless suppressed to unrealistically low rates, Ostwald ripening supports slow, tunable transport across the DIB over extended timescales. 
Coalescence, though more complex to model explicitly, has been shown to yield the same scaling behavior as Ostwald ripening and is, therefore, not expected to qualitatively alter the main results.
\begin{figure}
\includegraphics[width=.5\textwidth]{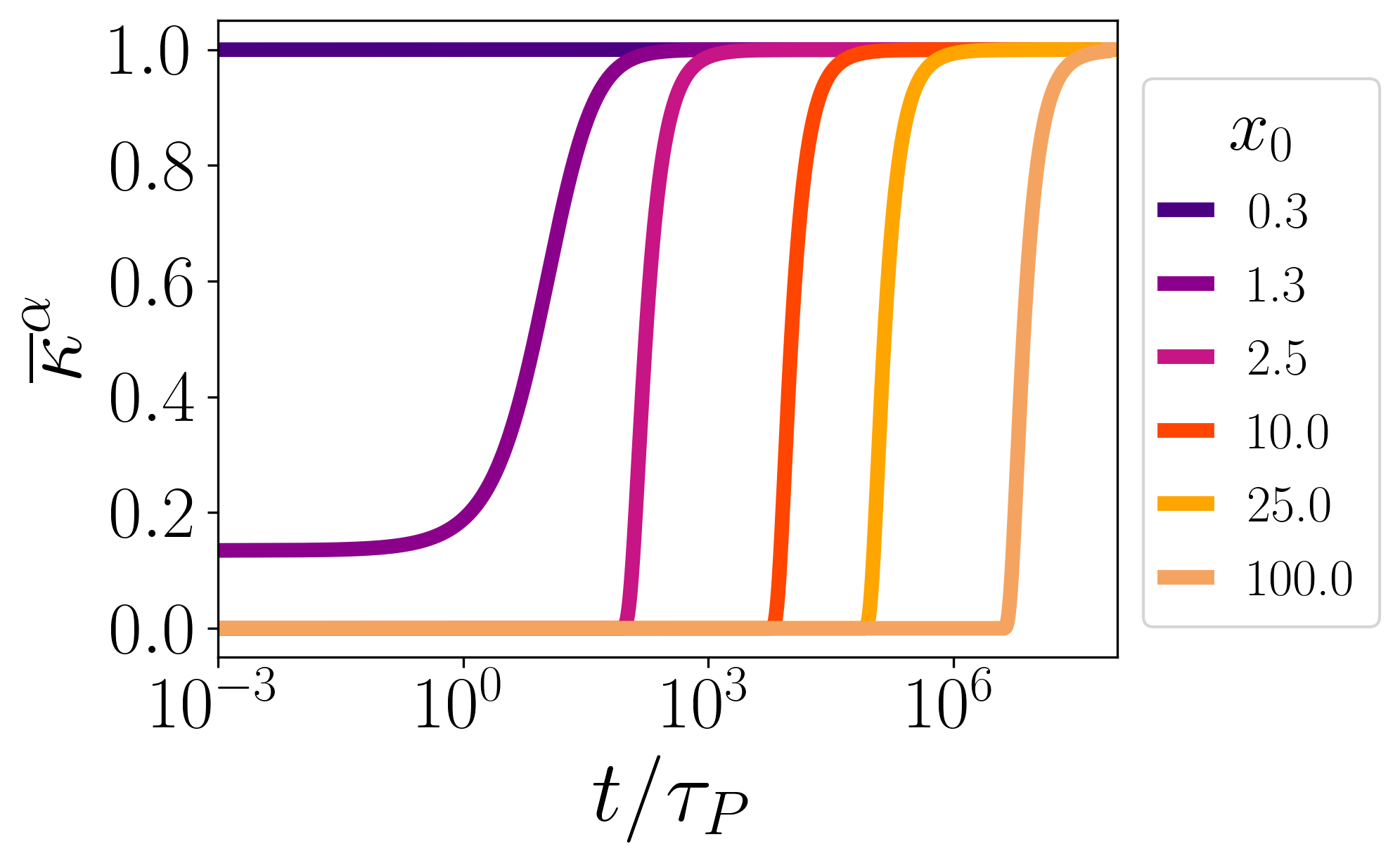}
	\caption{The non-dimensional membrane permeability, $\overline{\kappa}^{\alpha}$, plotted as a function of time in units of the pore-growth time scale, $t/\tau_p$, for dye sizes measured against the initial pore radius $x_0=r^{\alpha}/R_0$ increasing across four orders of magnitude. In the Ostwald ripening mechanism, the area fraction of pores remains fixed. Therefore, since $\overline{\kappa}^{\alpha}$ is non-dimensionalized by its value if all initial pores are permeable, all curves plateau at unity.}
	\label{fig:permeability}
\end{figure} 

\begin{figure*}
	\centering
\includegraphics[width=1\textwidth]{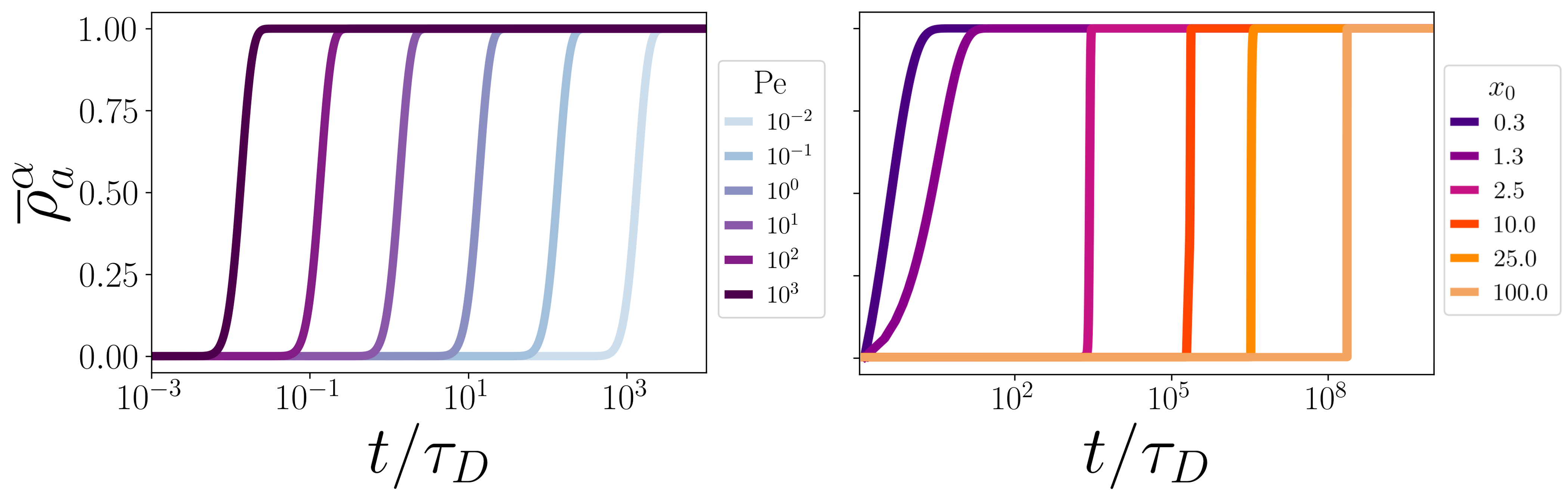}
	\caption{The non-dimensional dye density in the acceptor droplet, $\bar{\rho}^{\alpha}_a$, plotted as a function of dimensionless time, $t/\tau_D$ where $\tau_D$ is the dye diffusion time scale. In the left panel, $\tau_D$ is varied relative to the pore growth time-scale \textit{via} $\mathrm{Pe} = \tau_D/\tau_P$, with the scaled dye size relative to the initial pore radius $x_0 = r^{\alpha}/R_0$ fixed at $x_0 = 1.5$. Here, we see that increasing $\mathrm{Pe}$ for fixed particle size shifts the density profile to earlier times. In the right panel, the scaled dye sizes, $x_0$, are varied while $\mathrm{Pe} = 1.0$ and we see that increasing dye size shifts the density profile to later times. }
	\label{fig:density}
\end{figure*} 

\section{Dynamic Permeability due to Ostwald Ripening}
\label{sec:Perm}
Having established the driving forces for pore formation and growth in the DIB, we use Eq.~\eqref{eq:dimensionalperm} to find the membrane permeability for species of different sizes in the case that pores grow through Ostwald ripening. 
We refer readers to the SI to find results for the cases that pores primarily grow through desorption.

We introduce the dimensionless particle size ${x_0 = r^\alpha/R_0}$ where, again, $R_0$ is the initial critical pore size. 
We also assume that between the time when pores nucleate and the time when pores reach the asymptotic form of the distribution that is governed by $\mathcal{N}(x)$, the change in the density of dye in the acceptor, $\rho_a$, is negligible for all species because nucleation and growth occur on vastly different timescales~\cite{Frolov2006EntropicNanodomains, Marqusee1984Theory, Lifshitz1961TheSolutions, Wagner1961TheorieOstwaldReifung}. 
The initial times discussed here therefore correspond to the point at which the pore distribution reaches its asymptotic form.

Using the self-similar ansatz for $n(R,t)$ [Eq.~\eqref{eq:ansatz}], we can rewrite the dimensionless permeability in terms of the universal scaling function:
\begin{equation} \label{eq:permeability_dx}
    \overline{\kappa}^{\alpha}( t/\tau_P) = \frac{1}{\phi_0} \displaystyle \int_{x^{\alpha}(t/\tau_P)}^{3/2} dx \ \mathcal{N}(x) \pi x^2, 
\end{equation}
where ${x^{\alpha}(t)= r^{\alpha}/R_c(t)}$ measures the ratio of the size of species $\alpha$ to the time-dependent critical radius in Eq.~\eqref{eq:2Dcriticalradius}. 
Although the bilayer is finite with radius $\ell$, we take the upper bound in Eq.~\eqref{eq:permeability_dx} to $3/2$ because this is the upper bound for the universal scaling function $\mathcal{N}(x)$ in the Ostwald ripening mechanism, and we expect the bilayer radius to be significantly larger than that of any pore present within it. 
The time-dependence of the permeability now comes from the lower bound of the integral.
For small $x^\alpha$ (particles significantly smaller than the average pore size at a particular time) the integral covers the entire area under the curve, which is equal to the total area fraction of pores in the bilayer. 
For that same $t/\tau_P$---and therefore the same distribution---larger particles increase the lower bound of Eq.~\eqref{eq:permeability_dx}, which entails integrating over a smaller fraction of the universal scaling function.
This reflects that there are fewer pores through which the larger particles can travel. 
As a result, the permeability, $\kappa^\alpha$, decreases with particle size.

Figure~\ref{fig:permeability} shows that the permeability is extremely sensitive to the particle size. 
At early times ${t \ll \tau_P}$ small particles have finite permeability while larger ones have zero permeability. 
Since the initial distribution only features pores up to $3/2$ times the size of the initial critical radius, the membrane permeability is only non-zero for species with ${x_0 < 3/2}$.
The time at which the permeability rises above zero increases with the particle size. 
This is expected given that Eq.~\eqref{eq:criticalradius} indicates the critical radius behaves like  ${R_c \sim (t/\tau_P)^{1/3}}$. 
The ``wait time'' for a particle to cross the membrane, which depends on the presence of pores in the bilayer that are larger than the particle, then should also grow $\sim x_0^{3}$. 
One can readily verify that this trend is borne out by examining ${x_0 = 2.5}$ and ${x_0 = 25}$ curves, where it is clear that increasing the species size by a factor of $10$ extends the wait time by approximately $10^{3}$. 
Finally, when the particle is significantly smaller than the average pore size (small $x_0$ or large $t/\tau_P$), the dimensionless permeability plateaus to unity. 
This is an effect of integrating Eq.~\eqref{eq:permeability_dx} over the entire distribution. 
At this limit, particle transport no longer depends on the pore growth rate.
Instead, under the assumption that the pores are evenly distributed throughout the DIB, the permeability of the bilayer is simply ${\kappa^{\alpha} =  D_0^{\alpha}}\phi_0/h$.

\subsection{Acceptor Dye Density} Next, we apply solutions to the permeability to solve Eq.~\eqref{eq:dimlessdens} for the density of each species in the acceptor droplet. 
From our previous discussion, we know that the density depends on both the P\'eclet number and the particle size.
Figure~\ref{fig:density} emphasizes that the density changes \textit{quantitatively} with $\mathrm{Pe}$ but \textit{qualitatively} with $x_0$. 
In the left panel, we observe that increasing $\mathrm{Pe}$ shifts the curves to smaller $t/\tau_D$. 
Since $\mathrm{Pe}$ is the ratio of the particle diffusion timescale to the pore growth timescale, increasing $\mathrm{Pe}$ for fixed $x_0$ (and therefore fixed $D^\alpha$ and $\tau_D$) involves decreasing $\tau_P$ so that the pore growth rate becomes faster than the particle diffusion rate. 
For example, if ${\mathrm{Pe = 1.0}}$ particle diffusion occurs at the same rate as pore growth. 
Increasing $\mathrm{Pe}$ in the Ostwald ripening mechanism, for which ${\tau_P = R_0^3/M_0 \sigma}$, entails decreasing the initial critical radius, $R_0$, or increasing the surfactant mobility, $M_0$, and line tension, $\sigma$, at the pore interface, each of which will increase the pore growth rate.
As discussed following our derivation of Eq.~\eqref{eq:velocity}, pores will grow faster with increasing $M_0$ and $\sigma$ because the surfactants will be moving more rapidly and there is a stronger driving force for pores to become larger and reduce the ratio of pore perimeter to area.
Additionally, a smaller initial pore set by $R_0$ grows faster because increasing the pore area by the size of a single surfactant results in a larger increase in the pore radius. 
This can also be seen by inspecting the short-time limit of Eq.~\eqref{eq:2Dcriticalradius}: ${\overline{R_c} \sim 1 + 4 t M_0 \sigma (R_0)^{-3}/3}$, 
from which we see that reducing $R_0$ increases the initial rate of change of the pore radius.
As $\mathrm{Pe}$ increases, pores grow more quickly than particles diffuse through the pores and transport is limited by the particle diffusion timescale. 
On the other hand, if $\mathrm{Pe} < 1.0$ then $\tau_P > \tau_D$ and pores grow more slowly than particles diffuse.
This shifts the curves to larger $t/\tau_D$ because pores with $R > r^\alpha$ do not exist until later times.
In this limit, regardless of how quickly a species moves through a droplet, its transport between droplets is limited by the availability of large enough pores in the DIB.

The right panel of Fig. \ref{fig:density} shows how species density varies at a fixed $\mathrm{Pe}$ for the same $x_0$ values displayed in Fig.~\ref{fig:permeability}. 
Particles for which the permeability is finite at early times result in \textit{exponential} early growth in the density while those with zero permeability are unable to pass the membrane until later times when the permeability becomes finite.
As with the permeability, the transition between zero and finite density corresponds to the time at which the pore distribution first includes pores larger than the particle size. 
We see that even particles up to $100$ times larger than the initial pores can cross the bilayer, but it will take nearly a billion times longer to observe this than for particles that fit within the initial distribution.
Notably, by increasing $x_0$ in the right panel of Fig.~\ref{fig:density}, we are also slowing down the diffusion of dye, thereby increasing $\tau_D$.
To maintain a fixed ratio ${\mathrm{Pe} = \tau_D/\tau_P}$ this necessarily increases $\tau_P$, which corresponds to decreasing the pore growth rate.
Therefore, these curves reflect a situation in which the pore growth timescale becomes slower as the particle size increases.

\begin{figure}
	\includegraphics[width=.49\textwidth]{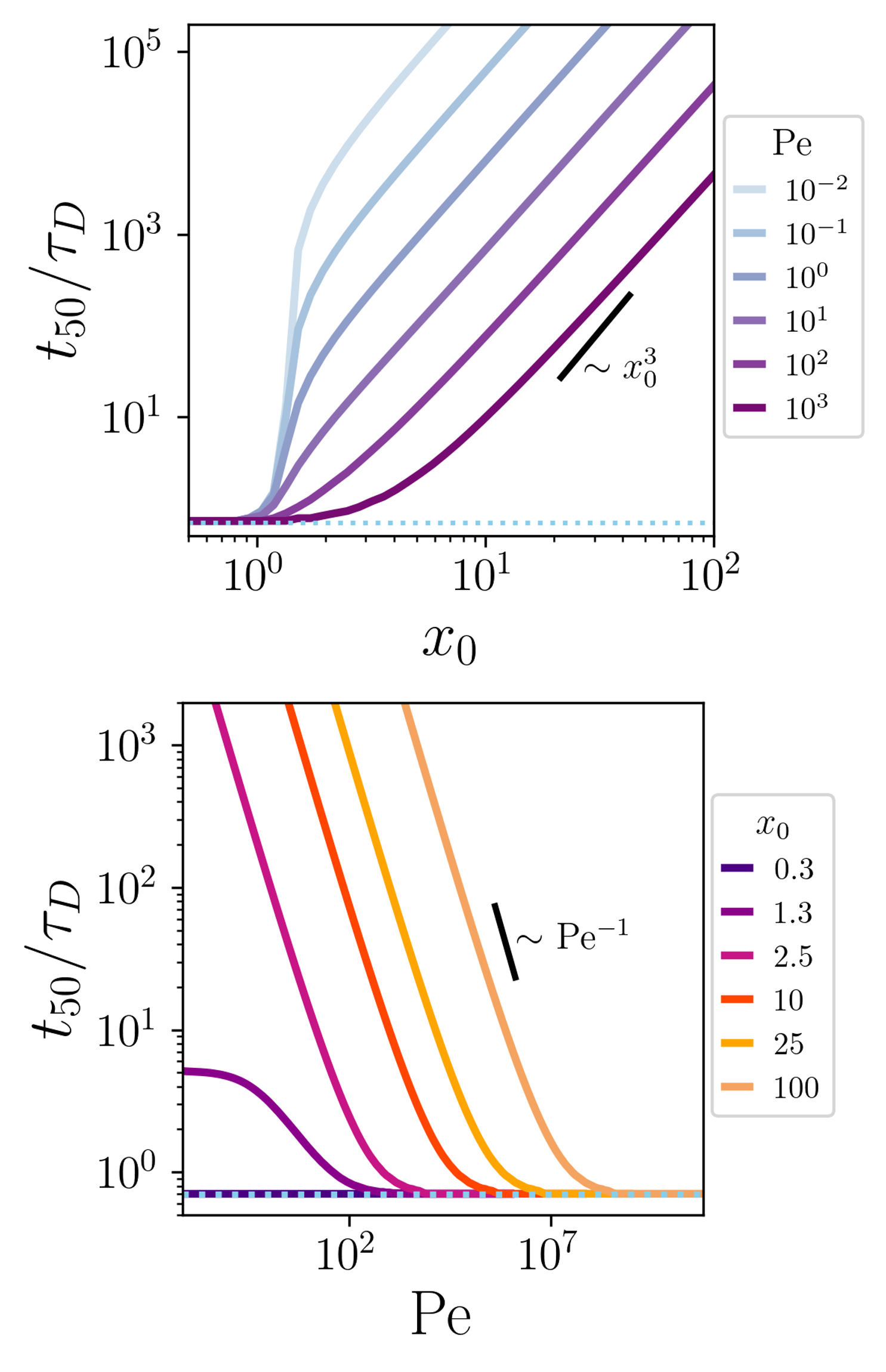}
	\caption{The time at which the dimensionless density in the acceptor drop reaches half its final value (${\overline{\rho}_a^\alpha = 0.5}$), relative to the dye diffusion timescale, $t_{50}/\tau_D$, is plotted for: (top panel) varying dye size relative to initial pore radius $x_0=r^{\alpha}/R_0$ and fixed pore growth rate relative to dye diffusion $\mathrm{Pe} = \tau_D/\tau_P$ and (bottom panel) fixed $\mathrm{Pe}$ and varying scaled dye size $x_0$. The dotted blue line in each panel lies at ${t_{50}/\tau_D = \ln(2)}$. }
	\label{fig:t50}
\end{figure}

The density curves in Fig.~\ref{fig:density} reveal signatures of the pore growth mechanism. 
We expect that if the pores are growing with $R_c \sim t^{1/3}$ then the time at which particles of size $x_0$ cross the bilayer should also scale $t \sim x_0^3$.
From the expression for the critical radius in Eq.~\eqref{eq:2Dcriticalradius}, this scaling only arises in the limit that the dye is much larger than the initial critical pore size.
We define ${t_{50}/\tau_D}$ as the time at which the dye density in the acceptor has increased to half its final value, ${\overline{\rho}_a^\alpha = 1/2}$, and confirm in Fig.~\ref{fig:t50} how this time-point changes with $x_0$ and $\mathrm{Pe}$. 
A full contour map of ${t_{50}/\tau_D}$ across ${10^{-1} < x_0 < 10^2}$ and ${10^{-1} < \mathrm{Pe} < 10^3}$ is provided in the SI. 
Figure~\ref{fig:t50} further highlights that the density variation depends more strongly on $x_0$ than on $\mathrm{Pe}$. 
We see that ${t_{50}/\tau_D \sim x_0^3}$ for large $x_0$ and ${t_{50}/\tau_D \sim \mathrm{Pe}^{-1}}$ for small $\mathrm{Pe}$. 
The cubic scaling with $x_0$ follows from the (approximate) Ostwald scaling growth of the pore radii found in Eq.~\eqref{eq:2Dcriticalradius}.
It also follows from $\mathrm{Pe} \sim \tau_D$ that ${t_{50}/\tau_D}$ should vary inversely with $\mathrm{Pe}$.
We emphasize that in this region $t_{50}$ is independent of $\tau_D$ and varies proportionally to $\tau_P.$

In the top panel of Fig.~\ref{fig:t50}, we see that all $\mathrm{Pe}$ curves plateau to a constant value for small $x_0$. 
This follows from the equation for the non-dimensionalized permeability Eq.~\eqref{eq:permeability_dx}, which approaches unity when the particles are smaller than the initial pores, $x_0 \to 0$. 
The resulting form of the density as ${\overline{\kappa}^\alpha \to 1}$ is: ${\overline{\rho}_a^{\alpha} = 1 - e^{-\overline{t}}}$. 
From this, it follows that the time for $\overline{\rho}_a^{\alpha}$ to reach $1/2$  approaches $\ln 2$.
Each of the $\mathrm{Pe}$ curves reaches this plateau at ${x_0 = 3/2}$, which is the upper bound for $x_0$ with finite permeability at early times (Fig.~\ref{fig:permeability}). 

The rate of pore growth does not affect ${t_{50}/\tau_D}$ for small $x_0$ because the initial distribution features pores large enough for species up to $3 R_0/2$ to pass through the bilayer. 
In the case that ${x_0 > 3/2}$, the time for any fraction of the particles to diffuse into the acceptor depends on the pore growth rate. 
Systems in which pores grow faster than particles diffuse (large $\mathrm{Pe}$) will therefore have lower ${t_{50}/\tau_D}$ for the same $x_0$. 
As a result, the transition to ${t_{50}/\tau_D \sim x_0^3}$ is steeper as $\mathrm{Pe}$ decreases. 
This is because when $\mathrm{Pe}$ is large, the pore distribution broadens at earlier times. 
By similar logic, when $\mathrm{Pe}$ is small,  the pore distribution grows more slowly so it takes longer for half of the particles with the same $x_0$ to diffuse into the acceptor. 
We expect that larger particles will have a diverging ${t_{50}/\tau_D}$ as ${\mathrm{Pe} \to 0}$ because ${\overline{\kappa}^\alpha = 0}$ when ${x > 3/2}$, and as ${\mathrm{Pe} \to 0}$ the pore distribution will never broaden to include larger $x_0$. 
In this case, even after infinite waiting time, particles will never transfer to the acceptor droplet.

The bottom panel shows that the curves also plateau to $\ln{2}$ for a fixed particle size as the pore growth rate increases. 
The $\mathrm{Pe}$ for which each curve plateaus increases with species size. 
This $\mathrm{Pe}$ corresponds to the pore growth rate at which, in the time it takes particles of different $r^\alpha$ to diffuse to the DIB, all the pores are larger than $r^\alpha$.
As $\mathrm{Pe}$ increases, particles cross the bilayer faster because large enough pores exist and transport is only limited by how quickly a species can diffuse between the donor and acceptor. 
As a result, larger particles have higher ${t_{50}/\tau_D}$ at the same $\mathrm{Pe}$. 
We also see that curves for ${x_0 < 3/2}$ plateau to a constant ${t_{50}/\tau_D}$ at low $\mathrm{Pe}$. 
This constant ${t_{50}/\tau_D}$, which decreases with $x_0$, is an effect of those species displaying finite permeability and exponential density growth at early times. 

In practice, one cannot vary $x_0$ and $\mathrm{Pe}$ independently.
For systems in which the particle diffusion constant takes the form of a Stokes--Einstein relation with ${D_0^{\alpha} \propto 1/r^{\alpha}}$, $\mathrm{Pe}$ will depend on $x_0$. 
To understand the expected behavior of these systems, we now assume a diffusion constant of a Stokes-Einstein form resulting in $\mathrm{Pe} \propto x_0$.
For a given system, we must therefore hold ${\mathrm{Pe}/x_0}$ fixed.
Since $\mathrm{Pe}$ is a ratio of timescales and $x_0$ is a ratio of length scales, $\mathrm{Pe}/x_0$ naturally is a ratio of velocities.
We define the ``pore growth velocity'' as $v_p = r^{\alpha}/\tau_p$.
This measures the velocity with which a pore must grow for it to increase in size by the radius of the dye particle in a time $\tau_P$.
We similarly define the ``dye diffusion velocity'' as $v_D = R_0/\tau_D$, which is the velocity with which a dye particle moves if it travels over the size of the initial pores in a single diffusion time $\tau_D$.
In terms of these quantities, we have ${\mathrm{Pe} / x_0 = v_P/v_D}$. 
Holding this ratio of velocities constant allows us to keep $\tau_P$ fixed.
Increasing ${v_P/v_D}$ corresponds to a faster pore growth rate, which allows us to understand how transport dynamics depend on the particle size alone.

Figure~\ref{fig:experiment} illustrates that when we fix the membrane properties, ${t_{50}/\tau_D}$ increases quadratically, rather than cubically, with the particle size. 
We stress that the pore growth rate still scales as $R_c \sim t^{1/3}$ and that this quadratic scaling is to be expected since we are effectively dividing the scaling in the upper panel of Fig.~\ref{fig:t50} by a single power of $x_0$.
As in Fig.~\ref{fig:t50}, we see that ${t_{50}/\tau_D \to \ln{2}}$ for particles that fit through the pores that are present in the initial distribution.

\section{Experimental Validation}
\label{sec:Exp}
We now have clear predictions that can be verified with experiments. 
To test our theory, we suggest an experiment that measures the density of differently sized dyes, using fluorescence intensity for traditional dyes or UV-vis spectroscopy for plasmonic nanoparticles, as they cross from a donor to an acceptor through a DIB~\cite{Bayley2008DropletBilayers, Schlicht2015Droplet-interface-bilayerNetworks, Huang2022PhysicochemicalBilayers, Allen2022Layer-by-layermulti-DIBs}. 
As discussed above, our theoretical framework is fully parameterized by the dimensionless dye size $x_0$ and the P\'eclet number $\mathrm{Pe}$. 
Since ${x_0 = r^\alpha/R_0}$, it is essential to fix $R_0$ across experiments in order to isolate effects of varying the dye size $r^\alpha$. 
To do so, we suggest making DIBs from monolayer-laden droplets that have the same surface coverage.
The surface coverage serves as the main means of controlling membrane properties such as the in-plane surfactant mobility and the membrane line tension, which in turn determine the initial critical radius. 
The simplest way to ensure consistent bilayer properties for dyes of different sizes is to simultaneously monitor the transport of all dyes, although this approach presents challenges in tracking the density of multiple particles simultaneously.

\begin{figure}
    \centering
    \includegraphics[width=1\linewidth]{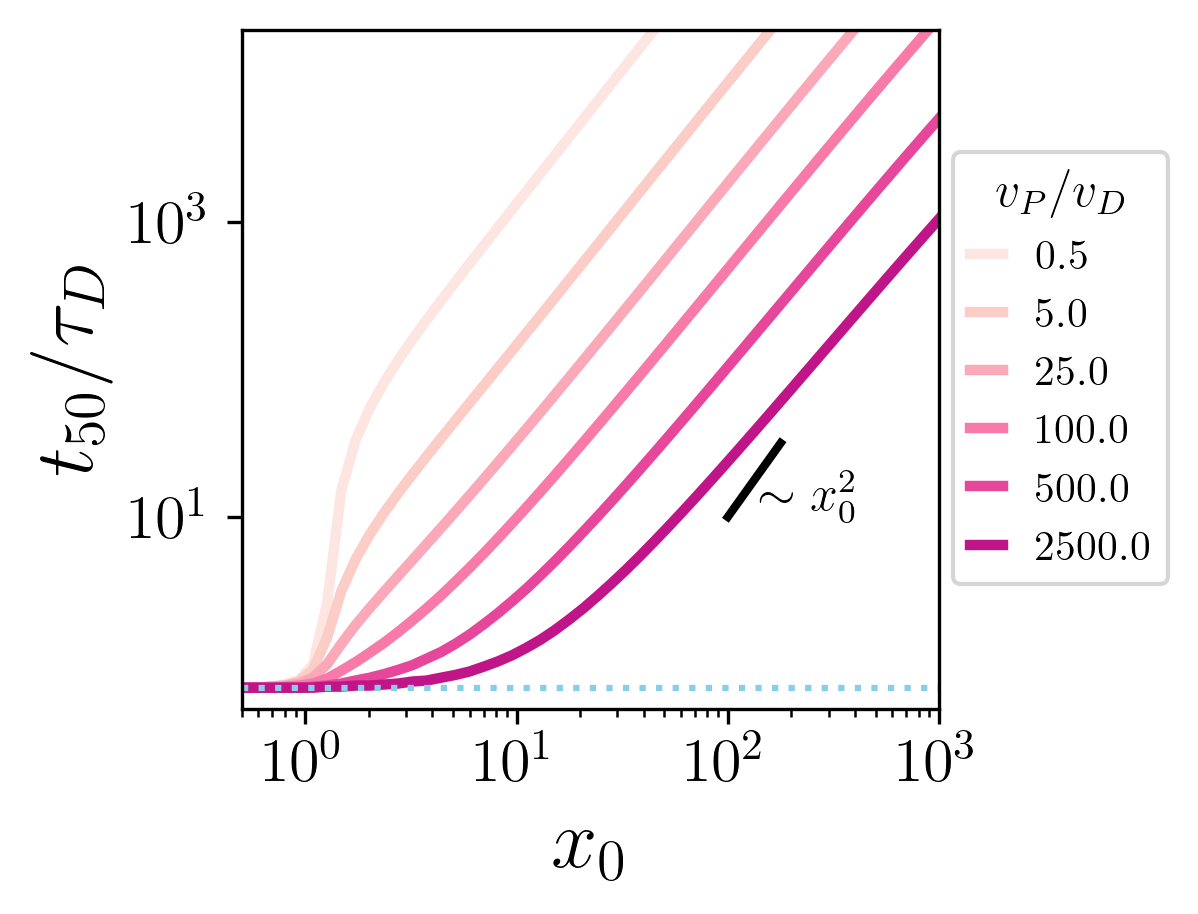}
    \caption{A plot of the time at which the dimensionless density in the acceptor reaches half its final value ($\overline{\rho}_a^\alpha = 0.5$), relative to the dye diffusion time scale $t_{50}/\tau_D$, plotted as a function of dye size relative to the initial pore radius $x_0 = r^{\alpha}/R_0$. Each line represents a different pore growth rate that is independent of the dye size, and is represented by the ratio between the pore growth velocity $v_P$ and the dye diffusion velocity $v_D$; the larger this ratio, the faster the growth of pores. The dotted blue line lies at $t_{50}/\tau_D = \ln(2)$.}
    \label{fig:experiment}
\end{figure}

After tracking how the density of each species in an acceptor droplet changes with time, one can compare the data to the  predictions for ${t_{50}/\tau_D}$ as a function of particle size.
Measuring the difference in ${t_{50}/\tau_D}$ for known particle sizes will suggest the dominant mechanism of pore growth in a particular DIB. 
For example, since the membrane properties (and therefore $\tau_P$) are held constant in this experiment, we expect that if the dominant mechanism of pore growth is either Ostwald ripening or coalescence, then ${t_{50}/\tau_D}$ will grow quadratically with the particle size as shown in Fig.~\ref{fig:experiment}.
On the other hand, if particles of vastly different sizes cross the DIB at similar times, then pore growth is likely dominated by desorption.

Performing these measurements offers both a way to test the theoretical predictions and a method to understand physical properties of the DIB. 
By systematically varying the dye size and tracking the resulting density changes, one may extract key physical parameters such as the initial pore size, the effective surfactant mobility in the bilayer, and even determine if the DIB stretches significantly over time. 
In this way, the theory provides insight into dynamic permeability that arises in metastable membranes and offers not only a predictive framework but also a means to infer structural properties that are otherwise challenging to access with experiments.

\section{Summary and Conclusions}
\label{sec:Conc}
We have shown how structural changes in a DIB can be investigated by studying how the bilayer regulates particle transport.
To do this, we considered a model system in which two surfactant-coated droplets are brought into contact, forming a bilayer at their interface.
One droplet is initially loaded with particles while the other is initially empty; the rate at which particles transfer between the droplets depends on the \textit{dynamic permeability} of the DIB.
This time-dependent permeability provides a window into the distribution of the number and sizes of pores which act as conduits for dye particles. 
We can thus infer structural properties of the membrane through dye-transport measurements.

To illustrate how various pore growth mechanisms affect transport across the DIB, we established a mechanical picture of pore growth kinetics.
We focused on growth via Ostwald ripening while also briefly discussing pore growth via coalescence and presenting the theory of desorption-driven growth in the SI.
We identified that Ostwald ripening and coalescence exhibit algebraic scaling in two dimensions while the desorption mechanism features an exponential pore growth rate.
From this we can conclude that Ostwald ripening and coalescence are appropriate descriptions of DIBs that are stable for several days while desorption better describes membranes that disintegrate quickly.
Using the pore distributions, we found how the membrane permeability and the species density on either side of the DIB varies with the species size and the pore growth rate.
Small particles have non-zero permeability at early times while larger particles must wait until the pore distribution includes large enough pores before they cross the bilayer.
The time until large enough pores exist depends on the pore growth mechanism.
Those particles with finite early permeability also exhibit an exponential increase in the density at early times.
Increasing the pore growth rate shifts the density curves to earlier times.

While we provided density solutions for fixed dye size $x_0$ and $\mathrm{Pe}$, we noted that one cannot vary these parameters independently because changing the dye size affects the dye diffusion timescale, which in turn varies $\mathrm{Pe}$.
We therefore also provided the scaling behavior for varying dye size for a fixed membrane, which will feature a constant pore growth rate.
We showed that when the pore growth rate $\tau_P$ is constant, the time for a given fraction of particles to enter the acceptor droplet scales quadratically rather than cubically with the particle size.
We proposed an experiment to verify these theoretical predictions.
Such a study would involve measuring the time evolution of dye particle density in the acceptor droplet.
The scaling behavior of this measurement will point to the dominant mechanism of pore growth in the bilayer and indicate the timescale of bilayer dissolution.
As DIBs are commonly applied as model systems for studying lipid bilayers in biological systems, we hope that our model can inform more specialized DIBs and provide insight into structural properties in synthetic models and biological systems alike.

While our model presents an important step forward, there are several key areas for improvement and expansion. 
For example, we have not accounted for geometric fluctuations in the DIB membrane, which are known to affect the growth kinetics of domains in lipid bilayers~\cite{Yu2025Pattern}. 
Although these fluctuations do not prevent pores from growing to cover the entire DIB when droplets fully coalesce, they can influence the late-time growth dynamics of the pores.
This could, in turn, alter the time-dependent permeability, especially at longer timescales, and warrants further investigation.
Additionally, while we have treated three mechanisms for pore growth (Ostwald ripening, coalescence, and desorption) separately, these processes are likely to occur concurrently in real systems.
A promising direction for future work would be to develop models that account for multiple growth pathways simultaneously.

We cannot ignore the possibility of hitherto unknown pore growth mechanisms. 
In such cases, there is \textit{no} pore distribution, $n(R,t)$, ready to be used in Eq.~\eqref{eq:dimensionalperm}.
However, our proposed constitutive equation for the current of dye across the DIB, used in Eq.~\eqref{eq:drhodt}, can be used to \textit{infer} $n(R,t)$ from experimental data. 
By measuring the density as a function of time, through methods discussed in Section~\ref{sec:Exp}, the dynamic permeability may be fitted from Eq.~\eqref{eq:drhodt}. 
If these steps are repeated for a pair of dye molecules with very similar sizes in identically prepared droplets, the difference between their resulting inferred permeabilities at the same point in time closely approximates the derivative of Eq.~\eqref{eq:dimensionalperm} with respect to $r^{\alpha}$. 
Therefore, if this experiment is repeated for many different pairs of dye molecules at many different times, a picture of the pore distribution and its evolution can be built up. 
Importantly, this method does not interfere with the natural evolution of the membrane, a risk run with other direct imaging methods.
As such, the framework presented here can provide unprecedented insight into the structure and dynamics of metastable membranes both theoretically and experimentally.

\acknowledgments 
We thank Luke Langford and Eric Weiner for helpful discussions.
This work is supported by the U.S. Department of Energy, Office of Science, Office of Basic Energy Sciences, Materials Sciences and Engineering Division under Contract No. DE-AC02-05-CH11231 within the Adaptive Interfacial Assemblies Towards Structuring Liquids program (KCTR16). 
N.A.S. acknowledges partial support from the NSF Graduate Research Fellowship under DGE 1752814 and DGE 2146752.

\clearpage
\onecolumngrid

\section*{Supporting Information}

\renewcommand{\thesection}{S\arabic{section}}
\renewcommand{\theequation}{S\arabic{equation}} 
\renewcommand{\thefigure}{S\arabic{figure}}
\renewcommand{\thepage}{S\arabic{page}}
\setcounter{equation}{0}
\setcounter{section}{0}
\setcounter{figure}{0}
\setcounter{page}{1}

\section{Finite normal stress difference near pore radius}

We show how, by assuming that the normal stress difference is sharply peaked at the pore interface, the jump in the radial across is the interface is related to line tension as in Eq.~\eqref{eq:stressjump}. 
First, we know from Eq.~\eqref{eq:stressjump} of the main text
\begin{equation}
    \frac{\partial \Sigma_{rr}}{\partial r} = \frac{2}{r}\left(\Sigma_{rr}-\Sigma_{tt}\right).
\end{equation}
Integrating both sides from $R^{\dagger}_{-}$ (inside the pore) to $R^{\dagger}_{+}$ (outside the pore) we obtain
\begin{equation}
    \Sigma_{rr}(R^{\dagger}_{+})-\Sigma_{rr}(R^{\dagger}_{-}) = \displaystyle \int_{R^{\dagger}_{-}}^{R^{\dagger}_{+}} dr \ \frac{2}{r} \left(\Sigma_{rr}-\Sigma_{tt}\right).
\end{equation}
In a spatially uniform system, the normal stress difference must vanish. 
Close to the pore interface, however, there is a sudden change in composition that results in the normal stress difference being sharply peaked around $R^{\dagger}$. 
This allows the integral be approximated following Ref.~\cite{Lindell1993DeltaMethod} by Taylor expanding $1/r$ about that radius, and truncating at desired order:
\begin{equation}
    \displaystyle\int_{R^{\dagger}_{-}}^{R^{\dagger}_{+}} dr \ \frac{2}{r} \left(\Sigma_{rr}-\Sigma_{tt}\right) = \frac{1}{R^{\dagger}}\sum_{n=0}^{\infty} (-1)^n  (R^{\dagger})^{-n} \displaystyle\int_{R^{\dagger}_{-}}^{R^{\dagger}_{+}} dr \ (r-R^{\dagger})^{n} \left(\Sigma_{rr}-\Sigma_{tt}\right).
\end{equation}
If we discard those terms $\mathcal{O}((R^{\dagger})^{-2})$ and greater, we obtain the result in the main text:
\begin{equation}
    \Sigma_{rr}(R^{\dagger}_{+})-\Sigma_{rr}(R^{\dagger}_{-}) = \frac{2}{R^{\dagger}} \sigma + \mathcal{O}((R^{\dagger})^{-2}),
\end{equation}
where the line tension is defined as
\begin{equation}
    \sigma = \displaystyle\int_{R^{\dagger}_{-}}^{R^{\dagger}_{+}} dr \ \left(\Sigma_{rr}-\Sigma_{tt}\right).
\end{equation}

\section{Linearization of In-plane Surfactant Dynamics}
Here we demonstrate, through linearizing the equations of motion for the membrane composition, that the composition current is directly proportional to the gradient in the effective surface tension [as in Eq.~\eqref{eq:dynamicmech}] and that, in steady state, the effective surface tension solves the Laplace equation [see Eq.~\eqref{eq:radialdynamicmech}]. 

We begin with the continuity equation
\begin{equation}
    \frac{\partial \phi}{\partial t} = - \bm{\nabla}\cdot\mathbf{j} = -\bm{\nabla} \cdot \left(M(\phi) \bm{\nabla} \cdot \bm{\Sigma} \right),
\end{equation}
where the concentration dependence of the mobility, $M$, has been made explicit.
We suppose that the composition varies weakly around the uniform value $\phi^{\rm mem}_{0}$, so that $\phi \approx \phi^{\rm mem}_0 + \delta \phi$. 
By weakly varying, we suppose that $1 \gg|\bm{\nabla} \delta \phi| \gg \nabla^2 \delta \phi \gg (\cdots)$ so that only the lowest derivatives of $\delta \phi$ need to be kept to leading order. 
This assumption greatly simplifies the stress tensor since, from Eq.~\eqref{eq:stress} of the main text, all terms other than the effective surface tension include derivatives of $\delta \phi$. 
It follows from this assumption that $\bm{\Sigma} \approx \upgamma_{\rm eff} \mathbf{I}$. 
To this order, the continuity equation is then written:
\begin{equation}
    \frac{\partial \phi}{\partial t} = -\bm{\nabla} \cdot \left(M(\phi) \bm{\nabla} \upgamma_{\rm eff}\right).
\end{equation}
Expanding the divergence on the right-hand side, we obtain $\bm{\nabla} M(\phi) \cdot \bm{\nabla} \upgamma_{\rm eff} + M(\phi) \nabla^2 \upgamma_{\rm eff}$. 
The first term is necessarily $\mathcal{O}(|\bm{\nabla} \delta \phi|^2)$, while the lowest order contribution to the second is $M(\phi^{\rm mem}_0) \nabla^2 \upgamma_{\rm eff}$. 
If we define the constant single-particle mobility in a uniform composition membrane as $M(\phi^{\rm mem}_0) = M_0$, the continuity equation becomes, to leading order, 
\begin{equation}
     \frac{\partial \phi}{\partial t} \approx - M(\phi^{\rm mem}_0) \nabla^2 \upgamma_{\rm eff}. 
 \end{equation}
Thus, if we neglect time-variations in the composition, the effective tension is \textit{harmonic}; $\nabla^2 \upgamma_{\rm eff} =0$. 
 Furthermore, we may extract the approximate current, \textit{post hoc}, $\mathbf{j} \approx M_0 \bm{\nabla} \upgamma_{\rm eff}$ as in Eq.~\eqref{eq:dynamicmech} of the main text.

\section{Details of universal scaling function}
We show that, under the assumptions that the bilayer size far exceeds any of the pores and that the system has reached the self-similar, stable, long-time distribution given by Eq.\eqref{eq:ansatz}, we obtain \textit{the same} universal scaling function as in the LSW theory. 
We also present some details of how to arrive at its form.

To derive the scaling function $\mathcal{N}(x)$, we begin with the continuity equation (absent a generation term):
\begin{equation}\label{eq:liouville}
    \frac{\partial n}{\partial t} + \frac{\partial}{\partial R}(v(R) n) = 0
\end{equation}
and use the relevant interface velocity and self-similarity ansatz $n(R,t) = \mathcal{N}(x)/R_c^{-(d+1)}$ where $x = R/R_c$.
For the Ostwald ripening mechanism, Eq.~\eqref{eq:liouville} becomes:
\begin{equation}\label{eq:liouville_ost}
\begin{array}{cc}
     &  \dfrac{\partial}{\partial t} (R_c ^{-(d+1)}\mathcal{N}(x)) + \dfrac{\partial}{\partial R} \left(\dfrac{2 M_0 \sigma}{R^2 \ln(\ell/R)}\left(\dfrac{R}{R_c} -1)\right)\right) (R_c ^{-(d+1)}\mathcal{N}(x))\\\\

    & R_c ^{-(d+2)} \dot{R_c}[(d+1) \mathcal{N}(x) + x \mathcal{N}'(x)] = \dfrac{1}{R_c^3 R_c^{-(d+1)}} \dfrac{1}{\ln{(R_c/\ell)}} \dfrac{\partial}{\partial x}\left(\dfrac{2M_0 \sigma}{x}\left(1 - \dfrac{1}{x}\right) \mathcal{N}(x)\right)\\\\

    & - R_c^2 \ln(R_c/\ell) \dot{R_c} = \dfrac{2M_0 \sigma \dfrac{\partial}{\partial x}\left( \dfrac{1}{x} - \dfrac{1}{x^2} \right)\mathcal{N}(x)}{[(d+1)\mathcal
    N(x) + x \mathcal{N}'(x)]} = \varepsilon.
\end{array}
\end{equation}
We note that since $R_c \ll \ell$, the logarithmic term in the denominator of the penultimate line is large and negative.
Furthermore, it varies logarithmically with $R_c$ and thus much more slowly than the algebraic terms $R_c^2$ and $R_c^{-3}$.
We therefore treat $\ln{R_c/\ell}$ as quasi-static and extract it from the derivative to make the problem analytically tractable.
Multiplying both sides by $R_c^2 \ln {R_c/\ell}$ provides the final line, where we have grouped those terms that depend on $R_c$ and $\mathcal{N}(x)$ on the left and right sides of the equation, respectively.
Since the LHS of Eq.~\eqref{eq:liouville_ost} depends on time through $R_c$ and the RHS depends only on $x$, both sides must be equal to a constant that is independent of both $x$ and $t$.

We can solve for $\mathcal{N}(x)$ and identify $\varepsilon$ by decomposing the integrand $\ln \mathcal{N}(x)$ into partial fractions and finding the roots of the denominator:
\begin{equation}\label{eq:ln_N}
    \begin{array}{cc}
         & \ln\mathcal{N}(x) = \displaystyle\int^{x} \dfrac{\mathcal{N}'(x)}{\mathcal{N}(x)}dx = \displaystyle\int^x \dfrac{dy}{y}\dfrac{(-\varepsilon(d+1) y^3 + 2 - y)}{(\varepsilon y^3 - y + 1)}.
    \end{array}
\end{equation}
The polynomial in the denominator of Eq.~\eqref{eq:ln_N} is a depressed cubic because it does not have a $y^2$ term.
One may apply Cardano's formula, which states that for a depressed cubic polynomial with the form $a y^3 + b y + c$, the discriminant is $\mathcal{D} = -4 b^3 a - 27 a^2 c^2$.
Applying the coefficients for the depressed cubic in Eq.~\eqref{eq:ln_N} gives
\begin{equation}
    \mathcal{D} = -4(-1)\varepsilon-27\varepsilon^2(1).
\end{equation}
Because $x$ is a ratio of length scales, we know the roots of the denominator in Eq.~\eqref{eq:ln_N} must be positive and real.
This restricts $\mathcal{D}\geq 0$, which in turn results in $
\varepsilon \leq 0$.
When $\varepsilon = 4/27, \mathcal{D} = 0$ and the polynomial has a repeated real root.
If $\varepsilon < 4/27, \mathcal{D} > 0$ and the polynomial has three distinct real roots.
LSW argue that $\varepsilon = 4/27$ is the only stable solution to the problem~\cite{Lifshitz1961TheSolutions}.
We proceed with this value of $\varepsilon$ to find the full form of $\mathcal{N}(x)$:
\begin{equation}\label{eq:N}
\begin{array}{cc}
     &  \ln \mathcal{N}(x) = \displaystyle\int^x dy \dfrac{(-\varepsilon(d+1)y^3 + 2 - y)}{y(\varepsilon y^3 - y + 1)} = \displaystyle\int^x dy \dfrac{2 - y -(4/27)(d+1) y^3}{y\left((4/27) y^3 - y + 1 \right)} = \displaystyle\int^x dy \dfrac{54 - 27y - 4(d+1)y^3}{4y \left(y - (3/2)\right)^2 (y + 3)}\\\\
     & = \displaystyle\int^x dy \dfrac{A}{y} + \dfrac{B}{y - (3/2)} + \dfrac{C}{\left(y - (3/2)\right)^2} + \dfrac{D}{y + 3},\\\\
     &\text{ where } 54 - 27 y - 4(d+1)y^3 = 4A\left(y - \dfrac{3}{2}\right)^2 (y + 3) + 4 B y \left(y - \dfrac{3}{2}\right) (y + 3) + 4 C y (y + 3) + 4 D y \left(y - \dfrac{3}{2}\right)^2.
\end{array}
\end{equation}
We solve for each of $A, B, C \text{ and } D$ by setting $y$ to the value that takes the denominator to $0$ (i. e. $y = 0$ to solve for $A$, $y = 3/2$ to solve for $C$, $y = -3$ to solve for $D$).
We then integrate and exponentiate Eq.~\eqref{eq:N} to obtain
\begin{equation}
    \mathcal{N}(x) = \mathcal{C}x^2 \left( \dfrac{3}{2} - x\right)^{-\left(2 + \dfrac{5d}{9}\right)} (3 + x)^{-\left(1 + \dfrac{4d}{9}\right)} \exp{\left( - \dfrac{d}{3 - 2x}\right)},
\end{equation}
which is general to any dimension.
We use the $d = 2$ case to find results for Ostwald ripening on $2$-dimensional membrane systems.

\section{Deriving the timescale of growth through desorption}
In this appendix, we discuss the pore growth dynamics in the case that growth is driven purely by desorption of surfactants from the pore perimeter.
We determine how the pore size distribution evolves with time and how this affects the dynamic permeability of the membrane.
Our starting assumption is that surfactant exclusively desorbs from the perimeter of the pore at a constant rate rate $\omega$, such that the radial velocity of the pore is
\begin{equation}
    v(R) = 2\pi \omega R.
\end{equation}
This form of the velocity shows that the radial growth rate is proportional to the pore perimeter.

We shall suppose that the initial distribution of pore radii is known, which allows us to determine the pore-size distribution per unit area, $n(R,t)$, from the continuity equation:
\begin{equation}\label{eq:desorption_continuity}
    \frac{\partial}{\partial t} n(R,t) + \frac{\partial}{\partial R}[v(R) n(R,t)] = 0.
\end{equation}
Note that there is no source term here, $\mathcal{C}=0$ in Eq.~\eqref{eq:nucleusdist}.
This is because, in the absence of mechanisms such as coalescence, desorption only changes the pore size continuously.
Furthermore, in contrast to growth mechanisms such as Ostwald ripening and coalescence, in which the number density $n$ decreases as smaller pores shrink or pores merge together, desorption-driven growth conserves the number of pores.
As the pores grow larger, the total pore area increases, making the area fraction and number density proportional to one another.

Equation~\eqref{eq:desorption_continuity} is a linear first-order PDE in $n(R,t)$, which may be solved exactly. To simplify our analysis, we define 
\begin{equation}\label{eq:desorption_velocity}
    y(R,t) = v(R) n(R,t),
\end{equation}
so that Eq.~\eqref{eq:desorption_continuity} becomes
\begin{equation}
    \dfrac{\partial y}{\partial t} + v(R)\dfrac{\partial y}{\partial R}= 0.
\end{equation}
Now we make a change of variables from $(t,R)$ to $(s,x)$ defined through
\begin{subequations}
\begin{equation}
    s(t,R)= t,
\end{equation}
\begin{equation}
    x(t,R)=t + f(R).
\end{equation}
\end{subequations}
The chain rule then gives us 
\begin{subequations}
\begin{equation}
    \frac{\partial y}{\partial t} = \frac{\partial y}{\partial s} + \frac{\partial y}{\partial x}
\end{equation} 
\begin{equation}
    \frac{\partial y}{\partial R} = f'(R) \frac{\partial y}{\partial x}.
\end{equation}
\end{subequations}
These allow us to write Eq.~\ref{eq:desorption_continuity} as 
\begin{equation}
    \frac{\partial y}{\partial s} + \frac{\partial y}{\partial x} + v(R) f'(R) \frac{\partial y}{\partial x}= 0. 
\end{equation}
If we choose $f'(R) = - 1/v(R)$, then the final two terms of the above equation cancel, leaving 
\begin{equation}
    \frac{\partial y}{\partial s} = 0. 
\end{equation}
Therefore, $y$ is a function of $x$ \textit{only}. 
Recalling our choice of $f(R) = - \int^{R} dR'/v(R')$ we have 
\begin{equation}\label{eqn:11}
     y(R,t) = Y(x) = Y\left(t - \int^R \dfrac{dR'}{v(R')}\right).
\end{equation}
We evaluate Eq.~\eqref{eqn:11} using the velocity defined in Eq.~\eqref{eq:desorption_velocity}:
\begin{equation}
    \int^R \dfrac{dR'}{v(R')} = \int^R \dfrac{dR'}{2\pi \omega R'} = \dfrac{1}{2\pi \omega}\log{R}.
\end{equation}

Defining $\tau = (2\pi \omega)^{-1} $ yields
\begin{equation}
    \begin{array}{cc}
         &  x = t/\tau - \log{R},\\\\
         & y(R,t) = Y(t/\tau - \log{R}),
    \end{array}
\end{equation}
which implies
\begin{equation}
    n(R,t) = \dfrac{1}{2\pi \omega R}Y(t/\tau - \log{R}).
\end{equation}

Given an initial distribution $n(R,t=0) = n_0(R)$, we find $Y$ by solving
\begin{equation}
    n_0 (R) = \dfrac{1}{2\pi \omega R} Y(-\log{R}).
\end{equation}
Let us suppose that the initial distribution is Gaussian with mean $\overline{R}_0$ and standard deviation $\sigma$ then,
\begin{equation}
    n_0(R) =  \frac{2\sigma^2}{1 + \text{erf}\left(\sigma \overline{R}_0/2\right)}\frac{1}{\sqrt{2 \pi}\sigma}\exp\left(-\dfrac{(R - \overline{R}_0)^2}{2\sigma^2} \right).
\end{equation}
Note the somewhat complicated normalisation constant. The distribution is only defined on the interval $R \in [0,\infty]$, since pores cannot have negative radii, so the distribution must satisfy $\int_0^{\infty} dR n_0(R)=1$.
We find
\begin{equation}
    Y(-\log{R}) = \frac{2\sigma^2}{1 + \text{erf}\left(\sigma \overline{R}_0/2\right)}\dfrac{2\pi\omega R}{\sqrt{2\pi}\sigma}\exp{\left(-\dfrac{(R - \overline{R}_0)^2}{2\sigma^2}\right)},
\end{equation}
from which we obtain
\begin{equation}\label{eq:desorption_distribution}
\begin{array}{cc}
     &  n(R,t) = \dfrac{2\sigma^2}{1 + \text{erf}\left(\sigma \overline{R}_0/2\right)} \dfrac{1}{\sqrt{2\pi}\sigma(t)} \exp\left(-\dfrac{(R - \overline{R}(t))^2}{2\sigma(t)^2}\right)\\\\
     & \overline{R}(t) = \overline{R}_0 e^{t/\tau}\\\\
     & \sigma(t) = \sigma e^{t/\tau}.
\end{array}
\end{equation}
This shows that, up to modifications due to the one-sided nature of the distribution, an initially Gaussian $n(R,t)$ remains so as the pores grow but \textit{both} the mean and variance increase exponentially with $e$-folding time $\tau$. 
\begin{figure}[H]
    \centering
    \includegraphics[width=0.5\linewidth]{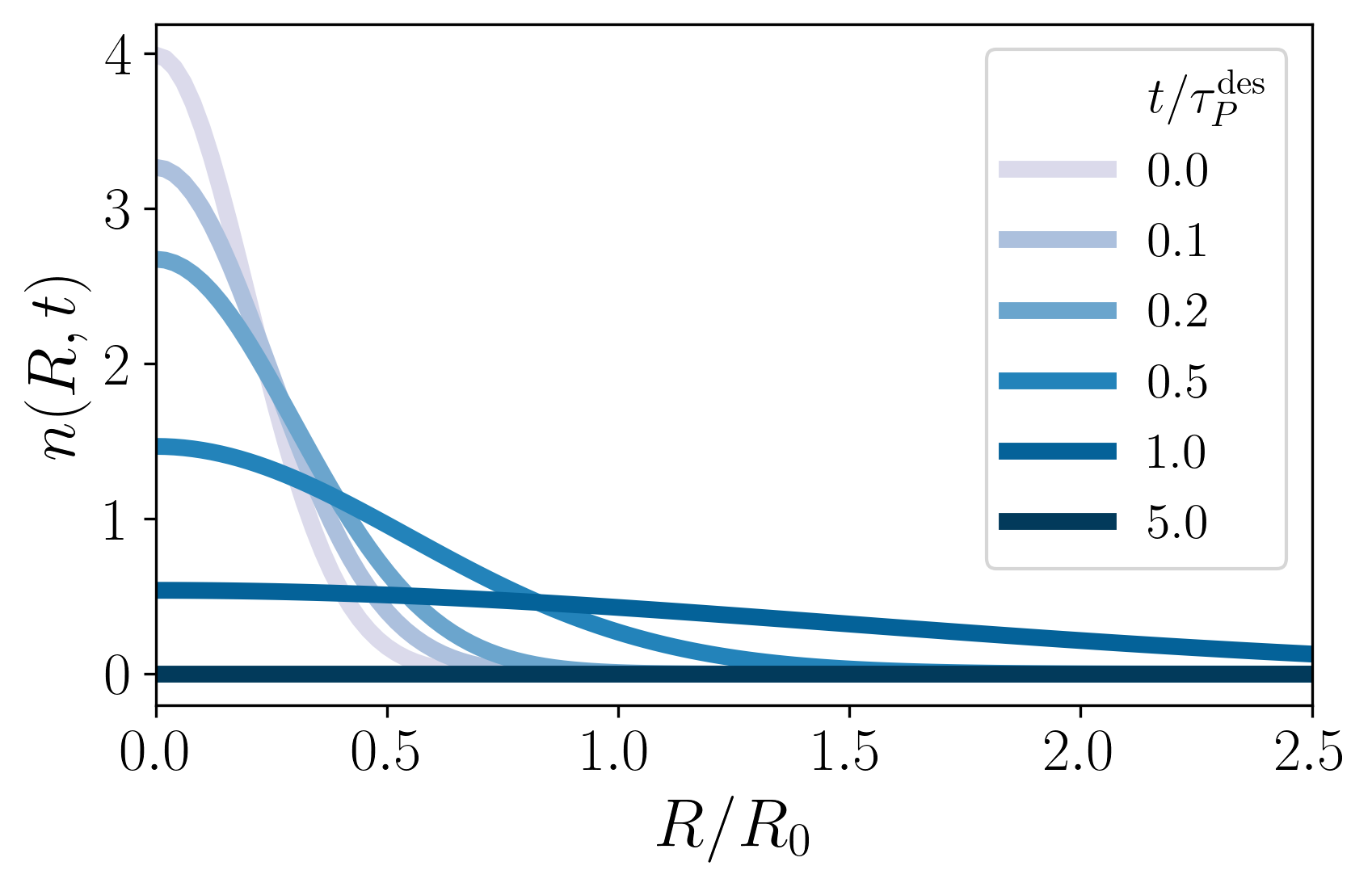} 
    \caption{Pore distribution for the desorption mechanism, plotted as a function of the pore size relative to the initial pore radius, $R/R_0$, for various different times in units of the characteristic desorption time, $t/\tau_P^{\rm des}$.}
    \label{fig:desorption_distribution}
\end{figure}
Figure~\ref{fig:desorption_distribution} shows how the pore size distribution evolves in the desorption mechanism.
The distribution remains Gaussian with a mean and variance that grow exponentially with time

\begin{figure}[H]
    \centering
    \includegraphics[width=1\linewidth]{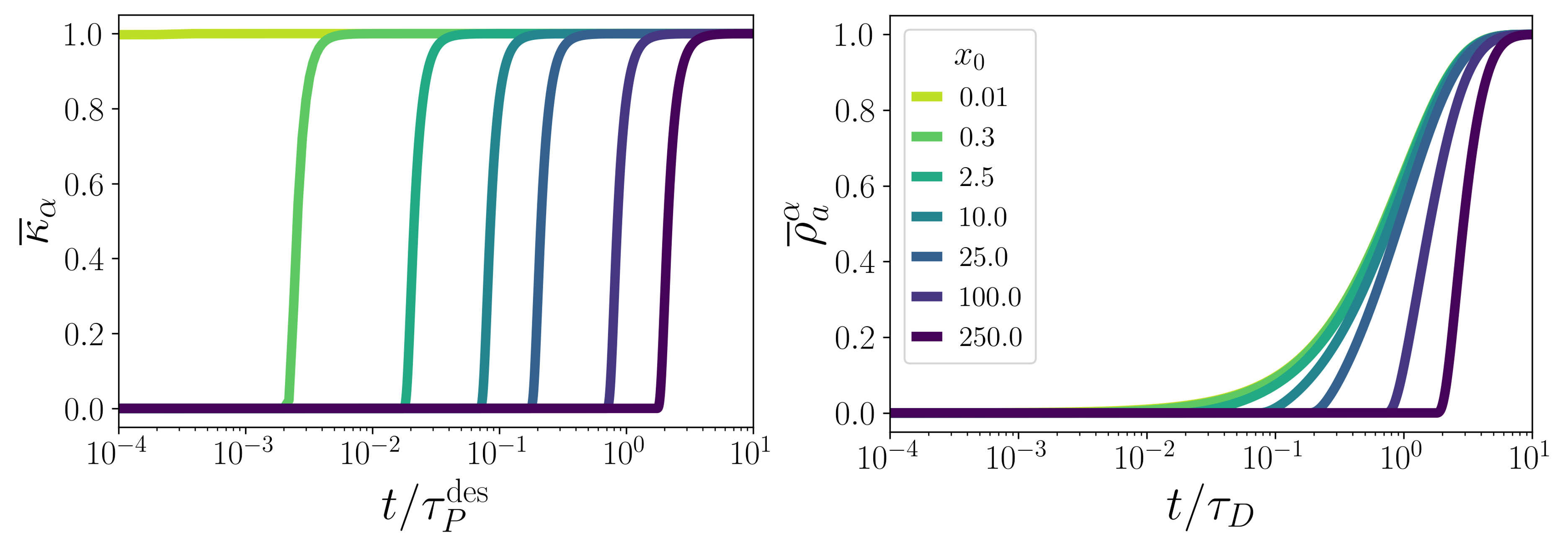}
    \caption{Evolution under the desorption mechanism of: the dimensionless membrane permeability, $\overline{\kappa}^{\alpha}$ as a function of time relative to the desorption time scale $t/\tau_P^{\rm des}$ (left) and the dimensionless acceptor density as a function of time relative to the dye diffusion time scale $t/\tau_D$ (right). }
    \label{fig:desorption_results}
\end{figure}
Figure~\ref{fig:desorption_results} shows the dynamic membrane permeability and the density change in the acceptor droplet for particles of various sizes.
In contrast to the Ostwald ripening results presented in Figs.~\ref{fig:permeability} and ~\ref{fig:density} in the main text, the desorption mechanism seems particles of vastly different sizes cross the membrane at roughly the same time.
This is an indication that for highly size-selective systems, pore growth likely occurs through Ostwald ripening.

We consider the rate at which particles must desorb to allow for size-selectivity comparable to the cases in which pores growth with algebraic scaling.
In order for particles of size $r_1$ and $2 r_1$ to cross the bilayer one order of magnitude part so that $t_{50}^{(r_1)} = 10 t_{50}^{(2 r_1)}$, we know from Eq.~\eqref{eq:desorption_distribution} that
\begin{equation}
    \begin{array}{cc}
         & R_0 \exp\left[2\pi \omega t_{50}^{(r_1)}\right] = r_1 \\\\
         & R_0 \exp\left[2\pi \omega t_{50}^{(2 r_1)}\right] = 2 r_1\\\\
         & 2 = \exp\left[2\pi \omega \left(t_{50}^{(2 r_1)} - t_{50}^{(r_1)}\right)\right] = \exp\left[{18 \pi \omega t_{50}^{(r_1)}}\right].
    \end{array}
\end{equation}
In the case that $t_{50}^{(r_1)} = 1$ hour, $\omega = 10^{-2}$ particles per hour, which corresponds to $1$ surfactant desorbing from the membrane every $100$ hours.
In the case that $t_{50}^{(r_1)} = 1$ day, $\omega = 10^{-2}$ particles per day, which corresponds to $1$ surfactant desorbing from the membrane every $100$ days. This implies that, any appreciable amount of desorption will lead to extremely rapid transport of dye of essentially any size. Therefore, desorption is unlikely to be the mechanism of pore growth in DIBs that prevent particles crossing for any extended period of time. 

\section{Contour map for Ostwald ripening}
Here, we provide the full contour map for ${t_{50}/\tau_D}$ in the Ostwald ripening mechanism across three decades of particle size and five decades of $\mathrm{Pe}$.
The time for $50\%$ of a particular species to enter the acceptor plateaus to $\ln{2}$ at low $x_0$ and large $\mathrm{Pe}$.
The transition to that plateau is steepest at low $\mathrm{Pe}$, for which ${t_{50}/\tau_D}$ dramatically drops to $\ln{2}$ for the particles that fit through the pores present in the initial distribution $(x_0 < 3/2)$.
As $\mathrm{Pe}$ increases, the plateau becomes smoother and occurs at larger $x_0$.
The figure provided in the main text is made from taking vertical and horizontal slices from this plot to observe how ${t_{50}/\tau_D}$ varies with $x_0$ for constant $\mathrm{Pe}$ and with $\mathrm{Pe}$ for constant $x_0$, respectively.

\begin{figure*}
    \centering
    \includegraphics[width=.8\linewidth]{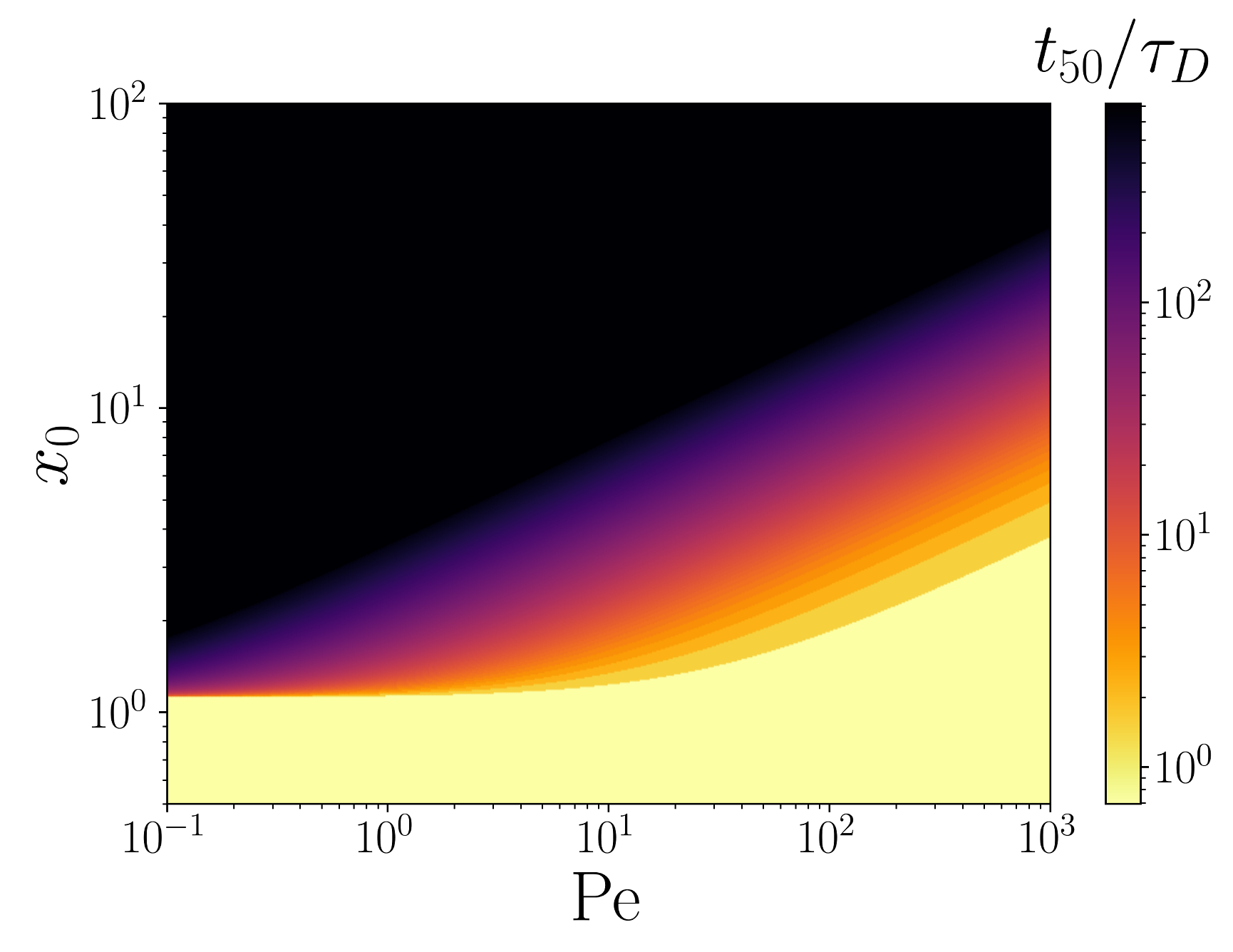}
    \caption{Full contour map for ${t_{50}/\tau_D}$ for the Ostwald ripening mechanism. We see that the time for half the particles of a given size to enter the acceptor droplet plateaus to $\ln{2}$ for $x_0 < 3/2$ and for increasing $\mathrm{Pe}$. The transition to this plateau becomes steeper as $\mathrm{Pe}$ decreases.}
    \label{fig:coarsening_contour}
\end{figure*}


\begin{thebibliography}{78}%
\makeatletter
\providecommand \@ifxundefined [1]{%
 \@ifx{#1\undefined}
}%
\providecommand \@ifnum [1]{%
 \ifnum #1\expandafter \@firstoftwo
 \else \expandafter \@secondoftwo
 \fi
}%
\providecommand \@ifx [1]{%
 \ifx #1\expandafter \@firstoftwo
 \else \expandafter \@secondoftwo
 \fi
}%
\providecommand \natexlab [1]{#1}%
\providecommand \enquote  [1]{``#1''}%
\providecommand \bibnamefont  [1]{#1}%
\providecommand \bibfnamefont [1]{#1}%
\providecommand \citenamefont [1]{#1}%
\providecommand \href@noop [0]{\@secondoftwo}%
\providecommand \href [0]{\begingroup \@sanitize@url \@href}%
\providecommand \@href[1]{\@@startlink{#1}\@@href}%
\providecommand \@@href[1]{\endgroup#1\@@endlink}%
\providecommand \@sanitize@url [0]{\catcode `\\12\catcode `\$12\catcode `\&12\catcode `\#12\catcode `\^12\catcode `\_12\catcode `\%12\relax}%
\providecommand \@@startlink[1]{}%
\providecommand \@@endlink[0]{}%
\providecommand \url  [0]{\begingroup\@sanitize@url \@url }%
\providecommand \@url [1]{\endgroup\@href {#1}{\urlprefix }}%
\providecommand \urlprefix  [0]{URL }%
\providecommand \Eprint [0]{\href }%
\providecommand \doibase [0]{https://doi.org/}%
\providecommand \selectlanguage [0]{\@gobble}%
\providecommand \bibinfo  [0]{\@secondoftwo}%
\providecommand \bibfield  [0]{\@secondoftwo}%
\providecommand \translation [1]{[#1]}%
\providecommand \BibitemOpen [0]{}%
\providecommand \bibitemStop [0]{}%
\providecommand \bibitemNoStop [0]{.\EOS\space}%
\providecommand \EOS [0]{\spacefactor3000\relax}%
\providecommand \BibitemShut  [1]{\csname bibitem#1\endcsname}%
\let\auto@bib@innerbib\@empty
%</preamble>
\bibitem [{\citenamefont {Tsofina}\ \emph {et~al.}(1966)\citenamefont {Tsofina}, \citenamefont {Liberman},\ and\ \citenamefont {Babakov}}]{Tsofina1966ProductionSolution}%
  \BibitemOpen
  \bibfield  {author} {\bibinfo {author} {\bibfnamefont {L.}~\bibnamefont {Tsofina}}, \bibinfo {author} {\bibfnamefont {E.}~\bibnamefont {Liberman}},\ and\ \bibinfo {author} {\bibfnamefont {A.}~\bibnamefont {Babakov}},\ }\href {https://doi.org/10.1038/212681a0} {\bibfield  {journal} {\bibinfo  {journal} {Nature}\ }\textbf {\bibinfo {volume} {212}},\ \bibinfo {pages} {681} (\bibinfo {year} {1966})}\BibitemShut {NoStop}%
\bibitem [{\citenamefont {Hwang}\ \emph {et~al.}(2008)\citenamefont {Hwang}, \citenamefont {Chen}, \citenamefont {Cronin}, \citenamefont {Holden},\ and\ \citenamefont {Bayley}}]{Hwang2008AsymmetricBilayers}%
  \BibitemOpen
  \bibfield  {author} {\bibinfo {author} {\bibfnamefont {W.~L.}\ \bibnamefont {Hwang}}, \bibinfo {author} {\bibfnamefont {M.}~\bibnamefont {Chen}}, \bibinfo {author} {\bibfnamefont {B.}~\bibnamefont {Cronin}}, \bibinfo {author} {\bibfnamefont {M.~A.}\ \bibnamefont {Holden}},\ and\ \bibinfo {author} {\bibfnamefont {H.}~\bibnamefont {Bayley}},\ }\href {https://doi.org/10.1021/ja802089s} {\bibfield  {journal} {\bibinfo  {journal} {Journal of the American Chemical Society}\ }\textbf {\bibinfo {volume} {130}},\ \bibinfo {pages} {5878} (\bibinfo {year} {2008})}\BibitemShut {NoStop}%
\bibitem [{\citenamefont {Holden}\ \emph {et~al.}(2007)\citenamefont {Holden}, \citenamefont {Needham},\ and\ \citenamefont {Bayley}}]{Holden2007FunctionalDroplets}%
  \BibitemOpen
  \bibfield  {author} {\bibinfo {author} {\bibfnamefont {M.~A.}\ \bibnamefont {Holden}}, \bibinfo {author} {\bibfnamefont {D.}~\bibnamefont {Needham}},\ and\ \bibinfo {author} {\bibfnamefont {H.}~\bibnamefont {Bayley}},\ }\href {https://doi.org/10.1021/ja072292a} {\bibfield  {journal} {\bibinfo  {journal} {Journal of the American Chemical Society}\ }\textbf {\bibinfo {volume} {129}},\ \bibinfo {pages} {8650} (\bibinfo {year} {2007})}\BibitemShut {NoStop}%
\bibitem [{\citenamefont {Mruetusatorn}\ \emph {et~al.}(2014)\citenamefont {Mruetusatorn}, \citenamefont {Boreyko}, \citenamefont {Venkatesan}, \citenamefont {Sarles}, \citenamefont {Hayes},\ and\ \citenamefont {Collier}}]{Mruetusatorn2014DynamicBilayers}%
  \BibitemOpen
  \bibfield  {author} {\bibinfo {author} {\bibfnamefont {P.}~\bibnamefont {Mruetusatorn}}, \bibinfo {author} {\bibfnamefont {J.~B.}\ \bibnamefont {Boreyko}}, \bibinfo {author} {\bibfnamefont {G.~A.}\ \bibnamefont {Venkatesan}}, \bibinfo {author} {\bibfnamefont {S.~A.}\ \bibnamefont {Sarles}}, \bibinfo {author} {\bibfnamefont {D.~G.}\ \bibnamefont {Hayes}},\ and\ \bibinfo {author} {\bibfnamefont {C.~P.}\ \bibnamefont {Collier}},\ }\href {https://doi.org/10.1039/C3SM53032A} {\bibfield  {journal} {\bibinfo  {journal} {Soft Matter}\ }\textbf {\bibinfo {volume} {10}},\ \bibinfo {pages} {2530} (\bibinfo {year} {2014})}\BibitemShut {NoStop}%
\bibitem [{\citenamefont {Guiselin}\ \emph {et~al.}(2018)\citenamefont {Guiselin}, \citenamefont {Law}, \citenamefont {Chakrabarti},\ and\ \citenamefont {Kusumaatmaja}}]{Guiselin2018DynamicBilayers}%
  \BibitemOpen
  \bibfield  {author} {\bibinfo {author} {\bibfnamefont {B.}~\bibnamefont {Guiselin}}, \bibinfo {author} {\bibfnamefont {J.~O.}\ \bibnamefont {Law}}, \bibinfo {author} {\bibfnamefont {B.}~\bibnamefont {Chakrabarti}},\ and\ \bibinfo {author} {\bibfnamefont {H.}~\bibnamefont {Kusumaatmaja}},\ }\href {https://doi.org/10.1103/PhysRevLett.120.238001} {\bibfield  {journal} {\bibinfo  {journal} {Physical Review Letters}\ }\textbf {\bibinfo {volume} {120}},\ \bibinfo {pages} {238001} (\bibinfo {year} {2018})}\BibitemShut {NoStop}%
\bibitem [{\citenamefont {Bird}\ \emph {et~al.}(2009)\citenamefont {Bird}, \citenamefont {Ristenpart}, \citenamefont {Belmonte},\ and\ \citenamefont {Stone}}]{Bird2009CriticalDroplets}%
  \BibitemOpen
  \bibfield  {author} {\bibinfo {author} {\bibfnamefont {J.~C.}\ \bibnamefont {Bird}}, \bibinfo {author} {\bibfnamefont {W.~D.}\ \bibnamefont {Ristenpart}}, \bibinfo {author} {\bibfnamefont {A.}~\bibnamefont {Belmonte}},\ and\ \bibinfo {author} {\bibfnamefont {H.~A.}\ \bibnamefont {Stone}},\ }\href {https://doi.org/10.1103/PhysRevLett.103.164502} {\bibfield  {journal} {\bibinfo  {journal} {Physical Review Letters}\ }\textbf {\bibinfo {volume} {103}},\ \bibinfo {pages} {164502} (\bibinfo {year} {2009})}\BibitemShut {NoStop}%
\bibitem [{\citenamefont {Almeida}\ \emph {et~al.}(2005)\citenamefont {Almeida}, \citenamefont {Pokorny},\ and\ \citenamefont {Hinderliter}}]{Almeida2005ThermodynamicsDomains}%
  \BibitemOpen
  \bibfield  {author} {\bibinfo {author} {\bibfnamefont {P.~F.}\ \bibnamefont {Almeida}}, \bibinfo {author} {\bibfnamefont {A.}~\bibnamefont {Pokorny}},\ and\ \bibinfo {author} {\bibfnamefont {A.}~\bibnamefont {Hinderliter}},\ }\href {https://doi.org/10.1016/J.BBAMEM.2005.12.004} {\bibfield  {journal} {\bibinfo  {journal} {Biochimica et Biophysica Acta (BBA) - Biomembranes}\ }\textbf {\bibinfo {volume} {1720}},\ \bibinfo {pages} {1} (\bibinfo {year} {2005})}\BibitemShut {NoStop}%
\bibitem [{\citenamefont {Yang}\ \emph {et~al.}(2016)\citenamefont {Yang}, \citenamefont {Kiessling},\ and\ \citenamefont {Tamm}}]{Yang2016LineFusion}%
  \BibitemOpen
  \bibfield  {author} {\bibinfo {author} {\bibfnamefont {S.~T.}\ \bibnamefont {Yang}}, \bibinfo {author} {\bibfnamefont {V.}~\bibnamefont {Kiessling}},\ and\ \bibinfo {author} {\bibfnamefont {L.~K.}\ \bibnamefont {Tamm}},\ }\href {https://doi.org/10.1038/ncomms11401} {\bibfield  {journal} {\bibinfo  {journal} {Nature Communications 2016 7:1}\ }\textbf {\bibinfo {volume} {7}},\ \bibinfo {pages} {1} (\bibinfo {year} {2016})}\BibitemShut {NoStop}%
\bibitem [{\citenamefont {Phillips}\ \emph {et~al.}(2012)\citenamefont {Phillips}, \citenamefont {Kondev},\ and\ \citenamefont {Theriot}}]{Phillips2012PhysicalCell}%
  \BibitemOpen
  \bibfield  {author} {\bibinfo {author} {\bibfnamefont {R.}~\bibnamefont {Phillips}}, \bibinfo {author} {\bibfnamefont {J.}~\bibnamefont {Kondev}},\ and\ \bibinfo {author} {\bibfnamefont {J.}~\bibnamefont {Theriot}},\ }\href {https://doi.org/https://doi.org/10.1201/9781134111589} {\emph {\bibinfo {title} {{Physical Biology of the Cell}}}},\ \bibinfo {edition} {2nd}\ ed.\ (\bibinfo  {publisher} {Garland Science},\ \bibinfo {address} {New York},\ \bibinfo {year} {2012})\BibitemShut {NoStop}%
\bibitem [{\citenamefont {Moscho}\ \emph {et~al.}(1996)\citenamefont {Moscho}, \citenamefont {Orwar}, \citenamefont {Chiu}, \citenamefont {Modi},\ and\ \citenamefont {Zare}}]{Moscho1996RapidVesicles}%
  \BibitemOpen
  \bibfield  {author} {\bibinfo {author} {\bibfnamefont {A.}~\bibnamefont {Moscho}}, \bibinfo {author} {\bibfnamefont {O.}~\bibnamefont {Orwar}}, \bibinfo {author} {\bibfnamefont {D.~T.}\ \bibnamefont {Chiu}}, \bibinfo {author} {\bibfnamefont {B.~P.}\ \bibnamefont {Modi}},\ and\ \bibinfo {author} {\bibfnamefont {R.~N.}\ \bibnamefont {Zare}},\ }\href {https://doi.org/https://doi.org/10.1073/pnas.93.21.11443} {\bibfield  {journal} {\bibinfo  {journal} {Proceedings of the National Academy of Sciences of the United States of America}\ }\textbf {\bibinfo {volume} {93}},\ \bibinfo {pages} {11443} (\bibinfo {year} {1996})}\BibitemShut {NoStop}%
\bibitem [{\citenamefont {Weso{\l}owska}\ \emph {et~al.}(2009)\citenamefont {Weso{\l}owska}, \citenamefont {Michalak}, \citenamefont {Maniewska},\ and\ \citenamefont {Hendrich}}]{Wesoowska2009GiantSystems}%
  \BibitemOpen
  \bibfield  {author} {\bibinfo {author} {\bibfnamefont {O.}~\bibnamefont {Weso{\l}owska}}, \bibinfo {author} {\bibfnamefont {K.}~\bibnamefont {Michalak}}, \bibinfo {author} {\bibfnamefont {J.}~\bibnamefont {Maniewska}},\ and\ \bibinfo {author} {\bibfnamefont {A.~B.}\ \bibnamefont {Hendrich}},\ }\href {https://doi.org/10.18388/abp.2009{\_}2514} {\bibfield  {journal} {\bibinfo  {journal} {Acta Biochimica Polonica}\ }\textbf {\bibinfo {volume} {56}},\ \bibinfo {pages} {33} (\bibinfo {year} {2009})}\BibitemShut {NoStop}%
\bibitem [{\citenamefont {Kahya}(2010)}]{Kahya2010Protein-proteinVesicles}%
  \BibitemOpen
  \bibfield  {author} {\bibinfo {author} {\bibfnamefont {N.}~\bibnamefont {Kahya}},\ }\href {https://doi.org/10.1016/j.bbamem.2010.02.028} {\bibfield  {journal} {\bibinfo  {journal} {Biochimica et Biophysica Acta - Biomembranes}\ }\textbf {\bibinfo {volume} {1798}},\ \bibinfo {pages} {1392} (\bibinfo {year} {2010})}\BibitemShut {NoStop}%
\bibitem [{\citenamefont {J{\o}rgensen}\ \emph {et~al.}(2017)\citenamefont {J{\o}rgensen}, \citenamefont {Kemmer},\ and\ \citenamefont {Pomorski}}]{Jrgensen2017MembraneTechniques}%
  \BibitemOpen
  \bibfield  {author} {\bibinfo {author} {\bibfnamefont {I.~L.}\ \bibnamefont {J{\o}rgensen}}, \bibinfo {author} {\bibfnamefont {G.~C.}\ \bibnamefont {Kemmer}},\ and\ \bibinfo {author} {\bibfnamefont {T.~G.}\ \bibnamefont {Pomorski}},\ }\href {https://doi.org/10.1007/s00249-016-1155-9} {\bibfield  {journal} {\bibinfo  {journal} {European Biophysics Journal}\ }\textbf {\bibinfo {volume} {46}},\ \bibinfo {pages} {103} (\bibinfo {year} {2017})}\BibitemShut {NoStop}%
\bibitem [{\citenamefont {Bach}\ and\ \citenamefont {Miller}(1980)}]{Bach1980GlycerylSolutions}%
  \BibitemOpen
  \bibfield  {author} {\bibinfo {author} {\bibfnamefont {D.}~\bibnamefont {Bach}}\ and\ \bibinfo {author} {\bibfnamefont {I.}~\bibnamefont {Miller}},\ }\href {https://doi.org/10.1016/S0006-3495(80)85125-3} {\bibfield  {journal} {\bibinfo  {journal} {Biophysical Journal}\ }\textbf {\bibinfo {volume} {29}},\ \bibinfo {pages} {183} (\bibinfo {year} {1980})}\BibitemShut {NoStop}%
\bibitem [{\citenamefont {Hladky}\ and\ \citenamefont {Gruen}(1982)}]{Hladky1982ThicknessMembranesb}%
  \BibitemOpen
  \bibfield  {author} {\bibinfo {author} {\bibfnamefont {S.}~\bibnamefont {Hladky}}\ and\ \bibinfo {author} {\bibfnamefont {D.}~\bibnamefont {Gruen}},\ }\href {https://doi.org/10.1016/S0006-3495(82)84556-6} {\bibfield  {journal} {\bibinfo  {journal} {Biophysical Journal}\ }\textbf {\bibinfo {volume} {38}},\ \bibinfo {pages} {251} (\bibinfo {year} {1982})}\BibitemShut {NoStop}%
\bibitem [{\citenamefont {Horn}\ and\ \citenamefont {Steinem}(2005)}]{Horn2005PhotocurrentsMembranesb}%
  \BibitemOpen
  \bibfield  {author} {\bibinfo {author} {\bibfnamefont {C.}~\bibnamefont {Horn}}\ and\ \bibinfo {author} {\bibfnamefont {C.}~\bibnamefont {Steinem}},\ }\href {https://doi.org/10.1529/biophysj.105.059550} {\bibfield  {journal} {\bibinfo  {journal} {Biophysical Journal}\ }\textbf {\bibinfo {volume} {89}},\ \bibinfo {pages} {1046} (\bibinfo {year} {2005})}\BibitemShut {NoStop}%
\bibitem [{\citenamefont {Nomoto}\ \emph {et~al.}(2018)\citenamefont {Nomoto}, \citenamefont {Takahashi}, \citenamefont {Fujii}, \citenamefont {Chiari}, \citenamefont {Toyota},\ and\ \citenamefont {Fujinami}}]{Nomoto2018EffectsMembranesb}%
  \BibitemOpen
  \bibfield  {author} {\bibinfo {author} {\bibfnamefont {T.}~\bibnamefont {Nomoto}}, \bibinfo {author} {\bibfnamefont {M.}~\bibnamefont {Takahashi}}, \bibinfo {author} {\bibfnamefont {T.}~\bibnamefont {Fujii}}, \bibinfo {author} {\bibfnamefont {L.}~\bibnamefont {Chiari}}, \bibinfo {author} {\bibfnamefont {T.}~\bibnamefont {Toyota}},\ and\ \bibinfo {author} {\bibfnamefont {M.}~\bibnamefont {Fujinami}},\ }\href {https://doi.org/10.2116/analsci.18P200} {\bibfield  {journal} {\bibinfo  {journal} {Analytical Sciences}\ }\textbf {\bibinfo {volume} {34}},\ \bibinfo {pages} {1237} (\bibinfo {year} {2018})}\BibitemShut {NoStop}%
\bibitem [{\citenamefont {Steinem}\ \emph {et~al.}(1996)\citenamefont {Steinem}, \citenamefont {Janshoff}, \citenamefont {Ulrich}, \citenamefont {Sieber},\ and\ \citenamefont {Galla}}]{Steinem1996ImpedanceTechniquesb}%
  \BibitemOpen
  \bibfield  {author} {\bibinfo {author} {\bibfnamefont {C.}~\bibnamefont {Steinem}}, \bibinfo {author} {\bibfnamefont {A.}~\bibnamefont {Janshoff}}, \bibinfo {author} {\bibfnamefont {W.-P.}\ \bibnamefont {Ulrich}}, \bibinfo {author} {\bibfnamefont {M.}~\bibnamefont {Sieber}},\ and\ \bibinfo {author} {\bibfnamefont {H.-J.}\ \bibnamefont {Galla}},\ }\href {https://doi.org/10.1016/0005-2736(95)00274-X} {\bibfield  {journal} {\bibinfo  {journal} {Biochimica et Biophysica Acta (BBA) - Biomembranes}\ }\textbf {\bibinfo {volume} {1279}},\ \bibinfo {pages} {169} (\bibinfo {year} {1996})}\BibitemShut {NoStop}%
\bibitem [{\citenamefont {Richter}\ \emph {et~al.}(2006)\citenamefont {Richter}, \citenamefont {B{\'{e}}rat},\ and\ \citenamefont {Brisson}}]{Richter2006FormationViewb}%
  \BibitemOpen
  \bibfield  {author} {\bibinfo {author} {\bibfnamefont {R.~P.}\ \bibnamefont {Richter}}, \bibinfo {author} {\bibfnamefont {R.}~\bibnamefont {B{\'{e}}rat}},\ and\ \bibinfo {author} {\bibfnamefont {A.~R.}\ \bibnamefont {Brisson}},\ }\href {https://doi.org/10.1021/la052687c} {\bibfield  {journal} {\bibinfo  {journal} {Langmuir}\ }\textbf {\bibinfo {volume} {22}},\ \bibinfo {pages} {3497} (\bibinfo {year} {2006})}\BibitemShut {NoStop}%
\bibitem [{\citenamefont {Seu}\ \emph {et~al.}(2007)\citenamefont {Seu}, \citenamefont {Pandey}, \citenamefont {Haque}, \citenamefont {Proctor}, \citenamefont {Ribbe},\ and\ \citenamefont {Hovis}}]{Seu2007EffectBilayersb}%
  \BibitemOpen
  \bibfield  {author} {\bibinfo {author} {\bibfnamefont {K.~J.}\ \bibnamefont {Seu}}, \bibinfo {author} {\bibfnamefont {A.~P.}\ \bibnamefont {Pandey}}, \bibinfo {author} {\bibfnamefont {F.}~\bibnamefont {Haque}}, \bibinfo {author} {\bibfnamefont {E.~A.}\ \bibnamefont {Proctor}}, \bibinfo {author} {\bibfnamefont {A.~E.}\ \bibnamefont {Ribbe}},\ and\ \bibinfo {author} {\bibfnamefont {J.~S.}\ \bibnamefont {Hovis}},\ }\href {https://doi.org/10.1529/biophysj.106.099721} {\bibfield  {journal} {\bibinfo  {journal} {Biophysical Journal}\ }\textbf {\bibinfo {volume} {92}},\ \bibinfo {pages} {2445} (\bibinfo {year} {2007})}\BibitemShut {NoStop}%
\bibitem [{\citenamefont {El~Kirat}\ \emph {et~al.}(2010)\citenamefont {El~Kirat}, \citenamefont {Morandat},\ and\ \citenamefont {Dufr{\^{e}}ne}}]{ElKirat2010NanoscaleMicroscopyb}%
  \BibitemOpen
  \bibfield  {author} {\bibinfo {author} {\bibfnamefont {K.}~\bibnamefont {El~Kirat}}, \bibinfo {author} {\bibfnamefont {S.}~\bibnamefont {Morandat}},\ and\ \bibinfo {author} {\bibfnamefont {Y.~F.}\ \bibnamefont {Dufr{\^{e}}ne}},\ }\href {https://doi.org/10.1016/j.bbamem.2009.07.026} {\bibfield  {journal} {\bibinfo  {journal} {Biochimica et Biophysica Acta (BBA) - Biomembranes}\ }\textbf {\bibinfo {volume} {1798}},\ \bibinfo {pages} {750} (\bibinfo {year} {2010})}\BibitemShut {NoStop}%
\bibitem [{\citenamefont {Cheney}\ \emph {et~al.}(2017)\citenamefont {Cheney}, \citenamefont {Weisgerber}, \citenamefont {Feuerbach},\ and\ \citenamefont {Knowles}}]{Cheney2017SingleCurvatureb}%
  \BibitemOpen
  \bibfield  {author} {\bibinfo {author} {\bibfnamefont {P.}~\bibnamefont {Cheney}}, \bibinfo {author} {\bibfnamefont {A.}~\bibnamefont {Weisgerber}}, \bibinfo {author} {\bibfnamefont {A.}~\bibnamefont {Feuerbach}},\ and\ \bibinfo {author} {\bibfnamefont {M.}~\bibnamefont {Knowles}},\ }\href {https://doi.org/10.3390/membranes7010015} {\bibfield  {journal} {\bibinfo  {journal} {Membranes}\ }\textbf {\bibinfo {volume} {7}},\ \bibinfo {pages} {15} (\bibinfo {year} {2017})}\BibitemShut {NoStop}%
\bibitem [{\citenamefont {Glazier}\ and\ \citenamefont {Salaita}(2017)}]{Glazier2017SupportedMechanobiologyb}%
  \BibitemOpen
  \bibfield  {author} {\bibinfo {author} {\bibfnamefont {R.}~\bibnamefont {Glazier}}\ and\ \bibinfo {author} {\bibfnamefont {K.}~\bibnamefont {Salaita}},\ }\href {https://doi.org/10.1016/j.bbamem.2017.05.005} {\bibfield  {journal} {\bibinfo  {journal} {Biochimica et Biophysica Acta (BBA) - Biomembranes}\ }\textbf {\bibinfo {volume} {1859}},\ \bibinfo {pages} {1465} (\bibinfo {year} {2017})}\BibitemShut {NoStop}%
\bibitem [{\citenamefont {Funakoshi}\ \emph {et~al.}(2006)\citenamefont {Funakoshi}, \citenamefont {Suzuki},\ and\ \citenamefont {Takeuchi}}]{Funakoshi2006LipidAnalysis}%
  \BibitemOpen
  \bibfield  {author} {\bibinfo {author} {\bibfnamefont {K.}~\bibnamefont {Funakoshi}}, \bibinfo {author} {\bibfnamefont {H.}~\bibnamefont {Suzuki}},\ and\ \bibinfo {author} {\bibfnamefont {S.}~\bibnamefont {Takeuchi}},\ }\href {https://doi.org/10.1021/AC0613479/ASSET/IMAGES/LARGE/AC0613479F00009.JPEG} {\bibfield  {journal} {\bibinfo  {journal} {Analytical Chemistry}\ }\textbf {\bibinfo {volume} {78}},\ \bibinfo {pages} {8169} (\bibinfo {year} {2006})}\BibitemShut {NoStop}%
\bibitem [{\citenamefont {Heron}\ \emph {et~al.}(2007)\citenamefont {Heron}, \citenamefont {Thompson}, \citenamefont {Mason},\ and\ \citenamefont {Wallace}}]{Heron2007DirectBilayers}%
  \BibitemOpen
  \bibfield  {author} {\bibinfo {author} {\bibfnamefont {A.~J.}\ \bibnamefont {Heron}}, \bibinfo {author} {\bibfnamefont {J.~R.}\ \bibnamefont {Thompson}}, \bibinfo {author} {\bibfnamefont {A.~E.}\ \bibnamefont {Mason}},\ and\ \bibinfo {author} {\bibfnamefont {M.~I.}\ \bibnamefont {Wallace}},\ }\href {https://doi.org/10.1021/JA075715H/SUPPL{\_}FILE/JA075715HSI20071022{\_}125035.PDF} {\bibfield  {journal} {\bibinfo  {journal} {Journal of the American Chemical Society}\ }\textbf {\bibinfo {volume} {129}},\ \bibinfo {pages} {16042} (\bibinfo {year} {2007})}\BibitemShut {NoStop}%
\bibitem [{\citenamefont {Bayley}\ \emph {et~al.}(2008)\citenamefont {Bayley}, \citenamefont {Cronin}, \citenamefont {Heron}, \citenamefont {Holden}, \citenamefont {Hwang}, \citenamefont {Syeda}, \citenamefont {Thompson},\ and\ \citenamefont {Wallace}}]{Bayley2008DropletBilayers}%
  \BibitemOpen
  \bibfield  {author} {\bibinfo {author} {\bibfnamefont {H.}~\bibnamefont {Bayley}}, \bibinfo {author} {\bibfnamefont {B.}~\bibnamefont {Cronin}}, \bibinfo {author} {\bibfnamefont {A.}~\bibnamefont {Heron}}, \bibinfo {author} {\bibfnamefont {M.~A.}\ \bibnamefont {Holden}}, \bibinfo {author} {\bibfnamefont {W.~L.}\ \bibnamefont {Hwang}}, \bibinfo {author} {\bibfnamefont {R.}~\bibnamefont {Syeda}}, \bibinfo {author} {\bibfnamefont {J.}~\bibnamefont {Thompson}},\ and\ \bibinfo {author} {\bibfnamefont {M.}~\bibnamefont {Wallace}},\ }\href {https://doi.org/10.1039/b808893d} {\bibfield  {journal} {\bibinfo  {journal} {Molecular BioSystems}\ }\textbf {\bibinfo {volume} {4}},\ \bibinfo {pages} {1191} (\bibinfo {year} {2008})}\BibitemShut {NoStop}%
\bibitem [{\citenamefont {Maglia}\ \emph {et~al.}(2009)\citenamefont {Maglia}, \citenamefont {Heron}, \citenamefont {Hwang}, \citenamefont {Holden}, \citenamefont {Mikhailova}, \citenamefont {Li}, \citenamefont {Cheley},\ and\ \citenamefont {Bayley}}]{Maglia2009DropletPropertiesb}%
  \BibitemOpen
  \bibfield  {author} {\bibinfo {author} {\bibfnamefont {G.}~\bibnamefont {Maglia}}, \bibinfo {author} {\bibfnamefont {A.~J.}\ \bibnamefont {Heron}}, \bibinfo {author} {\bibfnamefont {W.~L.}\ \bibnamefont {Hwang}}, \bibinfo {author} {\bibfnamefont {M.~A.}\ \bibnamefont {Holden}}, \bibinfo {author} {\bibfnamefont {E.}~\bibnamefont {Mikhailova}}, \bibinfo {author} {\bibfnamefont {Q.}~\bibnamefont {Li}}, \bibinfo {author} {\bibfnamefont {S.}~\bibnamefont {Cheley}},\ and\ \bibinfo {author} {\bibfnamefont {H.}~\bibnamefont {Bayley}},\ }\href {https://doi.org/10.1038/nnano.2009.121} {\bibfield  {journal} {\bibinfo  {journal} {Nature Nanotechnology}\ }\textbf {\bibinfo {volume} {4}},\ \bibinfo {pages} {437} (\bibinfo {year} {2009})}\BibitemShut {NoStop}%
\bibitem [{\citenamefont {Stephenson}\ \emph {et~al.}(2022)\citenamefont {Stephenson}, \citenamefont {Korner},\ and\ \citenamefont {Elvira}}]{Stephenson2022ChallengesBilayers}%
  \BibitemOpen
  \bibfield  {author} {\bibinfo {author} {\bibfnamefont {E.~B.}\ \bibnamefont {Stephenson}}, \bibinfo {author} {\bibfnamefont {J.~L.}\ \bibnamefont {Korner}},\ and\ \bibinfo {author} {\bibfnamefont {K.~S.}\ \bibnamefont {Elvira}},\ }\href {https://doi.org/10.1038/s41557-022-00989-y} {\bibfield  {journal} {\bibinfo  {journal} {Nature Chemistry 2022 14:8}\ }\textbf {\bibinfo {volume} {14}},\ \bibinfo {pages} {862} (\bibinfo {year} {2022})}\BibitemShut {NoStop}%
\bibitem [{\citenamefont {Villar}\ \emph {et~al.}(2011)\citenamefont {Villar}, \citenamefont {Heron},\ and\ \citenamefont {Bayley}}]{Villar2011FormationEnvironmentsb}%
  \BibitemOpen
  \bibfield  {author} {\bibinfo {author} {\bibfnamefont {G.}~\bibnamefont {Villar}}, \bibinfo {author} {\bibfnamefont {A.~J.}\ \bibnamefont {Heron}},\ and\ \bibinfo {author} {\bibfnamefont {H.}~\bibnamefont {Bayley}},\ }\href {https://doi.org/10.1038/nnano.2011.183} {\bibfield  {journal} {\bibinfo  {journal} {Nature Nanotechnology}\ }\textbf {\bibinfo {volume} {6}},\ \bibinfo {pages} {803} (\bibinfo {year} {2011})}\BibitemShut {NoStop}%
\bibitem [{\citenamefont {Jalali}\ \emph {et~al.}(2024)\citenamefont {Jalali}, \citenamefont {Nowroozi}, \citenamefont {Moradi},\ and\ \citenamefont {Shahlaei}}]{Jalali2024ExplorationSimulation}%
  \BibitemOpen
  \bibfield  {author} {\bibinfo {author} {\bibfnamefont {P.}~\bibnamefont {Jalali}}, \bibinfo {author} {\bibfnamefont {A.}~\bibnamefont {Nowroozi}}, \bibinfo {author} {\bibfnamefont {S.}~\bibnamefont {Moradi}},\ and\ \bibinfo {author} {\bibfnamefont {M.}~\bibnamefont {Shahlaei}},\ }\href {https://doi.org/https://doi.org/10.1016/j.abb.2024.110151} {\bibfield  {journal} {\bibinfo  {journal} {Archives of Biochemistry and Biophysics}\ }\textbf {\bibinfo {volume} {761}},\ \bibinfo {pages} {110151} (\bibinfo {year} {2024})}\BibitemShut {NoStop}%
\bibitem [{\citenamefont {Huang}\ \emph {et~al.}(2024)\citenamefont {Huang}, \citenamefont {Chandran~Suja}, \citenamefont {Yang}, \citenamefont {Malkovskiy}, \citenamefont {Tandon}, \citenamefont {Colom}, \citenamefont {Qin},\ and\ \citenamefont {Fuller}}]{Huang2024InterfacialMicroscopy}%
  \BibitemOpen
  \bibfield  {author} {\bibinfo {author} {\bibfnamefont {Y.}~\bibnamefont {Huang}}, \bibinfo {author} {\bibfnamefont {V.}~\bibnamefont {Chandran~Suja}}, \bibinfo {author} {\bibfnamefont {M.}~\bibnamefont {Yang}}, \bibinfo {author} {\bibfnamefont {A.~V.}\ \bibnamefont {Malkovskiy}}, \bibinfo {author} {\bibfnamefont {A.}~\bibnamefont {Tandon}}, \bibinfo {author} {\bibfnamefont {A.}~\bibnamefont {Colom}}, \bibinfo {author} {\bibfnamefont {J.}~\bibnamefont {Qin}},\ and\ \bibinfo {author} {\bibfnamefont {G.~G.}\ \bibnamefont {Fuller}},\ }\href {https://doi.org/https://doi.org/10.1016/j.jcis.2023.09.092} {\bibfield  {journal} {\bibinfo  {journal} {Journal of Colloid and Interface Science}\ }\textbf {\bibinfo {volume} {653}},\ \bibinfo {pages} {1196} (\bibinfo {year} {2024})}\BibitemShut {NoStop}%
\bibitem [{\citenamefont {Forth}\ \emph {et~al.}(2018)\citenamefont {Forth}, \citenamefont {Liu}, \citenamefont {Hasnain}, \citenamefont {Toor}, \citenamefont {Miszta}, \citenamefont {Shi}, \citenamefont {Geissler}, \citenamefont {Emrick}, \citenamefont {Helms},\ and\ \citenamefont {Russell}}]{Forth2018ReconfigurableLiquidsb}%
  \BibitemOpen
  \bibfield  {author} {\bibinfo {author} {\bibfnamefont {J.}~\bibnamefont {Forth}}, \bibinfo {author} {\bibfnamefont {X.}~\bibnamefont {Liu}}, \bibinfo {author} {\bibfnamefont {J.}~\bibnamefont {Hasnain}}, \bibinfo {author} {\bibfnamefont {A.}~\bibnamefont {Toor}}, \bibinfo {author} {\bibfnamefont {K.}~\bibnamefont {Miszta}}, \bibinfo {author} {\bibfnamefont {S.}~\bibnamefont {Shi}}, \bibinfo {author} {\bibfnamefont {P.~L.}\ \bibnamefont {Geissler}}, \bibinfo {author} {\bibfnamefont {T.}~\bibnamefont {Emrick}}, \bibinfo {author} {\bibfnamefont {B.~A.}\ \bibnamefont {Helms}},\ and\ \bibinfo {author} {\bibfnamefont {T.~P.}\ \bibnamefont {Russell}},\ }\href {https://doi.org/10.1002/adma.201707603} {\bibfield  {journal} {\bibinfo  {journal} {Advanced Materials}\ }\textbf {\bibinfo {volume} {30}},\ \bibinfo {pages} {1} (\bibinfo {year} {2018})}\BibitemShut {NoStop}%
\bibitem [{\citenamefont {Huang}\ \emph {et~al.}(2017)\citenamefont {Huang}, \citenamefont {Cui}, \citenamefont {Sun}, \citenamefont {Liu}, \citenamefont {Helms},\ and\ \citenamefont {Russell}}]{Huang2017Self-RegulatedLiquidsb}%
  \BibitemOpen
  \bibfield  {author} {\bibinfo {author} {\bibfnamefont {C.}~\bibnamefont {Huang}}, \bibinfo {author} {\bibfnamefont {M.}~\bibnamefont {Cui}}, \bibinfo {author} {\bibfnamefont {Z.}~\bibnamefont {Sun}}, \bibinfo {author} {\bibfnamefont {F.}~\bibnamefont {Liu}}, \bibinfo {author} {\bibfnamefont {B.~A.}\ \bibnamefont {Helms}},\ and\ \bibinfo {author} {\bibfnamefont {T.~P.}\ \bibnamefont {Russell}},\ }\href {https://doi.org/10.1021/acs.langmuir.7b01685} {\bibfield  {journal} {\bibinfo  {journal} {Langmuir}\ }\textbf {\bibinfo {volume} {33}},\ \bibinfo {pages} {7994} (\bibinfo {year} {2017})}\BibitemShut {NoStop}%
\bibitem [{\citenamefont {Toor}\ \emph {et~al.}(2017)\citenamefont {Toor}, \citenamefont {Helms},\ and\ \citenamefont {Russell}}]{Toor2017EffectDropletsb}%
  \BibitemOpen
  \bibfield  {author} {\bibinfo {author} {\bibfnamefont {A.}~\bibnamefont {Toor}}, \bibinfo {author} {\bibfnamefont {B.~A.}\ \bibnamefont {Helms}},\ and\ \bibinfo {author} {\bibfnamefont {T.~P.}\ \bibnamefont {Russell}},\ }\href {https://doi.org/10.1021/acs.nanolett.7b00556} {\bibfield  {journal} {\bibinfo  {journal} {Nano Letters}\ }\textbf {\bibinfo {volume} {17}},\ \bibinfo {pages} {3119} (\bibinfo {year} {2017})}\BibitemShut {NoStop}%
\bibitem [{\citenamefont {Wu}\ \emph {et~al.}(2023)\citenamefont {Wu}, \citenamefont {Bordia}, \citenamefont {Streubel}, \citenamefont {Hasnain}, \citenamefont {Pedroso}, \citenamefont {Cohen}, \citenamefont {Rad}, \citenamefont {Ashby}, \citenamefont {Omar}, \citenamefont {Geissler}, \citenamefont {Wang}, \citenamefont {Xue}, \citenamefont {Wang},\ and\ \citenamefont {Russell}}]{Wu2023BallisticAssemblies}%
  \BibitemOpen
  \bibfield  {author} {\bibinfo {author} {\bibfnamefont {X.}~\bibnamefont {Wu}}, \bibinfo {author} {\bibfnamefont {G.}~\bibnamefont {Bordia}}, \bibinfo {author} {\bibfnamefont {R.}~\bibnamefont {Streubel}}, \bibinfo {author} {\bibfnamefont {J.}~\bibnamefont {Hasnain}}, \bibinfo {author} {\bibfnamefont {C.~C.}\ \bibnamefont {Pedroso}}, \bibinfo {author} {\bibfnamefont {B.~E.}\ \bibnamefont {Cohen}}, \bibinfo {author} {\bibfnamefont {B.}~\bibnamefont {Rad}}, \bibinfo {author} {\bibfnamefont {P.}~\bibnamefont {Ashby}}, \bibinfo {author} {\bibfnamefont {A.~K.}\ \bibnamefont {Omar}}, \bibinfo {author} {\bibfnamefont {P.~L.}\ \bibnamefont {Geissler}}, \bibinfo {author} {\bibfnamefont {D.}~\bibnamefont {Wang}}, \bibinfo {author} {\bibfnamefont {H.}~\bibnamefont {Xue}}, \bibinfo {author} {\bibfnamefont {J.}~\bibnamefont {Wang}},\ and\ \bibinfo {author} {\bibfnamefont {T.~P.}\ \bibnamefont {Russell}},\ }\href {https://doi.org/10.1002/ADFM.202213844} {\bibfield  {journal} {\bibinfo  {journal} {Advanced Functional
  Materials}\ }\textbf {\bibinfo {volume} {33}},\ \bibinfo {pages} {2213844} (\bibinfo {year} {2023})}\BibitemShut {NoStop}%
\bibitem [{\citenamefont {Yang}\ \emph {et~al.}(2022)\citenamefont {Yang}, \citenamefont {Xia}, \citenamefont {Luo}, \citenamefont {Wu}, \citenamefont {Shi},\ and\ \citenamefont {Russell}}]{Yang2022ReconfigurableLiquidsb}%
  \BibitemOpen
  \bibfield  {author} {\bibinfo {author} {\bibfnamefont {Y.}~\bibnamefont {Yang}}, \bibinfo {author} {\bibfnamefont {Z.}~\bibnamefont {Xia}}, \bibinfo {author} {\bibfnamefont {Y.}~\bibnamefont {Luo}}, \bibinfo {author} {\bibfnamefont {Z.}~\bibnamefont {Wu}}, \bibinfo {author} {\bibfnamefont {S.}~\bibnamefont {Shi}},\ and\ \bibinfo {author} {\bibfnamefont {T.~P.}\ \bibnamefont {Russell}},\ }\href {https://doi.org/10.1016/j.supmat.2022.100013} {\bibfield  {journal} {\bibinfo  {journal} {Supramolecular Materials}\ }\textbf {\bibinfo {volume} {1}},\ \bibinfo {pages} {100013} (\bibinfo {year} {2022})}\BibitemShut {NoStop}%
\bibitem [{\citenamefont {Guan}\ \emph {et~al.}(2025)\citenamefont {Guan}, \citenamefont {Liu}, \citenamefont {Li}, \citenamefont {Kwok}, \citenamefont {Ding}, \citenamefont {Jiang},\ and\ \citenamefont {Ngai}}]{Guan2025DynamicInterfaces}%
  \BibitemOpen
  \bibfield  {author} {\bibinfo {author} {\bibfnamefont {X.}~\bibnamefont {Guan}}, \bibinfo {author} {\bibfnamefont {Y.}~\bibnamefont {Liu}}, \bibinfo {author} {\bibfnamefont {L.}~\bibnamefont {Li}}, \bibinfo {author} {\bibfnamefont {M.}~\bibnamefont {Kwok}}, \bibinfo {author} {\bibfnamefont {M.}~\bibnamefont {Ding}}, \bibinfo {author} {\bibfnamefont {H.}~\bibnamefont {Jiang}},\ and\ \bibinfo {author} {\bibfnamefont {T.}~\bibnamefont {Ngai}},\ }\href {https://doi.org/10.1002/advs.202415642} {\bibfield  {journal} {\bibinfo  {journal} {Advanced Science}\ }\textbf {\bibinfo {volume} {12}},\ \bibinfo {pages} {1} (\bibinfo {year} {2025})}\BibitemShut {NoStop}%
\bibitem [{\citenamefont {Fink}\ \emph {et~al.}(2024)\citenamefont {Fink}, \citenamefont {Wu}, \citenamefont {Kim}, \citenamefont {McGlasson}, \citenamefont {Abdelsamie}, \citenamefont {Emrick}, \citenamefont {Sutter‐Fella}, \citenamefont {Ashby}, \citenamefont {Helms},\ and\ \citenamefont {Russell}}]{Fink2024MixedInterfaceb}%
  \BibitemOpen
  \bibfield  {author} {\bibinfo {author} {\bibfnamefont {Z.}~\bibnamefont {Fink}}, \bibinfo {author} {\bibfnamefont {X.}~\bibnamefont {Wu}}, \bibinfo {author} {\bibfnamefont {P.~Y.}\ \bibnamefont {Kim}}, \bibinfo {author} {\bibfnamefont {A.}~\bibnamefont {McGlasson}}, \bibinfo {author} {\bibfnamefont {M.}~\bibnamefont {Abdelsamie}}, \bibinfo {author} {\bibfnamefont {T.}~\bibnamefont {Emrick}}, \bibinfo {author} {\bibfnamefont {C.~M.}\ \bibnamefont {Sutter‐Fella}}, \bibinfo {author} {\bibfnamefont {P.~D.}\ \bibnamefont {Ashby}}, \bibinfo {author} {\bibfnamefont {B.~A.}\ \bibnamefont {Helms}},\ and\ \bibinfo {author} {\bibfnamefont {T.~P.}\ \bibnamefont {Russell}},\ }\href {https://doi.org/10.1002/smll.202308560} {\bibfield  {journal} {\bibinfo  {journal} {Small}\ }\textbf {\bibinfo {volume} {20}} (\bibinfo {year} {2024})}\BibitemShut {NoStop}%
\bibitem [{\citenamefont {Wu}\ \emph {et~al.}(2024)\citenamefont {Wu}, \citenamefont {Xue}, \citenamefont {Fink}, \citenamefont {Helms}, \citenamefont {Ashby}, \citenamefont {Omar},\ and\ \citenamefont {Russell}}]{Wu2024OversaturatingNanoparticleSurfactantsb}%
  \BibitemOpen
  \bibfield  {author} {\bibinfo {author} {\bibfnamefont {X.}~\bibnamefont {Wu}}, \bibinfo {author} {\bibfnamefont {H.}~\bibnamefont {Xue}}, \bibinfo {author} {\bibfnamefont {Z.}~\bibnamefont {Fink}}, \bibinfo {author} {\bibfnamefont {B.~A.}\ \bibnamefont {Helms}}, \bibinfo {author} {\bibfnamefont {P.~D.}\ \bibnamefont {Ashby}}, \bibinfo {author} {\bibfnamefont {A.~K.}\ \bibnamefont {Omar}},\ and\ \bibinfo {author} {\bibfnamefont {T.~P.}\ \bibnamefont {Russell}},\ }\href {https://doi.org/10.1002/anie.202403790} {\bibfield  {journal} {\bibinfo  {journal} {Angewandte Chemie International Edition}\ }\textbf {\bibinfo {volume} {63}},\ \bibinfo {pages} {1} (\bibinfo {year} {2024})}\BibitemShut {NoStop}%
\bibitem [{\citenamefont {Chai}\ \emph {et~al.}(2020)\citenamefont {Chai}, \citenamefont {Hasnain}, \citenamefont {Bahl}, \citenamefont {Wong}, \citenamefont {Li}, \citenamefont {Geissler}, \citenamefont {Kim}, \citenamefont {Jiang}, \citenamefont {Gu}, \citenamefont {Li}, \citenamefont {Lei}, \citenamefont {Helms}, \citenamefont {Russell},\ and\ \citenamefont {Ashby}}]{Chai2020DirectInterfacec}%
  \BibitemOpen
  \bibfield  {author} {\bibinfo {author} {\bibfnamefont {Y.}~\bibnamefont {Chai}}, \bibinfo {author} {\bibfnamefont {J.}~\bibnamefont {Hasnain}}, \bibinfo {author} {\bibfnamefont {K.}~\bibnamefont {Bahl}}, \bibinfo {author} {\bibfnamefont {M.}~\bibnamefont {Wong}}, \bibinfo {author} {\bibfnamefont {D.}~\bibnamefont {Li}}, \bibinfo {author} {\bibfnamefont {P.}~\bibnamefont {Geissler}}, \bibinfo {author} {\bibfnamefont {P.~Y.}\ \bibnamefont {Kim}}, \bibinfo {author} {\bibfnamefont {Y.}~\bibnamefont {Jiang}}, \bibinfo {author} {\bibfnamefont {P.}~\bibnamefont {Gu}}, \bibinfo {author} {\bibfnamefont {S.}~\bibnamefont {Li}}, \bibinfo {author} {\bibfnamefont {D.}~\bibnamefont {Lei}}, \bibinfo {author} {\bibfnamefont {B.~A.}\ \bibnamefont {Helms}}, \bibinfo {author} {\bibfnamefont {T.~P.}\ \bibnamefont {Russell}},\ and\ \bibinfo {author} {\bibfnamefont {P.~D.}\ \bibnamefont {Ashby}},\ }\href {https://doi.org/10.1126/sciadv.abb8675} {\bibfield  {journal} {\bibinfo  {journal} {Science Advances}\ }\textbf {\bibinfo
  {volume} {6}},\ \bibinfo {pages} {1} (\bibinfo {year} {2020})}\BibitemShut {NoStop}%
\bibitem [{\citenamefont {Wu}\ \emph {et~al.}(2025)\citenamefont {Wu}, \citenamefont {Xue}, \citenamefont {Fink}, \citenamefont {Xia}, \citenamefont {Sarma}, \citenamefont {Gan}, \citenamefont {Katsaras}, \citenamefont {Ercius}, \citenamefont {Rad}, \citenamefont {Helms}, \citenamefont {Ashby}, \citenamefont {Omar}, \citenamefont {Collier},\ and\ \citenamefont {Russell}}]{Wu2025Submitted}%
  \BibitemOpen
  \bibfield  {author} {\bibinfo {author} {\bibfnamefont {X.}~\bibnamefont {Wu}}, \bibinfo {author} {\bibfnamefont {H.}~\bibnamefont {Xue}}, \bibinfo {author} {\bibfnamefont {Z.}~\bibnamefont {Fink}}, \bibinfo {author} {\bibfnamefont {Z.}~\bibnamefont {Xia}}, \bibinfo {author} {\bibfnamefont {N.~A.}\ \bibnamefont {Sarma}}, \bibinfo {author} {\bibfnamefont {X.}~\bibnamefont {Gan}}, \bibinfo {author} {\bibfnamefont {J.}~\bibnamefont {Katsaras}}, \bibinfo {author} {\bibfnamefont {P.}~\bibnamefont {Ercius}}, \bibinfo {author} {\bibfnamefont {B.}~\bibnamefont {Rad}}, \bibinfo {author} {\bibfnamefont {B.~A.}\ \bibnamefont {Helms}}, \bibinfo {author} {\bibfnamefont {P.~D.}\ \bibnamefont {Ashby}}, \bibinfo {author} {\bibfnamefont {A.~K.}\ \bibnamefont {Omar}}, \bibinfo {author} {\bibfnamefont {C.~P.}\ \bibnamefont {Collier}},\ and\ \bibinfo {author} {\bibfnamefont {T.~P.}\ \bibnamefont {Russell}},\ }\href@noop {} {\bibfield  {journal} {\bibinfo  {journal} {Submitted}\ } (\bibinfo {year} {2025})}\BibitemShut {NoStop}%
\bibitem [{\citenamefont {Glaser}\ \emph {et~al.}(1988)\citenamefont {Glaser}, \citenamefont {Leikin}, \citenamefont {Chernomordik}, \citenamefont {Pastushenko},\ and\ \citenamefont {Sokirko}}]{Glaser1988ReversiblePores}%
  \BibitemOpen
  \bibfield  {author} {\bibinfo {author} {\bibfnamefont {R.~W.}\ \bibnamefont {Glaser}}, \bibinfo {author} {\bibfnamefont {S.~L.}\ \bibnamefont {Leikin}}, \bibinfo {author} {\bibfnamefont {L.~V.}\ \bibnamefont {Chernomordik}}, \bibinfo {author} {\bibfnamefont {V.~F.}\ \bibnamefont {Pastushenko}},\ and\ \bibinfo {author} {\bibfnamefont {A.~I.}\ \bibnamefont {Sokirko}},\ }\href {https://doi.org/https://doi.org/10.1016/0005-2736(88)90202-7} {\bibfield  {journal} {\bibinfo  {journal} {Biochimica et Biophysica Acta (BBA) - Biomembranes}\ }\textbf {\bibinfo {volume} {940}},\ \bibinfo {pages} {275} (\bibinfo {year} {1988})}\BibitemShut {NoStop}%
\bibitem [{\citenamefont {Zhelev}\ and\ \citenamefont {Needham}(1993)}]{Zhelev1993Tension-stabilizedTension}%
  \BibitemOpen
  \bibfield  {author} {\bibinfo {author} {\bibfnamefont {D.~V.}\ \bibnamefont {Zhelev}}\ and\ \bibinfo {author} {\bibfnamefont {D.}~\bibnamefont {Needham}},\ }\href {https://doi.org/https://doi.org/10.1016/0005-2736(93)90319-U} {\bibfield  {journal} {\bibinfo  {journal} {Biochimica et Biophysica Acta (BBA) - Biomembranes}\ }\textbf {\bibinfo {volume} {1147}},\ \bibinfo {pages} {89} (\bibinfo {year} {1993})}\BibitemShut {NoStop}%
\bibitem [{\citenamefont {Marqusee}\ and\ \citenamefont {Dill}(1986)}]{Marqusee1986SoluteMicellesb}%
  \BibitemOpen
  \bibfield  {author} {\bibinfo {author} {\bibfnamefont {J.~A.}\ \bibnamefont {Marqusee}}\ and\ \bibinfo {author} {\bibfnamefont {K.~A.}\ \bibnamefont {Dill}},\ }\href {https://doi.org/10.1063/1.451621} {\bibfield  {journal} {\bibinfo  {journal} {The Journal of Chemical Physics}\ }\textbf {\bibinfo {volume} {85}},\ \bibinfo {pages} {433} (\bibinfo {year} {1986})}\BibitemShut {NoStop}%
\bibitem [{\citenamefont {Mitragotri}\ \emph {et~al.}(1999)\citenamefont {Mitragotri}, \citenamefont {Johnson}, \citenamefont {Blankschtein},\ and\ \citenamefont {Langer}}]{Mitragotri1999AnBilayers}%
  \BibitemOpen
  \bibfield  {author} {\bibinfo {author} {\bibfnamefont {S.}~\bibnamefont {Mitragotri}}, \bibinfo {author} {\bibfnamefont {M.~E.}\ \bibnamefont {Johnson}}, \bibinfo {author} {\bibfnamefont {D.}~\bibnamefont {Blankschtein}},\ and\ \bibinfo {author} {\bibfnamefont {R.}~\bibnamefont {Langer}},\ }\href {https://doi.org/10.1016/S0006-3495(99)76978-X} {\bibfield  {journal} {\bibinfo  {journal} {Biophysical Journal}\ }\textbf {\bibinfo {volume} {77}},\ \bibinfo {pages} {1268} (\bibinfo {year} {1999})}\BibitemShut {NoStop}%
\bibitem [{\citenamefont {Xiang}\ and\ \citenamefont {Anderson}(1997)}]{Xiang1997PermeabilityTheoryb}%
  \BibitemOpen
  \bibfield  {author} {\bibinfo {author} {\bibfnamefont {T.~X.}\ \bibnamefont {Xiang}}\ and\ \bibinfo {author} {\bibfnamefont {B.~D.}\ \bibnamefont {Anderson}},\ }\href {https://doi.org/10.1016/S0006-3495(97)78661-2} {\bibfield  {journal} {\bibinfo  {journal} {Biophysical Journal}\ }\textbf {\bibinfo {volume} {72}},\ \bibinfo {pages} {223} (\bibinfo {year} {1997})}\BibitemShut {NoStop}%
\bibitem [{\citenamefont {Lifshitz}\ and\ \citenamefont {Slyozov}(1961)}]{Lifshitz1961TheSolutions}%
  \BibitemOpen
  \bibfield  {author} {\bibinfo {author} {\bibfnamefont {I.~M.}\ \bibnamefont {Lifshitz}}\ and\ \bibinfo {author} {\bibfnamefont {V.~V.}\ \bibnamefont {Slyozov}},\ }\href {https://doi.org/https://doi.org/10.1016/0022-3697(61)90054-3} {\bibfield  {journal} {\bibinfo  {journal} {J. Phys. Chem. Solids Pergamon Press}\ }\textbf {\bibinfo {volume} {19}},\ \bibinfo {pages} {35} (\bibinfo {year} {1961})}\BibitemShut {NoStop}%
\bibitem [{\citenamefont {Wagner}(1961)}]{Wagner1961TheorieOstwaldReifung}%
  \BibitemOpen
  \bibfield  {author} {\bibinfo {author} {\bibfnamefont {C.}~\bibnamefont {Wagner}},\ }\href {https://doi.org/10.1002/BBPC.19610650704} {\bibfield  {journal} {\bibinfo  {journal} {Zeitschrift f{\"{u}}r Elektrochemie, Berichte der Bunsengesellschaft f{\"{u}}r physikalische Chemie}\ }\textbf {\bibinfo {volume} {65}},\ \bibinfo {pages} {581} (\bibinfo {year} {1961})}\BibitemShut {NoStop}%
\bibitem [{\citenamefont {Kozlov}\ \emph {et~al.}(1989)\citenamefont {Kozlov}, \citenamefont {Leikin}, \citenamefont {Chernomordik}, \citenamefont {Markin},\ and\ \citenamefont {Chizmadzhev}}]{Kozlov1989EuropeanContents}%
  \BibitemOpen
  \bibfield  {author} {\bibinfo {author} {\bibfnamefont {M.~M.}\ \bibnamefont {Kozlov}}, \bibinfo {author} {\bibfnamefont {S.~L.}\ \bibnamefont {Leikin}}, \bibinfo {author} {\bibfnamefont {L.~V.}\ \bibnamefont {Chernomordik}}, \bibinfo {author} {\bibfnamefont {V.~S.}\ \bibnamefont {Markin}},\ and\ \bibinfo {author} {\bibfnamefont {Y.~A.}\ \bibnamefont {Chizmadzhev}},\ }\href {https://doi.org/10.1007/BF00254765} {\bibfield  {journal} {\bibinfo  {journal} {Eur Biophys J}\ }\textbf {\bibinfo {volume} {17}},\ \bibinfo {pages} {121} (\bibinfo {year} {1989})}\BibitemShut {NoStop}%
\bibitem [{\citenamefont {Kozlovsky}\ \emph {et~al.}(2002)\citenamefont {Kozlovsky}, \citenamefont {Chernomordik},\ and\ \citenamefont {Kozlov}}]{Kozlovsky2002LipidDiaphragm}%
  \BibitemOpen
  \bibfield  {author} {\bibinfo {author} {\bibfnamefont {Y.}~\bibnamefont {Kozlovsky}}, \bibinfo {author} {\bibfnamefont {L.~V.}\ \bibnamefont {Chernomordik}},\ and\ \bibinfo {author} {\bibfnamefont {M.~M.}\ \bibnamefont {Kozlov}},\ }\href {https://doi.org/10.1016/S0006-3495(02)75274-0} {\bibfield  {journal} {\bibinfo  {journal} {Biophysical Journal}\ }\textbf {\bibinfo {volume} {83}},\ \bibinfo {pages} {2634} (\bibinfo {year} {2002})}\BibitemShut {NoStop}%
\bibitem [{\citenamefont {Kozlovsky}\ and\ \citenamefont {Kozlov}(2002)}]{Kozlovsky2002StalkCrisis}%
  \BibitemOpen
  \bibfield  {author} {\bibinfo {author} {\bibfnamefont {Y.}~\bibnamefont {Kozlovsky}}\ and\ \bibinfo {author} {\bibfnamefont {M.~M.}\ \bibnamefont {Kozlov}},\ }\href {https://doi.org/10.1016/S0006-3495(02)75450-7} {\bibfield  {journal} {\bibinfo  {journal} {Biophysical Journal}\ }\textbf {\bibinfo {volume} {82}},\ \bibinfo {pages} {882} (\bibinfo {year} {2002})}\BibitemShut {NoStop}%
\bibitem [{\citenamefont {Markin}\ \emph {et~al.}(1984)\citenamefont {Markin}, \citenamefont {Kozlov},\ and\ \citenamefont {Borovjagin}}]{Markin1984OnMechanism}%
  \BibitemOpen
  \bibfield  {author} {\bibinfo {author} {\bibfnamefont {V.~S.}\ \bibnamefont {Markin}}, \bibinfo {author} {\bibfnamefont {M.~M.}\ \bibnamefont {Kozlov}},\ and\ \bibinfo {author} {\bibfnamefont {V.~L.}\ \bibnamefont {Borovjagin}},\ }\href {https://pubmed.ncbi.nlm.nih.gov/6510702/} {\bibfield  {journal} {\bibinfo  {journal} {Gen. Physiol. Biophys}\ }\textbf {\bibinfo {volume} {5}},\ \bibinfo {pages} {361} (\bibinfo {year} {1984})}\BibitemShut {NoStop}%
\bibitem [{\citenamefont {Markin}\ and\ \citenamefont {Albanesi}(2002)}]{Markin2002MembraneRevisited}%
  \BibitemOpen
  \bibfield  {author} {\bibinfo {author} {\bibfnamefont {V.~S.}\ \bibnamefont {Markin}}\ and\ \bibinfo {author} {\bibfnamefont {J.~P.}\ \bibnamefont {Albanesi}},\ }\href {https://doi.org/10.1016/S0006-3495(02)75432-5} {\bibfield  {journal} {\bibinfo  {journal} {Biophysical Journal}\ }\textbf {\bibinfo {volume} {82}},\ \bibinfo {pages} {693} (\bibinfo {year} {2002})}\BibitemShut {NoStop}%
\bibitem [{\citenamefont {Ting}\ \emph {et~al.}(2011)\citenamefont {Ting}, \citenamefont {Appel{\"{o}}},\ and\ \citenamefont {Wang}}]{Ting2011MinimumRupture}%
  \BibitemOpen
  \bibfield  {author} {\bibinfo {author} {\bibfnamefont {C.~L.}\ \bibnamefont {Ting}}, \bibinfo {author} {\bibfnamefont {D.}~\bibnamefont {Appel{\"{o}}}},\ and\ \bibinfo {author} {\bibfnamefont {Z.~G.}\ \bibnamefont {Wang}},\ }\href {https://doi.org/10.1103/PHYSREVLETT.106.168101/FIGURES/3/MEDIUM} {\bibfield  {journal} {\bibinfo  {journal} {Physical Review Letters}\ }\textbf {\bibinfo {volume} {106}},\ \bibinfo {pages} {168101} (\bibinfo {year} {2011})}\BibitemShut {NoStop}%
\bibitem [{\citenamefont {Frolov}\ \emph {et~al.}(2006)\citenamefont {Frolov}, \citenamefont {Chizmadzhev}, \citenamefont {Cohen},\ and\ \citenamefont {Zimmerberg}}]{Frolov2006EntropicNanodomains}%
  \BibitemOpen
  \bibfield  {author} {\bibinfo {author} {\bibfnamefont {V.~A.~J.}\ \bibnamefont {Frolov}}, \bibinfo {author} {\bibfnamefont {Y.~A.}\ \bibnamefont {Chizmadzhev}}, \bibinfo {author} {\bibfnamefont {F.~S.}\ \bibnamefont {Cohen}},\ and\ \bibinfo {author} {\bibfnamefont {J.}~\bibnamefont {Zimmerberg}},\ }\href {https://doi.org/10.1529/biophysj.105.068502} {\bibfield  {journal} {\bibinfo  {journal} {Biophysical Journal}\ }\textbf {\bibinfo {volume} {91}},\ \bibinfo {pages} {189} (\bibinfo {year} {2006})}\BibitemShut {NoStop}%
\bibitem [{\citenamefont {Yu}\ and\ \citenamefont {Ko{\v{s}}}(2025)}]{Yu2025Pattern}%
  \BibitemOpen
  \bibfield  {author} {\bibinfo {author} {\bibfnamefont {Q.}~\bibnamefont {Yu}}\ and\ \bibinfo {author} {\bibfnamefont {A.}~\bibnamefont {Ko{\v{s}}}},\ }\href {https://doi.org/10.1039/d5sm00276a} {\bibfield  {journal} {\bibinfo  {journal} {Soft Matter}\ ,\ \bibinfo {pages} {4288}} (\bibinfo {year} {2025})}\BibitemShut {NoStop}%
\bibitem [{\citenamefont {Zwanzig}(1992)}]{Zwanzig1992DiffusionBarrier}%
  \BibitemOpen
  \bibfield  {author} {\bibinfo {author} {\bibfnamefont {R.}~\bibnamefont {Zwanzig}},\ }\href {https://doi.org/10.1021/j100189a004} {\bibfield  {journal} {\bibinfo  {journal} {The Journal of Physical Chemistry}\ }\textbf {\bibinfo {volume} {96}},\ \bibinfo {pages} {3926} (\bibinfo {year} {1992})}\BibitemShut {NoStop}%
\bibitem [{\citenamefont {Kalinay}\ and\ \citenamefont {Percus}(2006)}]{Kalinay2006CorrectionsEquation}%
  \BibitemOpen
  \bibfield  {author} {\bibinfo {author} {\bibfnamefont {P.}~\bibnamefont {Kalinay}}\ and\ \bibinfo {author} {\bibfnamefont {J.~K.}\ \bibnamefont {Percus}},\ }\href {https://doi.org/10.1103/PhysRevE.74.041203} {\bibfield  {journal} {\bibinfo  {journal} {Physical Review E}\ }\textbf {\bibinfo {volume} {74}},\ \bibinfo {pages} {041203} (\bibinfo {year} {2006})}\BibitemShut {NoStop}%
\bibitem [{\citenamefont {Kalinay}\ and\ \citenamefont {Percus}(2008)}]{Kalinay2008ApproximationsEquation}%
  \BibitemOpen
  \bibfield  {author} {\bibinfo {author} {\bibfnamefont {P.}~\bibnamefont {Kalinay}}\ and\ \bibinfo {author} {\bibfnamefont {J.~K.}\ \bibnamefont {Percus}},\ }\href {https://doi.org/10.1103/PhysRevE.78.021103} {\bibfield  {journal} {\bibinfo  {journal} {Physical Review E}\ }\textbf {\bibinfo {volume} {78}},\ \bibinfo {pages} {021103} (\bibinfo {year} {2008})}\BibitemShut {NoStop}%
\bibitem [{\citenamefont {Martens}\ \emph {et~al.}(2011)\citenamefont {Martens}, \citenamefont {Schmid}, \citenamefont {Schimansky-Geier},\ and\ \citenamefont {H{\"{a}}nggi}}]{Martens2011EntropicEquation}%
  \BibitemOpen
  \bibfield  {author} {\bibinfo {author} {\bibfnamefont {S.}~\bibnamefont {Martens}}, \bibinfo {author} {\bibfnamefont {G.}~\bibnamefont {Schmid}}, \bibinfo {author} {\bibfnamefont {L.}~\bibnamefont {Schimansky-Geier}},\ and\ \bibinfo {author} {\bibfnamefont {P.}~\bibnamefont {H{\"{a}}nggi}},\ }\href {https://doi.org/10.1103/PhysRevE.83.051135} {\bibfield  {journal} {\bibinfo  {journal} {Physical Review E}\ }\textbf {\bibinfo {volume} {83}},\ \bibinfo {pages} {051135} (\bibinfo {year} {2011})}\BibitemShut {NoStop}%
\bibitem [{\citenamefont {Dorfman}\ and\ \citenamefont {Yariv}(2014)}]{Dorfman2014AssessingEquation}%
  \BibitemOpen
  \bibfield  {author} {\bibinfo {author} {\bibfnamefont {K.~D.}\ \bibnamefont {Dorfman}}\ and\ \bibinfo {author} {\bibfnamefont {E.}~\bibnamefont {Yariv}},\ }\href {https://doi.org/10.1063/1.4890740} {\bibfield  {journal} {\bibinfo  {journal} {The Journal of Chemical Physics}\ }\textbf {\bibinfo {volume} {141}},\ \bibinfo {pages} {1} (\bibinfo {year} {2014})}\BibitemShut {NoStop}%
\bibitem [{\citenamefont {Valero~Valdes}\ and\ \citenamefont {Herrera~Guzman}(2014)}]{ValeroValdes2014Fick-JacobsCurves}%
  \BibitemOpen
  \bibfield  {author} {\bibinfo {author} {\bibfnamefont {C.}~\bibnamefont {Valero~Valdes}}\ and\ \bibinfo {author} {\bibfnamefont {R.}~\bibnamefont {Herrera~Guzman}},\ }\href {https://doi.org/10.1103/PhysRevE.90.052141} {\bibfield  {journal} {\bibinfo  {journal} {Physical Review E}\ }\textbf {\bibinfo {volume} {90}},\ \bibinfo {pages} {052141} (\bibinfo {year} {2014})}\BibitemShut {NoStop}%
\bibitem [{\citenamefont {Ledesma-Dur{\'{a}}n}\ \emph {et~al.}(2016)\citenamefont {Ledesma-Dur{\'{a}}n}, \citenamefont {Hern{\'{a}}ndez-Hern{\'{a}}ndez},\ and\ \citenamefont {Santamar{\'{i}}a-Holek}}]{Ledesma-Duran2016GeneralizedConditions}%
  \BibitemOpen
  \bibfield  {author} {\bibinfo {author} {\bibfnamefont {A.}~\bibnamefont {Ledesma-Dur{\'{a}}n}}, \bibinfo {author} {\bibfnamefont {S.~I.}\ \bibnamefont {Hern{\'{a}}ndez-Hern{\'{a}}ndez}},\ and\ \bibinfo {author} {\bibfnamefont {I.}~\bibnamefont {Santamar{\'{i}}a-Holek}},\ }\href {https://doi.org/10.1021/acs.jpcc.5b12145} {\bibfield  {journal} {\bibinfo  {journal} {The Journal of Physical Chemistry C}\ }\textbf {\bibinfo {volume} {120}},\ \bibinfo {pages} {7810} (\bibinfo {year} {2016})}\BibitemShut {NoStop}%
\bibitem [{\citenamefont {Langford}\ and\ \citenamefont {Omar}(2025)}]{Langford2025TheMatterb}%
  \BibitemOpen
  \bibfield  {author} {\bibinfo {author} {\bibfnamefont {L.}~\bibnamefont {Langford}}\ and\ \bibinfo {author} {\bibfnamefont {A.~K.}\ \bibnamefont {Omar}},\ }\href {https://arxiv.org/abs/2407.06462} {\bibfield  {journal} {\bibinfo  {journal} {arXiv:2407.06462}\ ,\ \bibinfo {pages} {1}} (\bibinfo {year} {2025})}\BibitemShut {NoStop}%
\bibitem [{\citenamefont {Desai}\ and\ \citenamefont {Kapral}(2009)}]{Desai2009DynamicsStructures}%
  \BibitemOpen
  \bibfield  {author} {\bibinfo {author} {\bibfnamefont {R.~C.}\ \bibnamefont {Desai}}\ and\ \bibinfo {author} {\bibfnamefont {R.}~\bibnamefont {Kapral}},\ }\href {https://doi.org/10.1017/CBO9780511609725.013} {\emph {\bibinfo {title} {{Dynamics of Self-Organized and Self-Assembled Structures}}}},\ \bibinfo {edition} {1st}\ ed.\ (\bibinfo  {publisher} {Cambridge University Press},\ \bibinfo {address} {Cambridge},\ \bibinfo {year} {2009})\ pp.\ \bibinfo {pages} {87--95}\BibitemShut {NoStop}%
\bibitem [{\citenamefont {Arnold}\ \emph {et~al.}(2023)\citenamefont {Arnold}, \citenamefont {Gubbala},\ and\ \citenamefont {Takatori}}]{Arnold2023ActiveDomains}%
  \BibitemOpen
  \bibfield  {author} {\bibinfo {author} {\bibfnamefont {D.~P.}\ \bibnamefont {Arnold}}, \bibinfo {author} {\bibfnamefont {A.}~\bibnamefont {Gubbala}},\ and\ \bibinfo {author} {\bibfnamefont {S.~C.}\ \bibnamefont {Takatori}},\ }\href {https://doi.org/10.1103/PHYSREVLETT.131.128402/MOV{\_}S6.MP4} {\bibfield  {journal} {\bibinfo  {journal} {Physical Review Letters}\ }\textbf {\bibinfo {volume} {131}},\ \bibinfo {pages} {128402} (\bibinfo {year} {2023})}\BibitemShut {NoStop}%
\bibitem [{\citenamefont {Marqusee}(1984{\natexlab{a}})}]{Marqusee1984DynamicsDimensionsb}%
  \BibitemOpen
  \bibfield  {author} {\bibinfo {author} {\bibfnamefont {J.~A.}\ \bibnamefont {Marqusee}},\ }\href {https://doi.org/10.1063/1.447698} {\bibfield  {journal} {\bibinfo  {journal} {J. Chem. Phys}\ }\textbf {\bibinfo {volume} {81}},\ \bibinfo {pages} {976} (\bibinfo {year} {1984}{\natexlab{a}})}\BibitemShut {NoStop}%
\bibitem [{\citenamefont {Marqusee}\ and\ \citenamefont {Ross}(1984)}]{Marqusee1984Theory}%
  \BibitemOpen
  \bibfield  {author} {\bibinfo {author} {\bibfnamefont {J.~A.}\ \bibnamefont {Marqusee}}\ and\ \bibinfo {author} {\bibfnamefont {J.}~\bibnamefont {Ross}},\ }\href {https://doi.org/10.1063/1.446427} {\bibfield  {journal} {\bibinfo  {journal} {J. Chem. Phys}\ }\textbf {\bibinfo {volume} {80}},\ \bibinfo {pages} {536} (\bibinfo {year} {1984})}\BibitemShut {NoStop}%
\bibitem [{\citenamefont {Strutt}(1879)}]{Strutt1879I.Drops}%
  \BibitemOpen
  \bibfield  {author} {\bibinfo {author} {\bibfnamefont {J.}~\bibnamefont {Strutt}},\ }\href {https://doi.org/10.1098/rspl.1878.0146} {\bibfield  {journal} {\bibinfo  {journal} {Proceedings of the Royal Society of London}\ }\textbf {\bibinfo {volume} {28}},\ \bibinfo {pages} {405} (\bibinfo {year} {1879})}\BibitemShut {NoStop}%
\bibitem [{\citenamefont {Rayleigh}(1882)}]{Rayleigh1882FurtherJets}%
  \BibitemOpen
  \bibfield  {author} {\bibinfo {author} {\bibfnamefont {L.}~\bibnamefont {Rayleigh}},\ }\href {https://doi.org/https://doi.org/10.1098/rspl.1882.0026} {\bibfield  {journal} {\bibinfo  {journal} {Proceedings of the Royal Society of London}\ }\textbf {\bibinfo {volume} {34}},\ \bibinfo {pages} {130} (\bibinfo {year} {1882})}\BibitemShut {NoStop}%
\bibitem [{\citenamefont {Blaschke}\ \emph {et~al.}(2012)\citenamefont {Blaschke}, \citenamefont {Lapp}, \citenamefont {Hof},\ and\ \citenamefont {Vollmer}}]{Blaschke2012BreathDroplets}%
  \BibitemOpen
  \bibfield  {author} {\bibinfo {author} {\bibfnamefont {J.}~\bibnamefont {Blaschke}}, \bibinfo {author} {\bibfnamefont {T.}~\bibnamefont {Lapp}}, \bibinfo {author} {\bibfnamefont {B.}~\bibnamefont {Hof}},\ and\ \bibinfo {author} {\bibfnamefont {J.}~\bibnamefont {Vollmer}},\ }\href {https://doi.org/10.1103/PhysRevLett.109.068701} {\bibfield  {journal} {\bibinfo  {journal} {Physical Review Letters}\ }\textbf {\bibinfo {volume} {109}},\ \bibinfo {pages} {068701} (\bibinfo {year} {2012})}\BibitemShut {NoStop}%
\bibitem [{\citenamefont {Meakin}(1992)}]{Meakin1992DropletCoalescence}%
  \BibitemOpen
  \bibfield  {author} {\bibinfo {author} {\bibfnamefont {P.}~\bibnamefont {Meakin}},\ }\href {https://doi.org/10.1088/0034-4885/55/2/002} {\bibfield  {journal} {\bibinfo  {journal} {Reports on Progress in Physics}\ }\textbf {\bibinfo {volume} {55}},\ \bibinfo {pages} {157} (\bibinfo {year} {1992})}\BibitemShut {NoStop}%
\bibitem [{\citenamefont {Marqusee}(1984{\natexlab{b}})}]{Marqusee1984DynamicsDimensions}%
  \BibitemOpen
  \bibfield  {author} {\bibinfo {author} {\bibfnamefont {J.~A.}\ \bibnamefont {Marqusee}},\ }\href {https://doi.org/10.1063/1.447698} {\bibfield  {journal} {\bibinfo  {journal} {J. Chem. Phys}\ }\textbf {\bibinfo {volume} {81}},\ \bibinfo {pages} {976} (\bibinfo {year} {1984}{\natexlab{b}})}\BibitemShut {NoStop}%
\bibitem [{\citenamefont {Stanich}\ \emph {et~al.}(2013)\citenamefont {Stanich}, \citenamefont {Honerkamp-Smith}, \citenamefont {Garb{\`{e}}~Putzel}, \citenamefont {Warth}, \citenamefont {Lamprecht}, \citenamefont {Mandal}, \citenamefont {Mann}, \citenamefont {Hua},\ and\ \citenamefont {Keller}}]{Stanich2013CoarseningMembranes}%
  \BibitemOpen
  \bibfield  {author} {\bibinfo {author} {\bibfnamefont {C.~A.}\ \bibnamefont {Stanich}}, \bibinfo {author} {\bibfnamefont {A.~R.}\ \bibnamefont {Honerkamp-Smith}}, \bibinfo {author} {\bibfnamefont {G.}~\bibnamefont {Garb{\`{e}}~Putzel}}, \bibinfo {author} {\bibfnamefont {C.~S.}\ \bibnamefont {Warth}}, \bibinfo {author} {\bibfnamefont {A.~K.}\ \bibnamefont {Lamprecht}}, \bibinfo {author} {\bibfnamefont {P.}~\bibnamefont {Mandal}}, \bibinfo {author} {\bibfnamefont {E.}~\bibnamefont {Mann}}, \bibinfo {author} {\bibfnamefont {T.-A.~D.}\ \bibnamefont {Hua}},\ and\ \bibinfo {author} {\bibfnamefont {S.~L.}\ \bibnamefont {Keller}},\ }\href {https://doi.org/10.1016/j.bpj.2013.06.013} {\bibfield  {journal} {\bibinfo  {journal} {Biophysj}\ }\textbf {\bibinfo {volume} {105}},\ \bibinfo {pages} {444} (\bibinfo {year} {2013})}\BibitemShut {NoStop}%
\bibitem [{\citenamefont {Schlicht}\ and\ \citenamefont {Zagnoni}(2015)}]{Schlicht2015Droplet-interface-bilayerNetworks}%
  \BibitemOpen
  \bibfield  {author} {\bibinfo {author} {\bibfnamefont {B.}~\bibnamefont {Schlicht}}\ and\ \bibinfo {author} {\bibfnamefont {M.}~\bibnamefont {Zagnoni}},\ }\href {https://doi.org/10.1038/srep09951} {\bibfield  {journal} {\bibinfo  {journal} {Scientific Reports}\ }\textbf {\bibinfo {volume} {5}},\ \bibinfo {pages} {1} (\bibinfo {year} {2015})}\BibitemShut {NoStop}%
\bibitem [{\citenamefont {Huang}\ \emph {et~al.}(2022)\citenamefont {Huang}, \citenamefont {Fuller},\ and\ \citenamefont {Chandran~Suja}}]{Huang2022PhysicochemicalBilayers}%
  \BibitemOpen
  \bibfield  {author} {\bibinfo {author} {\bibfnamefont {Y.}~\bibnamefont {Huang}}, \bibinfo {author} {\bibfnamefont {G.}~\bibnamefont {Fuller}},\ and\ \bibinfo {author} {\bibfnamefont {V.}~\bibnamefont {Chandran~Suja}},\ }\href {https://doi.org/10.1016/j.cis.2022.102666} {\bibfield  {journal} {\bibinfo  {journal} {Advances in Colloid and Interface Science}\ }\textbf {\bibinfo {volume} {304}},\ \bibinfo {pages} {1} (\bibinfo {year} {2022})}\BibitemShut {NoStop}%
\bibitem [{\citenamefont {Allen}\ \emph {et~al.}(2022)\citenamefont {Allen}, \citenamefont {Albon},\ and\ \citenamefont {Elani}}]{Allen2022Layer-by-layermulti-DIBs}%
  \BibitemOpen
  \bibfield  {author} {\bibinfo {author} {\bibfnamefont {M.~E.}\ \bibnamefont {Allen}}, \bibinfo {author} {\bibfnamefont {J.}~\bibnamefont {Albon}},\ and\ \bibinfo {author} {\bibfnamefont {Y.}~\bibnamefont {Elani}},\ }\href {https://doi.org/10.1039/D1CC05155E} {\bibfield  {journal} {\bibinfo  {journal} {Chemical Communications}\ }\textbf {\bibinfo {volume} {58}},\ \bibinfo {pages} {60} (\bibinfo {year} {2022})}\BibitemShut {NoStop}%
\bibitem [{\citenamefont {Lindell}(1993)}]{Lindell1993DeltaMethod}%
  \BibitemOpen
  \bibfield  {author} {\bibinfo {author} {\bibfnamefont {I.~V.}\ \bibnamefont {Lindell}},\ }\href {https://doi.org/10.1119/1.17238} {\bibfield  {journal} {\bibinfo  {journal} {American Journal of Physics}\ }\textbf {\bibinfo {volume} {61}},\ \bibinfo {pages} {438} (\bibinfo {year} {1993})}\BibitemShut {NoStop}%
\end{thebibliography}
\end{document}